\documentclass[traditabstract]{aa} 
\usepackage{time}
\usepackage{color}
\usepackage{natbib}
\usepackage[percent]{overpic}
\usepackage{amstext,amsmath,mathtools}
\usepackage{graphicx}
\usepackage{epsfig}
\usepackage{epstopdf}
\usepackage[colorlinks=true,urlcolor=blue,citecolor=blue,linkcolor=blue]{hyperref}

\usepackage{textcomp}
\usepackage{txfonts}
\usepackage{url}

\begin{document}
\titlerunning{Opposite polarity field and convective downflows in a sunspot penumbra}
\title{Opposite polarity field with convective downflow and its relation to magnetic spines in a sunspot penumbra}

\author{G.B Scharmer\inst{1,2}
\and
J. de la Cruz Rodriguez\inst{3}
\and
P. S\"utterlin\inst{1,2}
\and 
V.M.J. Henriques\inst{1,2}}

\institute{Institute for Solar Physics, Royal Swedish Academy of Sciences,
AlbaNova University Center, SE 106\,91 Stockholm, Sweden \and
Stockholm Observatory, Dept. of Astronomy, Stockholm University,
AlbaNova University Center, SE 106\,91 Stockholm, Sweden \and
Department of Physics and Astronomy, Uppsala University, Box 516, SE 751\,20 Uppsala, Sweden}
\date{Draft: \now\ \today}
\frenchspacing

\abstract{We discuss NICOLE inversions of \ion{Fe}{I} 630.15~nm and 630.25~nm Stokes spectra from a sunspot penumbra recorded with the CRISP imaging spectropolarimeter on the Swedish 1-m Solar Telescope at a spatial resolution close to 0\farcs15. We report on narrow radially extended lanes of opposite polarity field, located at the boundaries between areas of relatively horizontal magnetic field (the intra-spines) and much more vertical field (the spines). These lanes harbor convective downflows of about 1~km~s$^{-1}$. The locations of these downflows close to the spines agree with predictions from the convective gap model (the ``gappy penumbra'') proposed six years ago, and more recent three-dimensional magnetohydrodynamic simulations. We also confirm the existence of strong convective flows throughout the entire penumbra, showing the expected correlation between temperature and vertical velocity, and having vertical root mean square velocities of about 1.2~km~s$^{-1}$.
}


\maketitle

\section{Introduction}
The strong, nearly horizontal and radial outflows discovered more than 100 years ago  \citep{1909MNRAS..69..454E}, have inspired 1-dimensional (1D) numerical modeling of the penumbral filaments as flux tubes, with a siphon flow indirectly driven by the assumed difference in magnetic field strength between the ascending (in the penumbra) and descending (outside the penumbra) parts of magnetic flux tubes \citep{1968MitAG..25..194M,
1989ApJ...337..977M, 1997Natur.390..485M}. \citet{1998ApJ...493L.121S, 1998A&A...337..897S} instead modeled these flows as transient flows triggered by heating of thin flux tubes embedded in a convectively unstable penumbral atmosphere of given properties. 

In parallel to the development of the theoretical models described above, spectropolarimetric data, initially at only 1\arcsec spatial resolution, were used to develop a plethora of 1-component (with vertical gradients) and 2-component (without vertical gradients) empirical models of flux tubes and similar models of the penumbral magnetic field and flows \citep[cf.][]{1993A&A...275..283S, 2002A&A...381..668S, 2002A&A...393..305M,2003A&A...403L..47B,2004A&A...427..319B, 2005A&A...436..333B, 2007A&A...471..967B, 2007ApJ...666L.133B, 2010ApJ...720.1417P}. The ability of these simplified models to represent observed data (most of which were recorded at low spatial resolution), as well as their apparent consistency with the simulations of \citet{1998A&A...337..897S, 1998ApJ...493L.121S} was interpreted to validate both the theoretical and observationally based models.

However, neither of the above models can provide the overall heat flux needed to explain the radiative losses of the penumbra \citep{2003A&A...411..257S, 2006A&A...447..343S}. For this, efficient convection is needed, with \emph{vertical} root mean square (RMS) velocities comparable to those of the quiet Sun. 
\citet{2006A&A...447..343S}, \citet{2006A&A...460..605S}, and \citet{2009SSRv..144..229S} concluded that the penumbral filamentary structure and complex magnetic field topology must be the result of convection opening up radially aligned essentially field-free gaps, just below the visible surface. \citet{2005ApJ...622.1292S} and \citet{2009A&A...508..963S} proposed to explain observed penumbral Stokes profiles in terms of spatially \emph{unresolved} fluctuations in velocity and magnetic field, the micro-structured magnetic atmospheres (MISMAs). 

Recent simulations \citep{2007ApJ...669.1390H, 2009Sci...325..171R, 2011ApJ...729....5R, 2012ApJ...750...62R} clearly support the presence of strong penumbral convection at \emph{observable} scales, and lead to the surprising conclusion that the Evershed flow corresponds to the horizontal and radially outward component of this convection \citep{2008ApJ...677L.149S, 2009Sci...325..171R}. This interpretation of penumbral fine structure is however not without controversy, the main argument being that observational evidence for convective \emph{downflows} well inside the outer boundary of the penumbra is missing \citep{2009A&A...508.1453F, 2011arXiv1107.2586F}. Recently, such evidence was reported in the \ion{C}{I} 538.03~nm line \citep{2011Sci...333..316S, 2011ApJ...734L..18J} and the \ion{Fe}{I} 630.15~nm line \citep{2012A&A...540A..19S}, based on observations with the Swedish 1-m Solar Telescope \citep[SST; ][]{2003SPIE.4853..341S} and its imaging spectropolarimeter CRISP \citep{2006A&A...447.1111S,2008ApJ...689L..69S}. Synthetic spectra calculated from simulations \citep{2011ApJ...729....5R} confirm that such downflows should be visible in the \ion{C}{I} 538.03~nm line at the spatial resolution of the SST \citep{2011ApJ...739...35B}.

An important question in the context of the present work concerns the existence of opposite polarity field in the penumbra. The existence of such field can be taken as strong evidence of (convective) downflows, dragging down some of the field. \citet{2011arXiv1107.2586F} investigated spectropolarimetric Hinode data from a sunspot close to disk center and concluded that 36\% of all penumbral downflows give rise to Stokes $V$ profiles showing an additional 3rd lobe in the red wing of the 630.25~nm line, indicating a hidden opposite polarity. However, nearly all the downflow pixels found by him are located at the outermost parts of the penumbra, where downflows are ubiquitous. The question of whether or not there is convection in the main body of the penumbra can therefore not be addressed with these data. Recently, \citet{2012arXiv1211.5776S}\footnote{Version 1 of this arXiv entry is an earlier version of the present manuscript originally submitted on 5 Jul. 2012.} and \citet{2013A&A...549L...4R} reported the detection of opposite polarity field also in the interior of the penumbra, thus providing independent support to earlier evidence demonstrating that the penumbra is fully convective \citep{2011Sci...333..316S, 2011ApJ...734L..18J,2012A&A...540A..19S}. 

In the present paper, we apply Stokes inversions jointly to \ion{Fe}{I} 630.15~nm and 630.25~nm spectra, recorded with CRISP/SST close to the diffraction limit of 0\farcs15. We describe the data reduction, the rationale for straylight compensation, and the determination of the CRISP transmission profile in Sect.~2. In Sect.~3, we describe the NICOLE inversions \citep{2000ApJ...530..977S,2011A&A...529A..37S}, allowing for gradients in both the line-of-sight (LOS) velocities and the magnetic field, and using an individual spectral transmission profile for each pixel. We validate the inversions of our Stokes spectra in Sect.~4, and use the inversions in Sect.~5 to establish the \emph{locations} of opposite polarity patches and their association with magnetic spines in the deep layers of a sunspot penumbra. We then analyze the azimuthal variations of the observed LOS velocities at the opposite polarity patches, and draw conclusions about their vertical and radial velocities.

\section{Observations and Data Reduction}
The observed data analyzed are of a reasonably symmetric sunspot at approximately 15\degr{} heliocentric distance, obtained on 23 May 2010 with the SST and its imaging spectropolarimeter CRISP. These data have been described before \citep{2011Sci...333..316S, 2012A&A...540A..19S}, but here we extend the analysis by including 15 wavelengths scanned in the 630.2~nm line and a continuum wavelength 60~pm redward of this line, in addition to the 15 wavelengths scanned for the \ion{Fe}{I} 630.15 line and analyzed by \citet{2012A&A...540A..19S}. A complete scan of the two \ion{Fe}{I} lines required only 16.5~s observing time.

We first compensated the images for darks and the ``raw'' flats similar to as described by \citet{2011A&A...534A..45S}, and then compensated the images for residual low-order aberrations with the multi-object multi-frame blind deconvolution (MOMFBD) code of \citet{2005SoPh..228..191V}. This was followed by a small-scale dewarping (small-scale geometric distortion correction) of the images to remove remaining alignment errors from high-altitude seeing at arc second scale, which cannot be compensated for by the $4\arcsec\times4\arcsec$ subfield MOMFBD processing (Henriques 2012). 

We also re-determined the polarization properties of the SST, using calibration data obtained with a 1-m rotating linear polarizer mounted in front of the SST in May, June and October 2011. After demodulation of the MOMFBD restored images with an improved version of the polarization model of \citet{selbing05sst}, we checked the data for residual cross-talk from $I$ to $Q$, $U$ and $V$ and from $V$ to $Q$ and $U$ using methods similar to those described by \citet{2002A&A...381..668S}. The cross-talk found was 0.3\% or less from $I$ to $Q$, $U$ and $V$, 2.7\% from $V$ to $U$, and much smaller from $V$ to $Q$. These cross-talks were compensated for.

The wavelength scale (used for establishing the LOS velocity reference) was set by comparing the observed line profiles for granulation surrounding the sunspot and void of strong fields, with those of 3D simulations \citep{2011A&A...528A.113D}. With this calibration, we found an average velocity of the umbra to be 60~m~s$^{-1}$ at $\tau_c=$~0.95 and -80~m~s$^{-1}$ at $\tau_c=$~0.01 from the inversions (described in Sect.~3), such that this wavelength scale can be considered quite accurate.
\begin{figure}
\center
\includegraphics[bb=55 70 570 710, angle=-90, width=0.4\textwidth,clip]{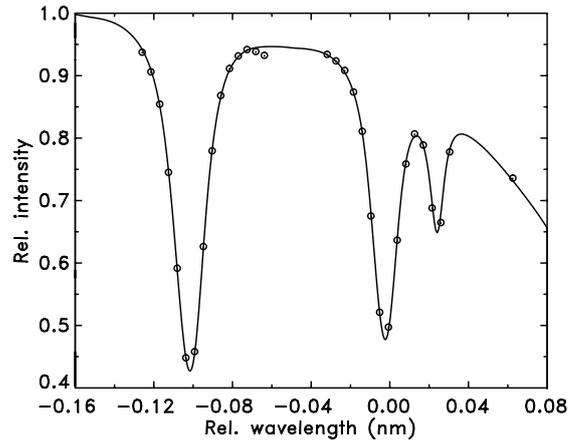}
 \caption{
Observed flatfield spectrum for a pixel near the center of the FOV (circles).
The solid line shows the FTS atlas spectrum for the same wavelength range, convolved with the fitted CRISP transmission profile and multiplied with the fitted pre-filter transmission curve. The strength and position of the telluric blend have been determined in a separate fit (see text).
}
\label{fig:CRISP}
\end{figure}

\subsection{Determination of the CRISP transmission profile}
We derived the transmission profile of CRISP and the pre-filter at each pixel, as well as a ``true'' flat-field image at each wavelength, by comparing the observed flat-field spectra, obtained close to disk center, to the FTS atlas spectrum \citep[Brault \& Neckel 1987; see][]{1999SoPh..184..421N}. Our fits go beyond the determination of ``cavity maps'' \citep{2008ApJ...689L..69S}, which characterize the transmission profile only in terms of a pure wavelength shift, by allowing also for variable \emph{asymmetries} in the transmission profile over the field-of-view (FOV). Such asymmetries come from \emph{relative} wavelength shifts of the transmission profiles of the two etalons. We implemented our fitting procedure using analytical expressions for the transmission profiles of the two etalons \citep[cf.][]{2006A&A...447.1111S}. The free parameters of the fit are the wavelength shifts of the two etalons, the reflectivity of the high-resolution (HR) etalon and the parameters of an assumed linear variation of the pre-filter transmission with wavelength (correcting for any deviations from the assumed theoretical shape, corresponding to a 2-cavity interference filter). Due to the low reflectivity of the low-resolution (LR) etalon (83.9\%), its reflectivity variations over the FOV have a negligible impact on the combined transmission profile of the two etalons, and this parameter was held constant at its nominal value. Our initial fits returned an average HR reflectivity that was somewhat below that given by the manufacturer of the etalon coating (93.56\%), suggesting the presence of spectral straylight of approximately 1\%. This is because spectral straylight results in a spectral line profile with reduced depth, which is compensated for in our fits by a widened transmission profile, i.e. a reduced reflectivity. However, including the first side lobe on each side of the peak for the HR etalon transmission profile returned excellent fits and an average reflectivity (93.44\%) that is very close to that expected (93.56\%). Our fits therefore give no reason for suspecting significant levels of spectral straylight beyond the approximately 0.8\% contributed from the first side lobes of the combined transmission profile. From the fits, we also determined and compensated for the telluric blend in the red wing of the 630.25~nm line, as follows: We first removed the telluric line from the FTS solar atlas and replaced it with a ``modeled'' narrow Lorentzian line profile of unknown strength and width, located at an unknown wavelength. To fit the parameters of this Lorentzian, we shifted the wavelength scales of all flat-field spectra in the FOV to a common wavelength scale according to the measured cavity errors. Using the spectra with their shifted wavelengths, and convolving the synthetic FTS spectrum (including the modeled telluric line), we then iteratively fitted all these spectra to a \emph{single} Lorentzian profile, characterized by its strength, width and center wavelength. A similar procedure was applied to the science data since the strength and wavelength (relative to the 630~nm lines) of the telluric line changes during the day. In general, the fits of the observed CRISP profiles to the convolved FTS spectrum are excellent, as illustrated in Fig.~\ref{fig:CRISP}. 

Our characterization of the CRISP transmission profile allows the inversions to take into account any asymmetries in the transmission profile resulting from the \emph{relative} wavelength shifts of the peak transmissions of the two Fabry-Perot etalons. In particular, such asymmetries could otherwise lead to errors in the LOS velocity gradient. 

\begin{figure*}[tbp]
\centering
\begin{overpic}[width=0.99\textwidth,angle=0]{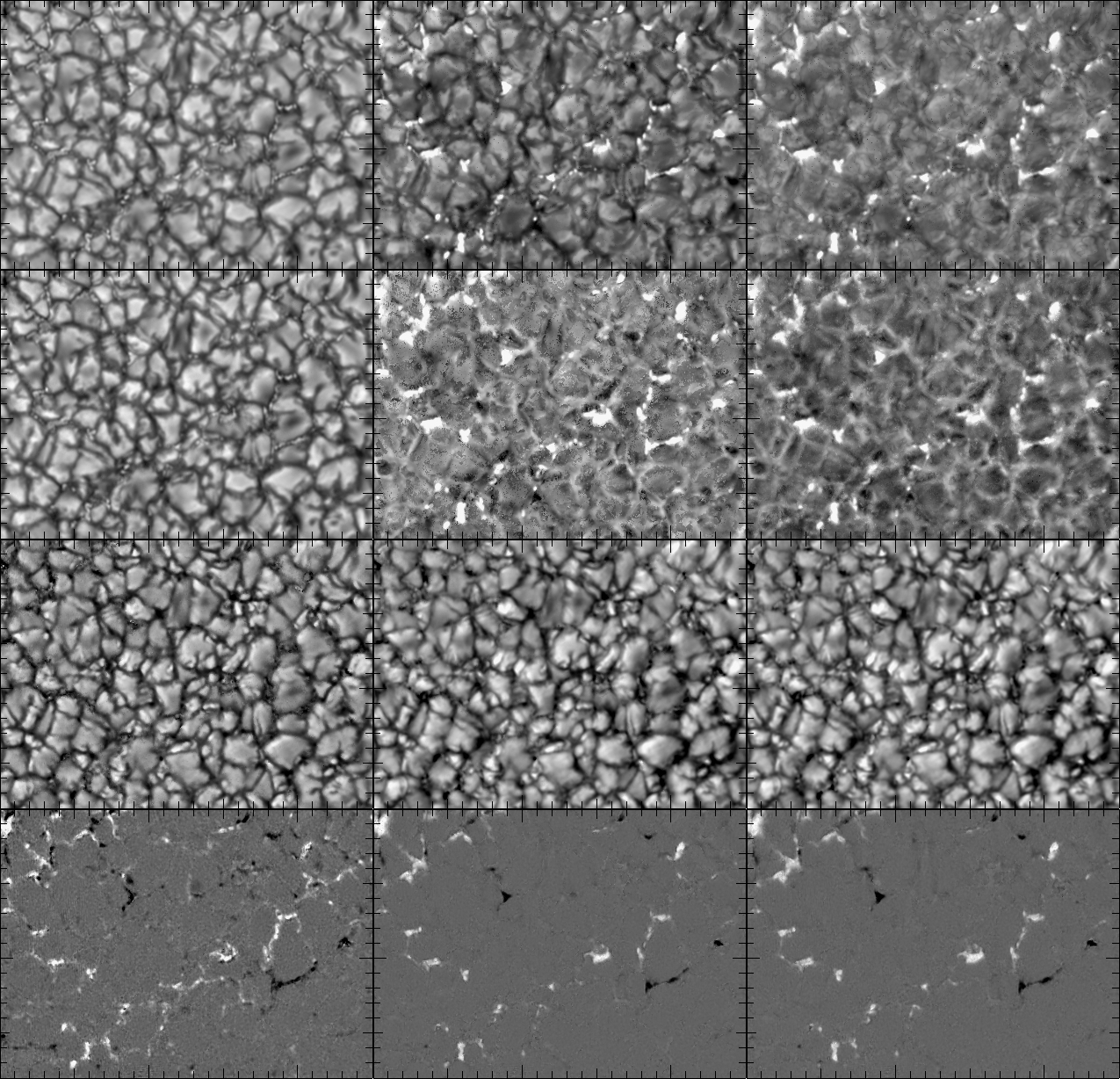} 
\put(1,74){\textcolor{black}{\textbf{\large a)}}}
\put(34.3,74){\textcolor{black}{\textbf{\large b)}}}
\put(67.6,74){\textcolor{black}{\textbf{\large c)}}}
\put(1,50){\textcolor{black}{\textbf{\large d)}}}
\put(34.3,50){\textcolor{black}{\textbf{\large e)}}}
\put(67.6,50){\textcolor{black}{\textbf{\large f)}}}\index{\footnote{}}
\put(1,26){\textcolor{black}{\textbf{\large g)}}}
\put(34.3,26){\textcolor{black}{\textbf{\large h)}}}
\put(67.6,26){\textcolor{black}{\textbf{\large i)}}}
\put(1,2){\textcolor{black}{\textbf{\large j)}}}
\put(34.3,2){\textcolor{black}{\textbf{\large k)}}}
\put(67.6,2){\textcolor{black}{\textbf{\large l)}}}
\end{overpic}
  \caption{
  Panels a-c show the temperature at $\tau_c=$~0.95, 0.05 and 0.01, clipped at (5500, 6700), (4600, 5300), and (4300, 5200)~K resp., for inversions made assuming depth-independent magnetic field and LOS velocity, and in panels d-f) the corresponding temperatures obtained with 2 nodes for the velocity and 3 nodes for $B_{\rm{LOS}}$. Note the absence of inverse granulation in panel b. Panels g and h show the LOS velocity at $\tau_c=$~0.95 and $\tau_c=$~0.05 from the inversions corresponding to panels d-f, and in panel i the LOS velocity from the inversions assuming constant LOS velocity (panels a-c). The LOS velocities have been clipped at (-4.5, 4.5), (-2.3,~2.3) and (-2.3,~2.3)~km~s$^{-1}$. The bottom row (panels j-l) show the LOS magnetic field (with 3 nodes for $B_{\rm{LOS}}$) at $\tau_c=$~0.95, 0.05 and 0.01, clipped at (-1000, 1500)~G. Note the increasing area coverage (expansion with height) of the network field at smaller optical depths. The FOV shown is 25$\times$18\arcsec.} 
  \label{fig:temperature}
\end{figure*}
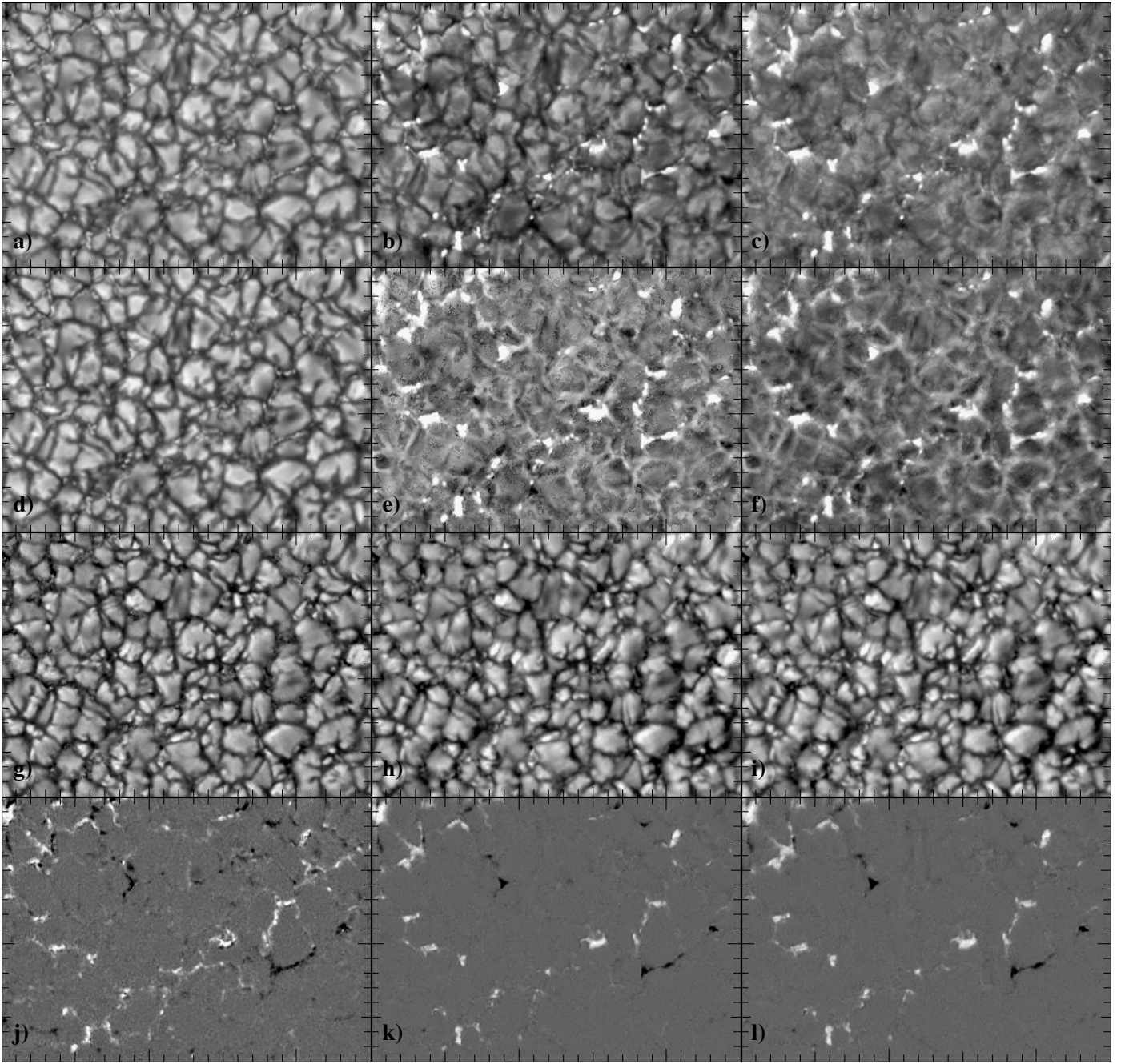

\subsection{Straylight compensation}
As the final step of our pre-processing, we compensated the data for straylight. The need and rationale for such compensation has been discussed extensively in previous publications \citep{2011Sci...333..316S, 2011A&A...529A..79N, 2012A&A...540A..19S}. Here, we summarize the arguments and some of the results, but do not repeat the detailed analysis presented earlier \citep[][their Supporting Online Material, SOM\footnote{{\small \url{http://www.sciencemag.org/content/333/6040/316/suppl/DC1}}}]{2011Sci...333..316S}.

Recent measurements of the Hinode/SOT point spread function (PSF) and RMS granulation contrasts from Hinode clearly demonstrate consistency with 3D simulations \citep{2008A&A...487..399W, 2008A&A...484L..17D, 2009A&A...503..225W, 2009A&A...501L..19M}, leading to the conclusion that there is no scientific ground for questioning the theoretically obtained granulation contrast. This implies that actual measurements of the granulation contrast can be used to constrain the PSF of other telescopes. For telescope diameters larger than approximately 0.5~m, almost all contributions to the granulation contrast come from spatial frequencies well within the diffraction limit, such that the obtained granulation contrast is determined (primarily) by the wings of the PSF, referred to below as ``straylight'', rather than by its core. This is not surprising, since granules have typical diameters of 1\farcs4. The present SST data contain both (nearly) non-magnetic granulation and a sunspot within the same FOV. The minimum umbra intensity in the sunspot provides a second important constraint on the width and strength of the wings of the PSF for our data.

A detailed analysis of the CRISP data obtained in the \ion{C}{I} 538.03~nm line \citep[][SOM]{2011Sci...333..316S} shows that the overall straylight must be more than 50\% at this wavelength to explain the granulation contrast. On the other hand, the measured minimum umbra intensity of only 15.6\% excludes the possibility that this straylight comes primarily from a PSF with very wide wings. In particular, the analysis excludes a PSF with Lorentzian wings, since that leads to negative intensities in the umbra when deconvolving the CRISP data. Trial deconvolutions with PSF's having Gaussian shapes lead to the result that the full width at half maximum of such a PSF must be less than about 2\farcs4 in order to reproduce the expected granulation contrast of 17\% at this wavelength and to give a minimum umbra intensity of more than 6\% \citep[][SOM, Table S2]{2011Sci...333..316S}. We cannot exclude the possibility of a quite narrow Gaussian straylight PSF from our data, however that would imply that the straylight fraction must be much higher than 50\%, which seems unlikely. Our analysis of the \ion{C}{I} 538.03~nm led us to adopt a straylight fraction of 58\% and a full width at half maximum (FWHM) of 1\farcs2 at that wavelength, but assuming a somewhat wider PSF of 1\farcs8 and a straylight fraction of about 54\% would reproduce the synthetic \ion{C}{I} 538.03~nm data just as well.  

The origin of this straylight has not been established with certainty, but is now strongly suspected to be mainly from phase errors (aberrations), such as residuals from MOMFBD processing, high-altitude seeing and possibly fixed high-order aberrations in the adaptive mirror or other optics \citep{2010A&A...521A..68S,2012A&A...537A..80L}. In particular, we note that MOMFBD processing (image restoration jointly of images recorded at many wavelengths) leads to only a marginal increase in the RMS granulation contrast, whereas speckle and multi-frame blind deconvolution (MFBD) processing (image restoration of images recorded at a single wavelength) lead to significantly stronger contrast enhancement \citep{2011A&A...533A..21P}. In addition, MOMFBD image reconstruction is always limited to accounting for a small number of atmospheric aberration modes (for our data, only 36 modes), leading to a tail of uncorrected high-order aberrations that gives a PSF with enhanced wings, qualitatively similar to the PSF modeled here \citep{2010A&A...521A..68S,2012A&A...537A..80L}.

We therefore compensate the present 630.15~nm and 630.25~nm data for straylight, using the same procedure as applied previously to the 538.03~nm data \citep{2011Sci...333..316S} and 630.15~nm data \citep{2012A&A...540A..19S}. We assume the following relation between the observed $I_o$ and ``true'' $I_t$ intensities at any wavelength and polarization state:
\begin{equation} 
I_o=(1-\alpha) I_t + \alpha I_t * G(W), 
\end{equation}
where $\alpha$ is the straylight fraction, `` * '' denotes convolution, and $G$ is a Gaussian straylight point-spread function (PSF), having a full width at half maximum of $W$. For the present inversions, we set $W$ to 1\farcs8 and $\alpha$ to 0.4, rather than to 1\farcs2 and 0.5 respectively, as used previously for the 630.15~nm data \citep{2012A&A...540A..19S}. Note that our model for the straylight is such that the diffraction limited core of the PSF is not compensated for beyond that already made in MOMFBD processing. In particular, straylight compensation, as described above, with a relatively wide PSF does not selectively amplify the high spatial frequencies of the images. 

Before straylight compensation our data give a granulation RMS contrast in the continuum of 7.5\% and a minimum umbra intensity of 18.4\%. Our present choice of straylight parameters increases the RMS contrast to 11.7\%, and reduces the minimum umbra intensity to 14.9\%, whereas our previous choice of setting $W$ to 1\farcs2 and $\alpha$ to 0.5 gave 12.5\% and 14.7\%, respectively. The present choice of straylight parameters gave fewer failures with the inversion code and less noisy maps for the inverted parameters, but otherwise very similar results as with the previously used values, justifying our choice. We note that our straylight compensation gives a granulation RMS contrast that is significantly below that obtained from simulations (14--14.5\%), such that the presently used data are under-compensated for straylight.

As mentioned earlier, telecentric Fabry-Perot systems, such as CRISP, give rise to wavelength shifts of the transmission profile over the FOV. Individual Stokes images recorded in the \emph{wings} of a spectral line therefore show intensity variations that are in part of solar origin and in part artificial from wavelength shifts of the transmission profile. To avoid amplifying any such artificial structures by straylight compensation, we used the cavity map to first shift each profile to a common wavelength grid, employing Hermitian interpolation \citep{1982PDAO...16...67H}. We then applied the straylight compensation to the data, and shifted back the profiles to the original wavelength grid. In the first interpolation, we expanded the wavelength grid by a factor 4 to ensure that the forward-backward interpolation did not introduce errors. This was checked by applying the forward-backward interpolation to the data without any straylight compensation.

We finally comment that straylight compensation of spectropolarimetric data has been implemented using widely different approaches in the past. For example, \citet{2008ApJ...687..668B} used an average \emph{global} Stokes $I$ profile (ignoring polarized straylight) averaged from non-magnetic granulation outside the spot. The relative contribution of this straylight profile to the measured profile in the sunspot was a free parameter of the inversions, fitted separately for each pixel. \citet{2007ApJ...662L..31O, 2007ApJ...670L..61O} instead employed a \emph{local} (unpolarized) straylight profile, obtained independently from 1$\times$1\arcsec{} boxes surrounding each pixel. Our approach is distinctly different from their approaches in that we use \emph{fixed} and identical straylight parameters for all pixels within the FOV and that we deconvolve all 4 Stokes images at each wavelength for the assumed straylight PSF. The observed granulation contrast and minimum umbra intensity do not allow us to firmly establish the strength and shape for the SST straylight PSF.

\begin{figure}
\center
\includegraphics[width=0.4\textwidth,clip]{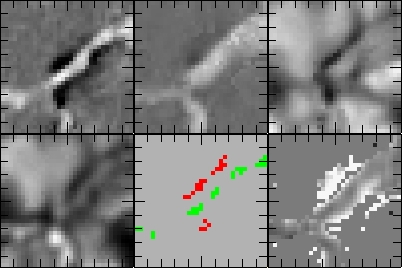}
\includegraphics[bb=201 35 508 705,angle=-90,width=0.235\textwidth,clip]{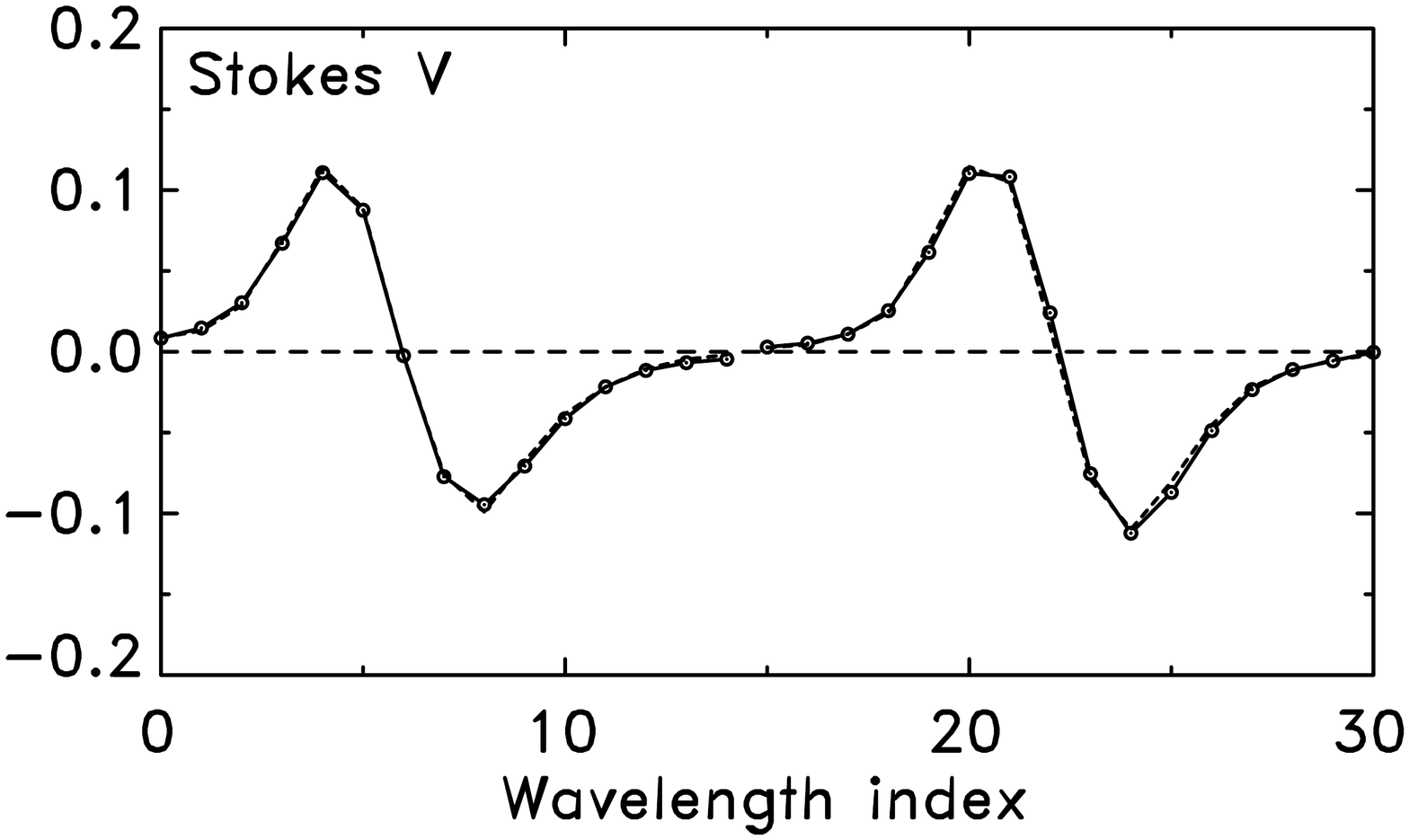}
\includegraphics[bb=201 35 508 705,angle=-90,width=0.235\textwidth,clip]{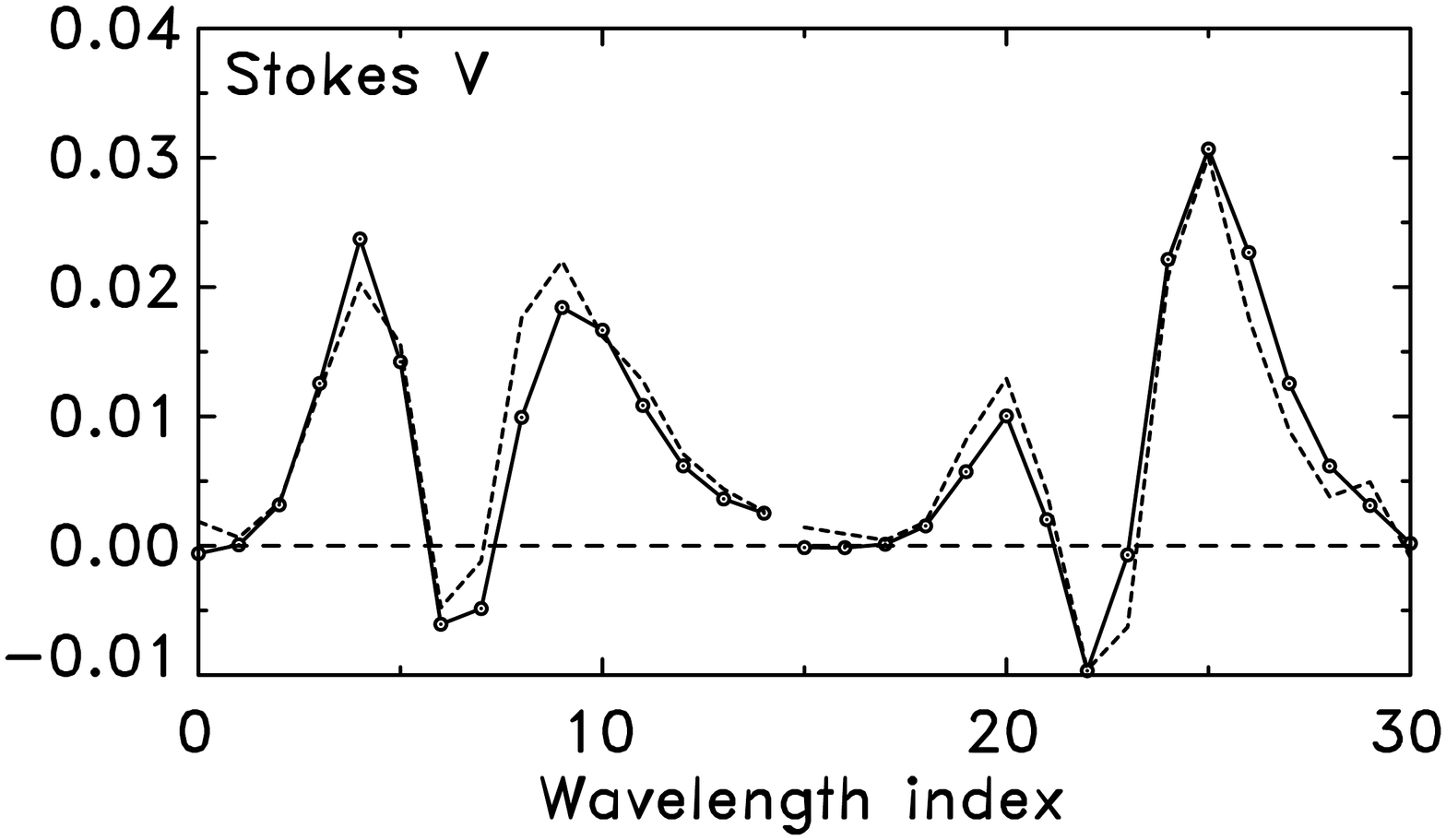}

  \caption{The top row of panels show the LOS magnetic field at $\tau_c=$~0.95, and 0.09 (both clipped at (-1500, 2000)~G), and the LOS velocity at $\tau_c=$~0.95, clipped at (-3.0, 2.0)~km~s$^{-1}$, from a 2\arcsec{}$\times$2\arcsec{} subfield showing a sheet-like network patch located in the lower-left corner of Figs.~\ref{fig:bz}, \ref{fig:bt} and \ref{fig:opp_pol}. The second row shows the temperature at $\tau_c=$~0.95, the green and red pixels used to produce the plots in the lower part of the Figure, and the Stokes $V$ area asymmetry ($\int V~d\lambda/\int |V|~d\lambda)$, clipped at (-0.5, 0.5). Note the presence of strong opposite polarity field at $\tau_c=$~0.95, the virtual absence of such field at $\tau_c=$~0.09 and the strongly increased width of the LOS magnetic field with height. Short tick marks are separated by 0\farcs{2}. The plots show the observed (dashed) and synthetic fitted (full) Stokes $V$ spectra averaged from the interior of the flux concentration (left, from green pixels) and the boundaries with opposite polarity field (right, from red pixels). Note the abnormal $V$ profile with 3 lobes from the opposite polarity field.}
\label{fig:network}
\end{figure}

\section{Inversions}
The inversions were applied jointly to the two \ion{Fe}{I} lines. Inversions were made assuming a single atmospheric component in each pixel, excluding straylight compensation beyond that of the pre-processing (Sect. 2). We used the inversion code NICOLE, which is an improved implementation of the inversion code of \citet{2000ApJ...530..977S}. NICOLE was modified to accept a separate transmission profile for each pixel. In order to carry out the convolution with the CRISP transmission profile, the synthetic Stokes spectra were calculated with a 4 times denser wavelength grid than that of the CRISP data. Our version of NICOLE computes the emerging intensity vector using a quadratic DELO-Bezier solution to the radiative transfer equation \citep{2013ApJ...764...33D}, which improves the accuracy of the synthetic spectra and response functions.

NICOLE was applied iteratively to the data with an increasing number of nodes, similar to as described by \citet{2011A&A...529A..37S}. We used the following final number of nodes: 3 each for temperature (using the HSRA model as the initial estimate) and the LOS magnetic field $B_{\rm{LOS}}$, 2 each for the LOS velocity and the transverse components of the magnetic field, and 1 for the micro-turbulence (in total 13 free parameters). Initially, we made the inversions with 2 nodes for $B_{\rm{LOS}}$, but decided to use 3 nodes for $B_{\rm{LOS}}$ instead in order to ensure that the inferred polarity reversals (Sect. 5) at continuum optical depth $\tau_c$=0.95, are robust results. We also made one inversion as above but with 3 nodes for the LOS velocity (and without micro-turbulence). The resulting velocity maps look good but are visibly noisier than with 2 nodes. We use that inversion only to verify one of our results obtained with 2 nodes. Unless otherwise stated, the results presented are based on inversions with 3 nodes for $B_{\rm{LOS}}$ and 2 nodes for the LOS velocity. 
Note that NICOLE models the \emph{global} variations of the atmospheric properties as being constant when using 1 node, as being linear in $\log \tau_c$ when using 2 nodes, and as being quadratic in $\log \tau_c$ when using 3 nodes. 

We note that the final number of nodes for our inversions is smaller than used by \citet{2011A&A...529A..37S}. This is a direct consequence of the spectral resolution of CRISP (about 6~pm at 630~nm) being smaller than for his SOT/Hinode data, reducing the ``height'' resolution along the LOS for our data through the broadening of the radiative transfer contribution functions within the passband of CRISP. We made a total of 49 inversion experiments
in order to explore the effects of improvements in the pre-processing of the data and changes in the number of nodes, and to verify the robustness of our results. However, these experiments do not allow us to draw firmer conclusions about the optimum choice of the straylight parameters than given by the constraints discussed in Sect. 2.2.

Abundances and atomic parameters used are identical to those of \citet{2011A&A...529A..37S}.

Using the angle between solar North and disk center and the heliocentric distance of the spot, we transformed the magnetic field to the local frame. We resolved the 180 degree ambiguity by defining an approximate center for the sunspot and choosing the azimuth angle offset (0 or 180 deg) for which the horizontal field at $\tau_c=$~0.09 is directed more outward than inward from the spot center. As discussed in Sect.~4.3, the so obtained penumbra field in the local frame on the average appears well aligned with the penumbral filaments, strongly indicating that the telescope polarization model and demodulation of the Stokes data gives good estimates of the direction of the transverse field. The procedure used for resolving the 180 degree ambiguity is of course inappropriate outside the sunspot and probably also in the lightbridge.

\section{Validation of Inversions}

To evaluate the quality of our inversions, we first discuss maps for the temperature, velocity and magnetic field returned by NICOLE over a $25\arcsec\times18\arcsec$ FOV, showing quiet Sun granulation and network field. We compare our inversion results to those of earlier high-resolution observations and to simulations showing both field-free convection and network fields. We also validate our obtained LOS magnetic field gradients from the overall penumbra spine structure, the light bridge, some of the brighter umbral dots and the surroundings of a pore close to the sunspot.

\begin{figure*}[tbp]
\centering
\begin{overpic}[width=0.99\textwidth,angle=0]{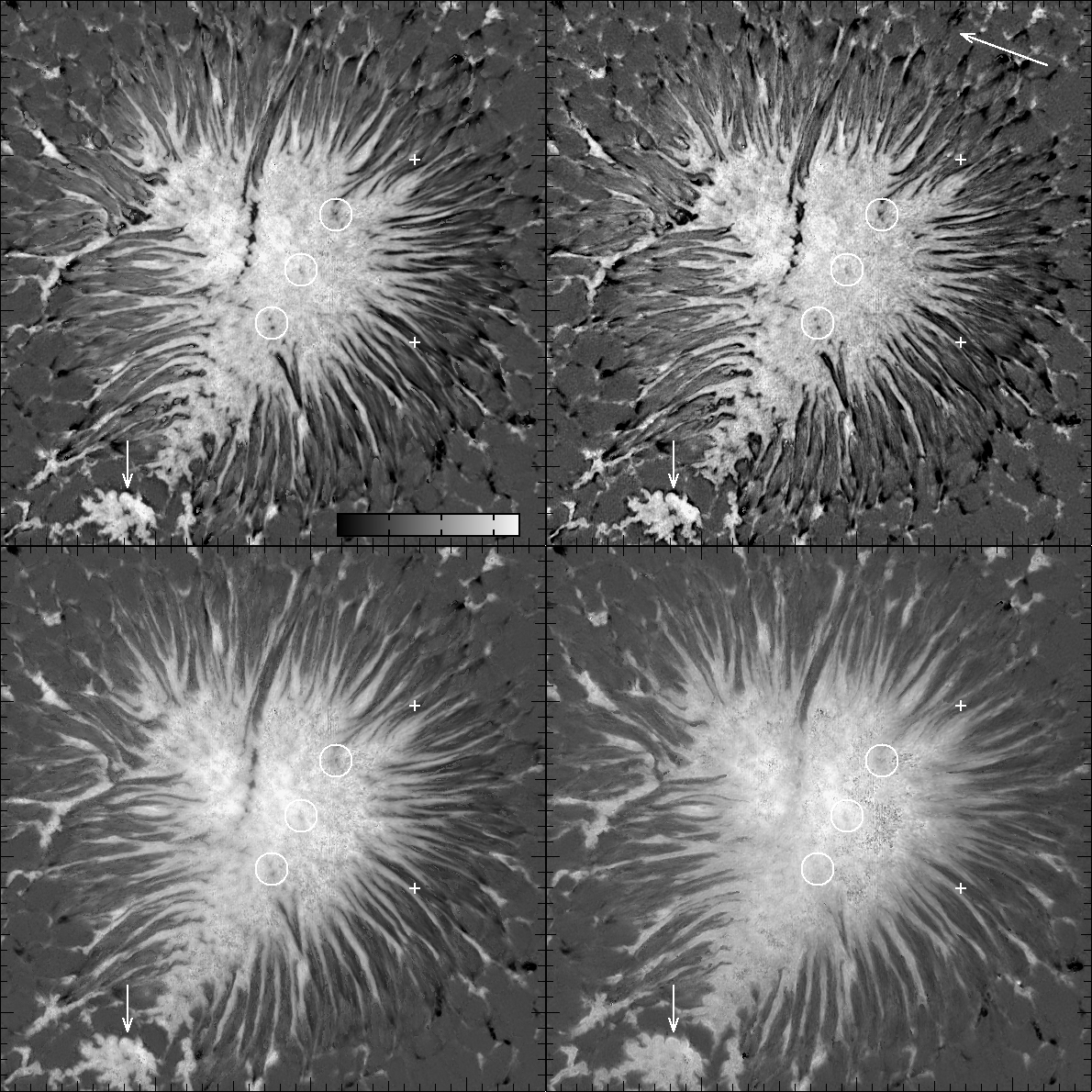} 
\put(1,52){\textcolor{white}{\textbf{\large a)}}}
\put(51,52){\textcolor{white}{\textbf{\large b)}}}
\put(1,2){\textcolor{white}{\textbf{\large c)}}}\index{\footnote{}}
\put(51,2){\textcolor{white}{\textbf{\large d)}}}
\put(28.6,53.5){\textcolor{white}{\textbf{\small -1000~~~~~~0~~~~~~1000~~~2000}}}
\end{overpic}

  \caption{
  Panels a and b show the LOS magnetic field (in the observers frame) at $\tau_c=$~0.95, for inversions made allowing for respectively 2 and 3 nodes for the LOS magnetic field, 2 nodes for the LOS velocity field and 3 nodes for the temperature. Panels c and d show the LOS magnetic field at $\tau_c=$~0.09 and 0.01 with 3 nodes for the LOS magnetic field. Note the presence of opposite polarity field at $\tau_c=$~0.95 in the penumbra, and the virtual absence of such opposite polarity field at smaller depths. The scaling of the magnetic field (in G) is according to the grayscale bar shown in panel a and is the same for panels b-d. The FOV shown is 35$\times$35\arcsec. The plus symbols indicate the first and last pixels of the plots shown in Fig.~\ref{fig:spines}. The long white arrow points in the direction of disk center, the shorter arrows to a pore discussed in Sect.~4.2.4.}
  \label{fig:bz}
\end{figure*}

\begin{figure}
\center
\includegraphics[bb=30 30 508 700,angle=-90,width=0.4\textwidth,clip]{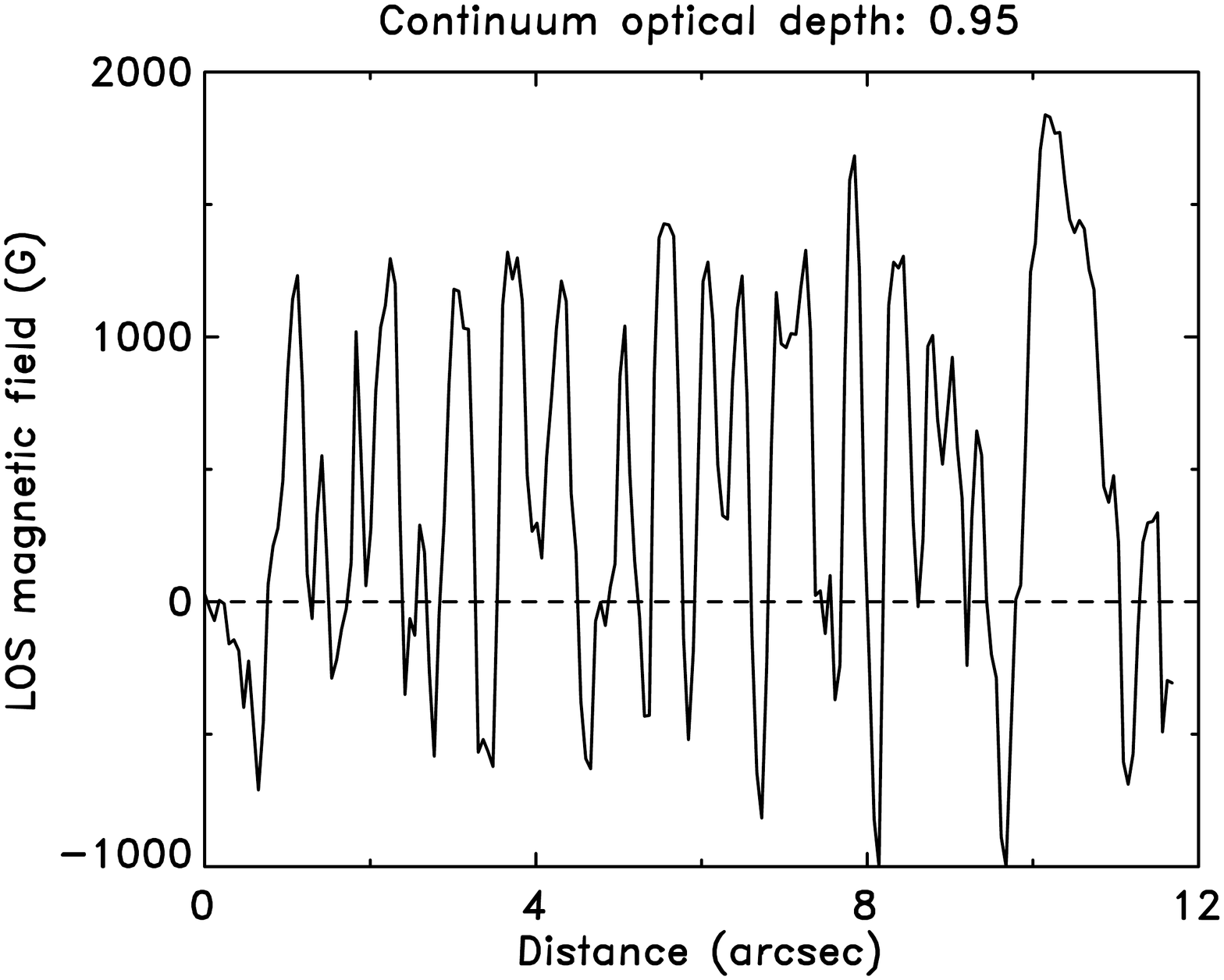}
\includegraphics[bb=30 30 508 700,angle=-90,width=0.4\textwidth,clip]{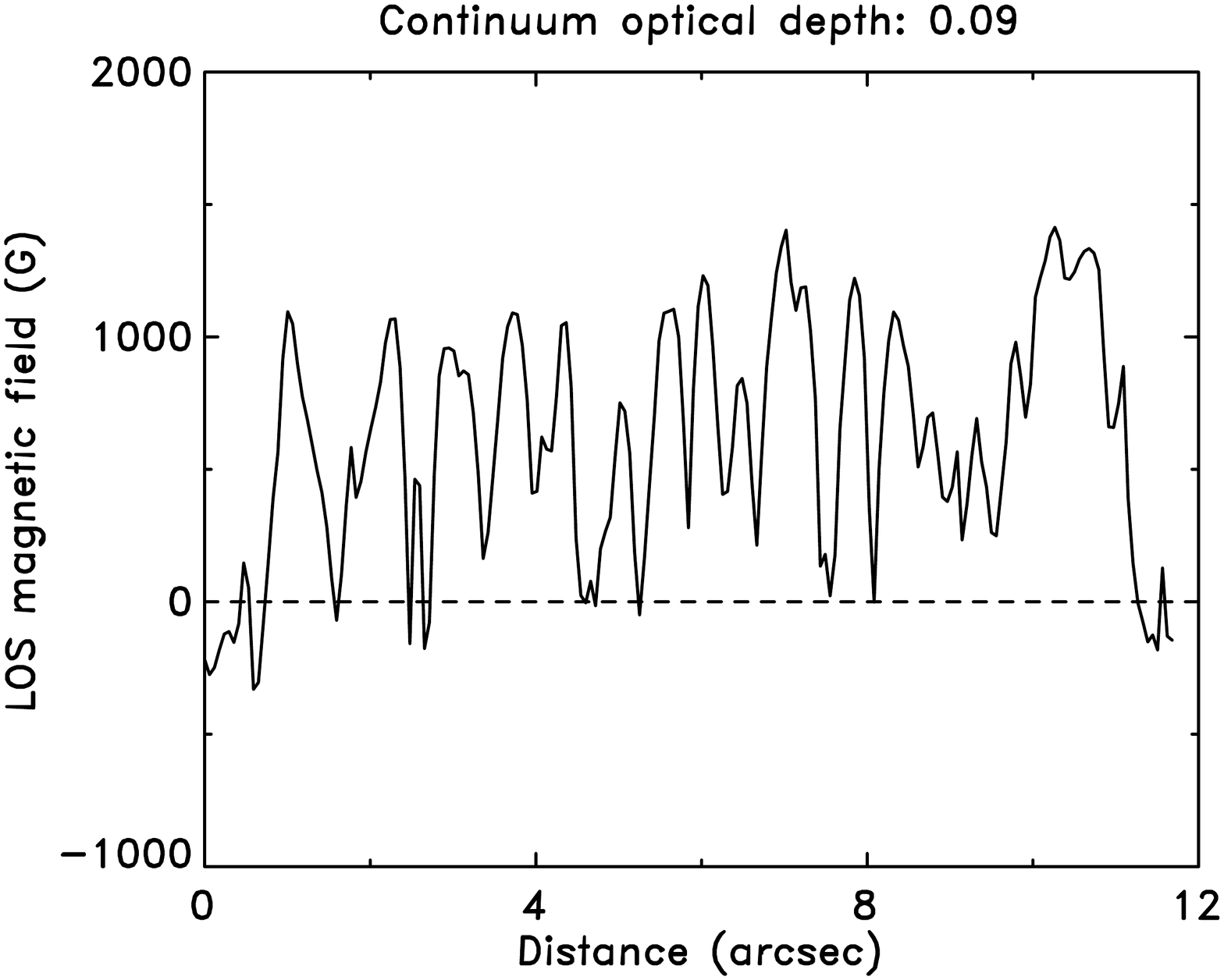}
\includegraphics[bb=30 30 565 700,angle=-90,width=0.4\textwidth,clip]{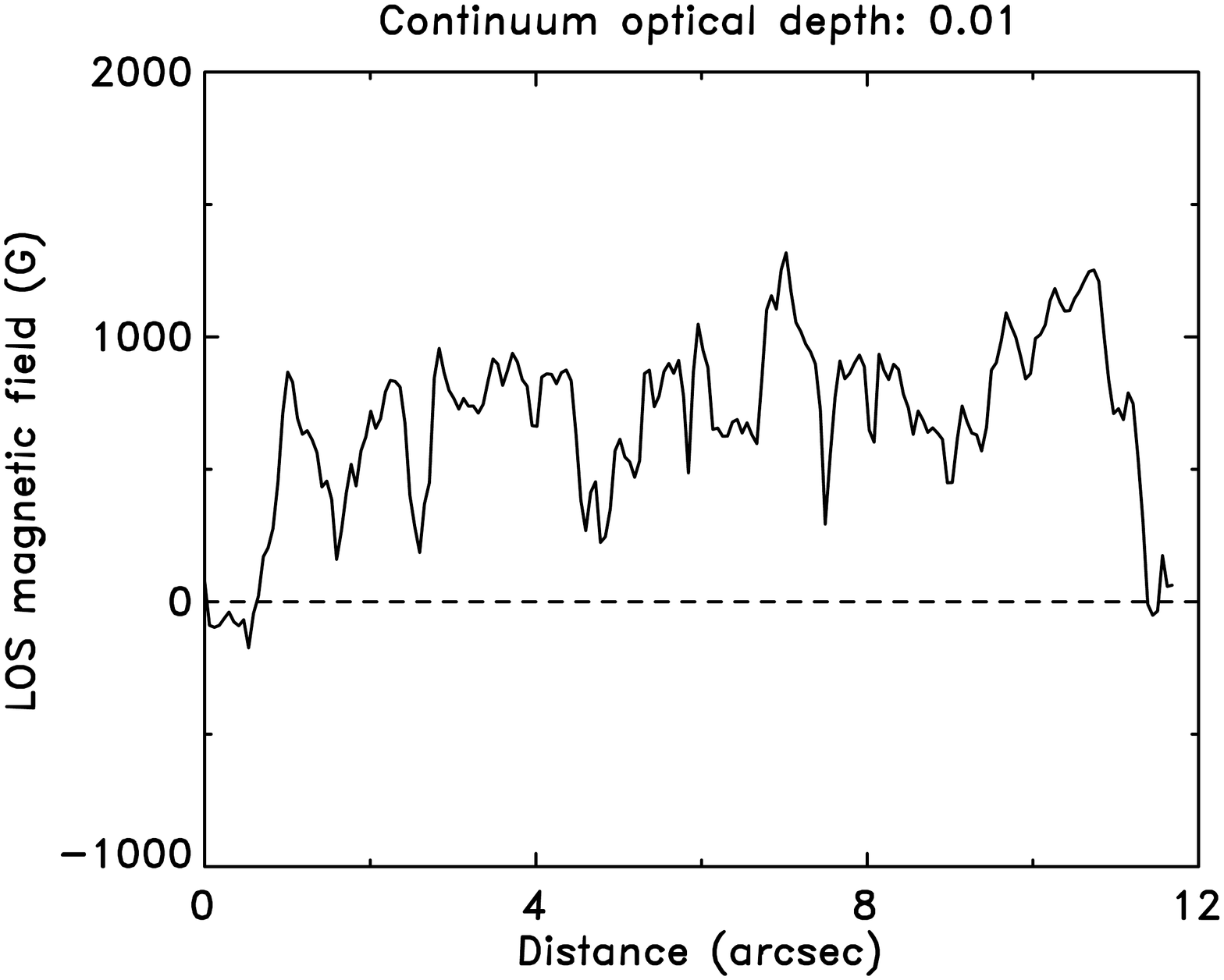}

  \caption{The plots show the LOS magnetic field (in the observers frame) along the line defined by the two plus symbols in Fig.~\ref{fig:bz} at $\tau_c=$~0.95, 0.09 and 0.01. Note the omnipresence of opposite polarity field at $\tau_c=$~0.95 and the virtual absence of such field in the upper layers.}
\label{fig:spines}
\end{figure}

\subsection{Temperature structure and velocity field} 
Figure~\ref{fig:temperature} shows in the top two rows the temperature of field-free granulation and patches of network field at $\tau_c=$~0.95, 0.05 and 0.01. These maps were obtained from inversions obtained without (panels a-c) and with allowance for gradients in the LOS velocity and magnetic field (panels d-f). The latter inversions (panels d-f) show reverse granulation at small optical depths, which is consistent with both observations in the wing of the \ion{Ca}{II} H-line for the same FOV \citep[][his Fig. 7]{2012A&A...548A.114H}, and numerical simulations of convection showing reverse granulation above $\tau_c=$~0.1 \citep{2007A&A...467..703C}. The absence of reverse granulation in the temperature maps obtained without allowance for gradients in the LOS velocity can be attributed to the intensities in the line wings depending strongly on a complicated correlation between temperature and LOS velocity at different heights. In the deep photosphere, this correlation is in the sense expected for convection. Above $\tau_c=$~0.1 this correlation reverses such that intergranular lanes (having downflows also at this height) are brighter than the interiors of granules, where overshooting convective upflows maintain their presence \citep{2007A&A...467..703C}. When the radiative response function peaks for the deeper layers, this enhances the granulation contrast in the red wing of strong spectral lines, whereas in the blue wing the intensity fluctuations from temperature and Doppler shifts largely cancel. Sufficiently close to line center, this pattern reverses. Without properly modeling both the gradients of the temperature and the LOS velocity, this leads to strong cross-talks and errors, in particular for the temperature stratifications in the upper layers of the atmosphere. 

Based on comparisons with both independent observations and simulations \citep{2005A&A...431..687L, 2007A&A...467..703C, 2011A&A...531A..17R, 2012A&A...548A.114H}, Fig.~\ref{fig:temperature} therefore demonstrates that we can have reasonable confidence in the stratifications and spatial variations of the temperature obtained from inversions allowing for LOS gradients in the Doppler velocity (and magnetic field). However, close inspection of panels e and f shows that this comes at the prize of enhanced noise in the measured temperatures in the upper layers of the photosphere. Using inversions with only 2 nodes for temperature instead of 3 nodes (these inversions are not shown in any of our Figures) removes this noise, but unfortunately also removes the inverse granulation pattern at $\tau_c=$~0.05. The conclusion is that \emph{3 nodes are needed for temperature and 2 nodes for the LOS velocity in order to disentangle LOS velocity and temperature gradients in the observed layers.} 

Figure~\ref{fig:temperature} also shows the LOS velocity at $\tau_c=$~0.95 and 0.05 from inversions allowing for gradients (panels g-h), and without allowance for such gradients (panel i). Note that the clippings are different for panel g and for panels h and i, and are scaled according to their RMS values (1.79 km~s$^{-1}$ at $\tau_c=$~0.95; 0.94~km~s$^{-1}$ at $\tau_c=$~0.05; and 0.93~km~s$^{-1}$ for panel i. The RMS velocity at $\tau_c=$~0.95 is very close to the value obtained from numerical simulations \citep[1.78~km~s$^{-1}$][their Table S2 in SOM]{2011Sci...333..316S}. Figure~\ref{fig:temperature} also demonstrates that the LOS velocity field at $\tau_c=$~0.05 is morphologically very similar to that at $\tau_c=$~0.95, but reduced in strength by nearly a factor of 2. This morphological similarity between the flow field at $\tau_c=$~0.95 and $\tau_c=$~0.05 is also consistent with numerical simulations of convection \citep{2007A&A...467..703C}. This comparison validates the magnitudes of the LOS velocities inferred from our inversions, and qualitatively the variation of these velocities with optical depth. Finally, this Figure demonstrates that making inversions that assume a constant LOS velocity (a Milne-Eddington like inversion) returns a velocity that corresponds to the velocity at $\tau_c\approx$~0.05 when allowing for gradients in the LOS velocity. This is in agreement with earlier comparisons between Milne-Eddington techniques and the SIR technique \citep[Stokes Inversion based on Response Functions,][]{1992ApJ...398..375R} \citep{1998ApJ...494..453W}.

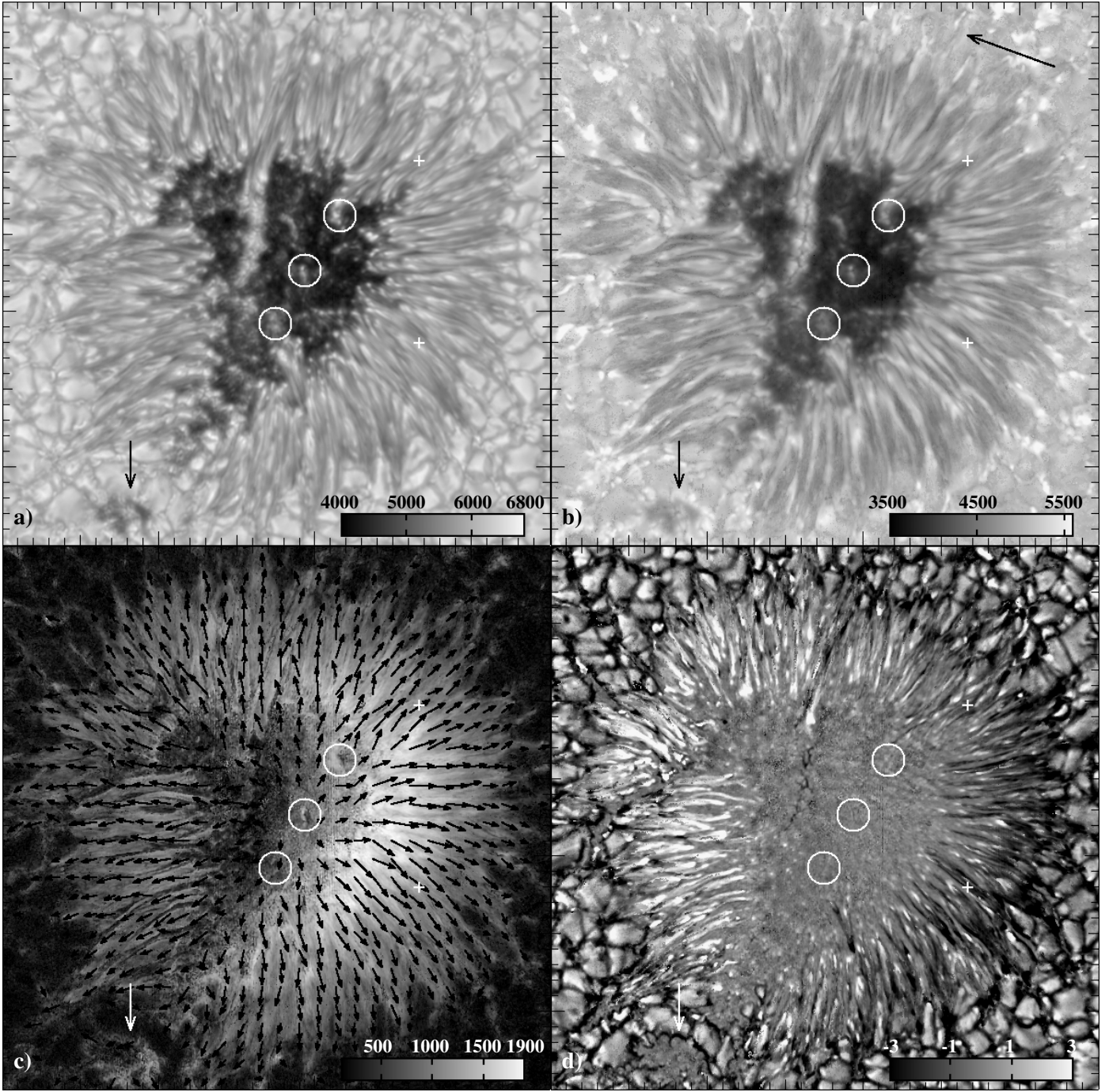
\begin{figure*}[tbp]
\centering
\begin{overpic}[width=0.99\textwidth,angle=0]{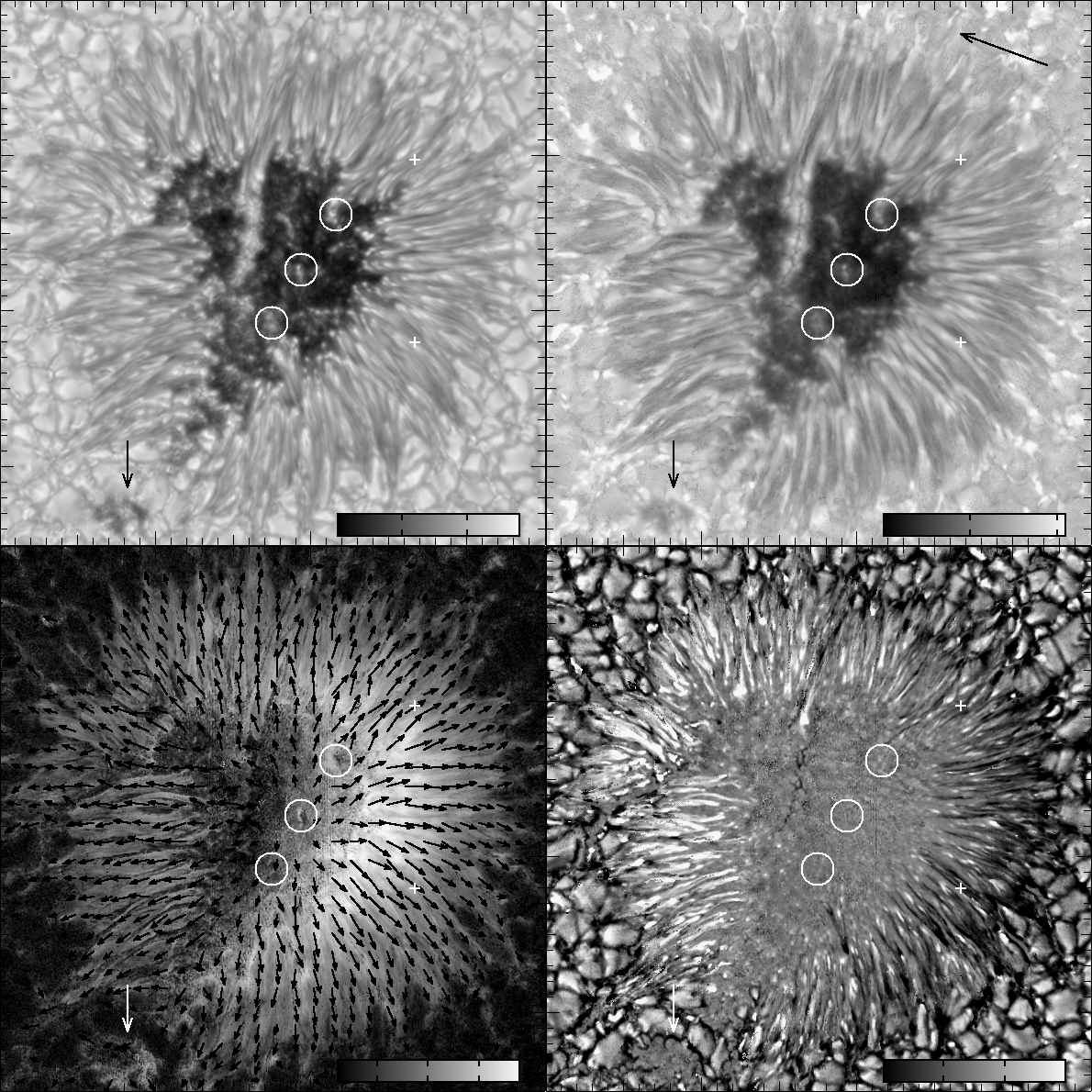} 
\put(1,52){\textcolor{black}{\textbf{\large a)}}}
\put(51,52){\textcolor{black}{\textbf{\large b)}}}
\put(1,2){\textcolor{white}{\textbf{\large c)}}}\index{\footnote{}}
\put(51,2){\textcolor{white}{\textbf{\large d)}}}
\put(29,53.5){\textcolor{black}{\textbf{\small 4000~~~~~~5000~~~~~~6000~~~6800}}}
\put(79,53.5){\textcolor{black}{\textbf{\small 3500~~~~~~~~~~~4500~~~~~~~~~~5500}}}
\put(29,3.5){\textcolor{white}{\textbf{\small ~~~~~~~~~500~~~~1000~~~1500~1900}}}\index{\footnote{}}
\put(79,3.5){\textcolor{white}{\textbf{\small~~~-3~~~~~~~~~-1~~~~~~~~~~1~~~~~~~~~~~3}}}

\end{overpic}

  \caption{
  Panels a, b, c and d show the temperature (in K) at $\tau_c=$~0.95 and 0.09, the transverse magnetic field (in G) at $\tau_c=$~0.09, and the LOS velocity (in km s$^{-1}$) at $\tau_c=$~0.95, respectively. In panel c is shown also the direction of the horizontal field (in the solar frame) at $\tau_c=$~0.09 for every 20th pixel. The FOV shown is 35$\times$35\arcsec. The circles mark examples of umbral dot-like structures that show strongly reduced transverse field on their \emph{limb} sides, consistent with a magnetic field folding over the dots. The long black arrow points in the direction of disk center, the shorter white and black arrows to a pore discussed in Sect. 4.2.4.} 
  \label{fig:bt}
\end{figure*}

\subsection{LOS magnetic field}
\subsubsection{Network}
We now turn to evaluating the measured LOS magnetic field. In panels j-l of Fig.~\ref{fig:temperature} we show $B_{\rm{LOS}}$ at $\tau_c=$~0.95, 0.05 and 0.01.  About a dozen of the flux concentrations shown in Fig.~\ref{fig:temperature} reach a peak LOS magnetic field stronger than 1.5~kG at $\tau_c=$~0.95, which is in good agreement with the field strengths for network field obtained from Sunrise data, about 1.45~kG \citep{2010ApJ...723L.164L}. The average LOS magnetic field for these pixels is 1.7, 1.3 and 1.1~kG at $\tau_c=$~0.95, 0.05 and 0.01, respectively. Near the \emph{boundaries} of these flux concentrations we find many examples where the LOS field \emph{increases} in strength by more than 300~G from $\tau_c=$~0.95 to $\tau_c=$~0.05. This is evident from the increasing size and ``fuzzier'' appearance of these concentrations at decreasing optical depths. The variations of $B_{\rm{LOS}}$ with optical depth both near the centers and boundaries of these flux concentrations are thus consistent with canopy fields expanding with height, as demonstrated indirectly from the center-to-limb variations of the LOS magnetic field of network field by \citet{2010A&A...518A..50P} and from Sunrise data by \citet{2012ApJ...758L..40M}. 

At the boundaries of a few of the larger flux concentrations, we see also tiny narrow sheets and small patches with opposite polarity field. Most Stokes $V$ profiles at these locations are normal with 2 lobes, but show strong asymmetries with a weak red lobe. However, we also find a few extreme examples with abnormal $V$ profiles having 3 lobes, as shown in Fig.~\ref{fig:network}. This Figure also demonstrates that our inversions give good fits even to abnormal $V$ profiles. The canonic interpretation of 2-lobed asymmetric $V$ profiles \citep[][and references therein]{2012ApJ...758L..40M} is that the canopy fields expanding with height produce a discontinuity in the magnetic field (the magnetopause), the height of which increases with increasing distance from the flux concentration. The field-free gas below the canopy flows downward, explaining the asymmetric $V$ profile. Our inversions, which assume a \emph{continuous} variation of the magnetic field with height, indicate in some cases the existence of opposite polarity field at deep layers adjacent to parts of some of the flux concentrations. We note that such opposite polarity field adjacent to small flux tubes and flux sheets can be found also in 3D magnetohydrodynamic (MHD) simulations \citep[][their Figs. 4 and 10]{2009A&A...504..595Y}. 

\subsubsection{Sunspot}
Figure~\ref{fig:bz} shows the sunspot LOS magnetic field obtained from the inversions. Panels a and b show the LOS field at $\tau_c=$~0.95, obtained with 2 nodes for $B_{\rm{LOS}}$ (linear variation with $\log \tau_c$) and 3 nodes (allowing for also a quadratic variation with $\log \tau_c$), respectively. In panels c and d are shown the LOS magnetic field at $\tau_c=$~0.09 and $\tau_c=$~0.01, respectively, obtained with inversions made with 3 nodes for $B_{\rm{LOS}}$. We note that the strong spine structure (the bright ``spokes'' in the penumbra LOS magnetic field maps) appears fundamentally different at  $\tau_c=$~0.95 and 0.01. In the deep layers, the penumbra shows numerous narrow radial structures, many of which are flanked with opposite polarity field. This spine structure is very distinct and ``sharp''. In the higher layers, most of the opposite polarity structures in the penumbra are absent, and the spines are wider and more fuzzy. Several intra-spines (the areas between two spines) can be seen to narrow strongly at small optical depths, and even disappear entirely at $\tau_c=$~0.01, especially in the innermost penumbra. It is also evident from Fig.~\ref{fig:bz} that the strength of the LOS field in the spines weakens with height. To further quantify these variations with height, we show in Fig.~\ref{fig:spines} plots of the LOS magnetic field (in the observers frame) across several filaments at the disk center side of the inner penumbra. 

The presence of opposite polarity field only in the \emph{deep} layers is obvious from these plots, as is the overall weakening of the strength and the fluctuations of the field in the upper layers. The detection of opposite polarity field in the deep layers of the inner parts of the penumbra is a fundamental result of our investigation. It is intriguing to note that the spine structure at $\tau_c=$~0.95 actually is even sharper and more well defined with 3 nodes than with 2 nodes for $B_{\rm{LOS}}$. We are confident that the detection of these opposite polarity patches well inside the penumbra is a robust result. However, the inversions with 3 nodes for $B_{\rm{LOS}}$ show enhanced levels of random noise. We discuss in the following only the results obtained with 3 nodes for $B_{\rm{LOS}}$. 

Noting that the spines represent azimuthal structures with much stronger \emph{vertical} field than in the intra-spines, as first described by \citet{1993ApJ...418..928L}, the observed spatial and LOS variations of the LOS magnetic field are perfectly understandable. The more vertical field in the spines expands with height to match the decreasing gas pressure. The \emph{weaker} spine field seen at smaller optical depths is qualitatively consistent with flux conservation of such structures spreading out with height. The ``sharp'' spine structures seen in the deep photosphere can be explained as (convective) flows shaping these structures. In the upper layers, the morphology of the spines can qualitatively be explained as a potential field ``extrapolation'' of this field structure controlled from below, such that the spine field folds over the intra-spines. Such field topologies in the penumbra were proposed by \citet{2006A&A...447..343S,2006A&A...460..605S}, and inferred from Hinode data by \citet{2008A&A...481L..13B}.

\subsubsection{Umbral dots and lightbridge}
Figure~\ref{fig:bz} shows several examples of umbral dots that have strongly reduced LOS magnetic field at $\tau_c=$~0.95, whereas at $\tau_c=$~0.01 these signatures are absent entirely. Evidently, the LOS magnetic field increases in strength with height \citep[as earlier described by][]{2008ApJ...678L.157R}. In Fig.~\ref{fig:bt} a few bright umbral dots (indicated with circles) can be seen that have strongly reduced transverse field at their \emph{limb} sides. These observed properties are consistent with a field that wraps around and closes above structures that have weak field, or are nearly field-free, deeper down. For many of the brighter umbral dots, we find upflows in the deepest layers. Similar upflows in the deep layers of peripheral (but not central) umbral dots have earlier been reported from Hinode data at lower spatial resolution than here \citep{2008ApJ...678L.157R}, and from SST observations of umbral dots \citep{2010ApJ...713.1282O, 2012ApJ...757...49W} and numerical umbra simulations \citep{2006ApJ...641L..73S}. However, we find no clear evidence of the expected downflows surrounding the upflows, as was the case also for the data of \citet{2008ApJ...678L.157R}. \citet{2010ApJ...713.1282O} and \citet{2012ApJ...757...49W} reported evidence of such downflows but \citet{2012ApJ...757...49W} found only patchy downflows in some cases and no downflows at all for many umbral dots.

The inner part of the light bridge shows a narrow irregular ridge with patches of opposite polarity field at $\tau_c=$~0.95; this feature is almost absent at $\tau_c=$~0.01. At these locations, panel d in Fig.~\ref{fig:bt} shows downflows. The transverse field is weak or absent at the limb side of this part of the light bridge, but not on its disk center side. As is the case for the umbral dots discussed previously, this is consistent with a (convective) field-free structure protruding into a magnetic field that closes over this structure higher up. 

\begin{figure}
\includegraphics[angle=0,width=0.485\textwidth,clip]{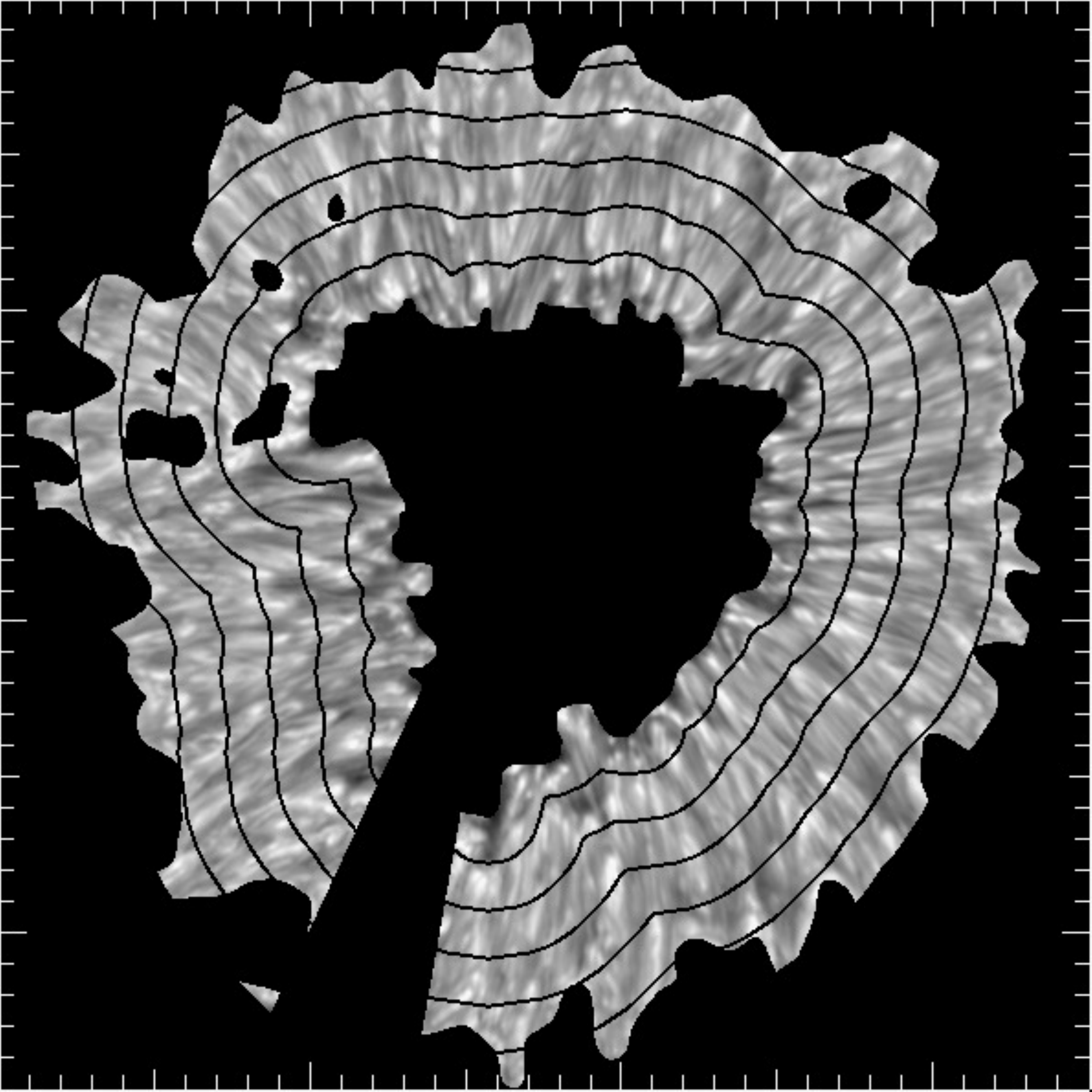}
 \centering
\caption{Masks used to divide the penumbra into 6 radial zones, with the temperature at $\tau_c=$~0.95 superimposed.}
\label{fig:masks}
\end{figure}

\begin{figure}
\includegraphics[bb=59 50 567 707,angle=-90,width=0.4\textwidth,clip]{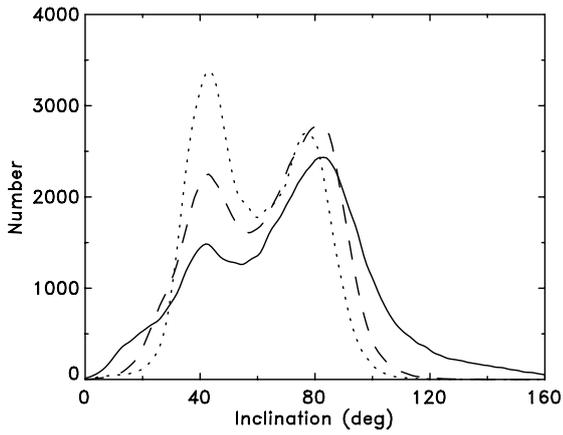}
 \centering
\caption{Histograms of the magnetic field inclination angle in the entire penumbra except the outermost radial zone (shown in Fig.~\ref{fig:masks}) at $\tau_c=$~0.95 (full), 0.09 (dashed), and 0.01 (dotted). The distributions are bimodal, with peaks at approx. 45\degr and 80\degr. We define spines/intra-spines as having inclinations smaller/larger than 60\degr. The histograms have been smoothed over 3\degr. }
\label{fig:incl_hist}
\end{figure}

\begin{figure}[tbp]
\center
 \includegraphics[angle=0,width=0.485\textwidth,clip]{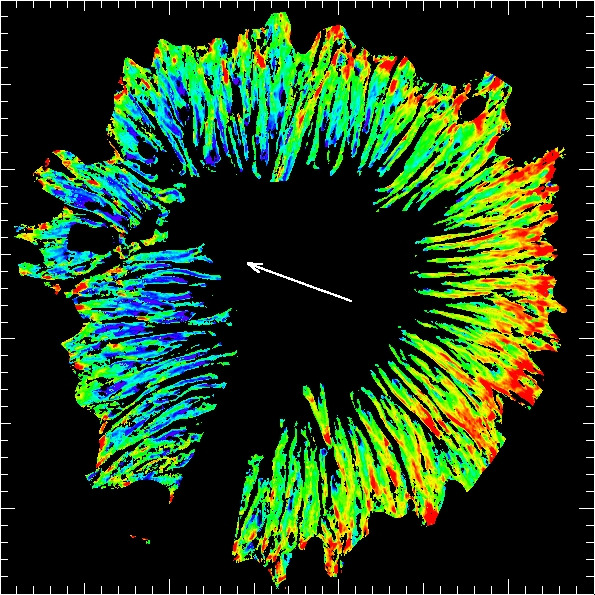}
 \includegraphics[angle=0,width=0.485\textwidth,clip]{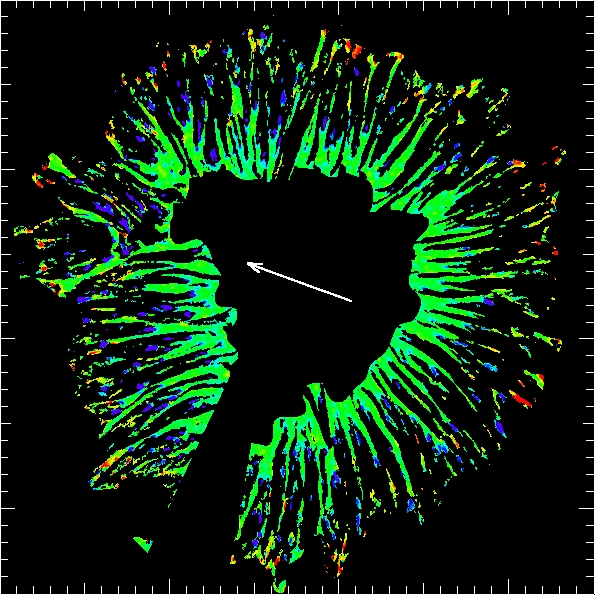}
\begin{overpic}[width=0.485\textwidth,angle=0]{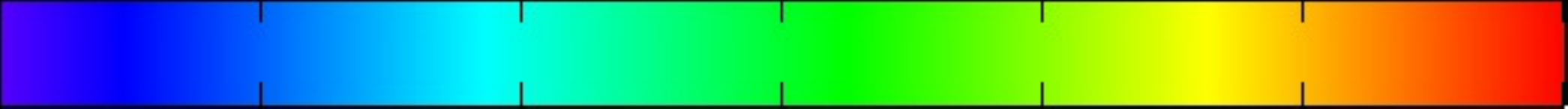} 
\put(-1,-4){\textcolor{black}{\textbf{-3}}}
\put(15.3,-4){\textcolor{black}{\textbf{-2}}}
\put(31.7,-4){\textcolor{black}{\textbf{-1}}}
\put(49,-4){\textcolor{black}{\textbf{0}}}
\put(64.3,-4){\textcolor{black}{\textbf{+1}}}
\put(80.7,-4){\textcolor{black}{\textbf{+2}}}
\put(97,-4){\textcolor{black}{\textbf{+3}}}
\end{overpic}
\vspace{0.02cm}\\
\caption{LOS velocity at $\tau_c=$~0.95 inferred from the inversions for intra-spines (top), and spines (bottom). The LOS velocities are color coded such that red and blue corresponds to velocities away from and towards the observer respectively, and are clipped at (-3.0, 3.0)~km~s$^{-1}$. The absence of systematic differences in LOS velocities between the disk center (direction of arrow) and limb sides of the penumbra demonstrates that \emph{the radial Evershed flow is absent in the spines}, as expected from the more vertical field there. The FOV is $\sim 35\times 35$ \arcsec{} and corresponds to that shown in Figs.~\ref{fig:bz} and \ref{fig:bt}. Tick marks are at 1\arcsec{} intervals.}
\label{fig:velocities}
\end{figure}

\begin{figure}[tbp]
\center
 \includegraphics[bb=24 95 509 709, angle=-90,width=0.24\textwidth,clip]{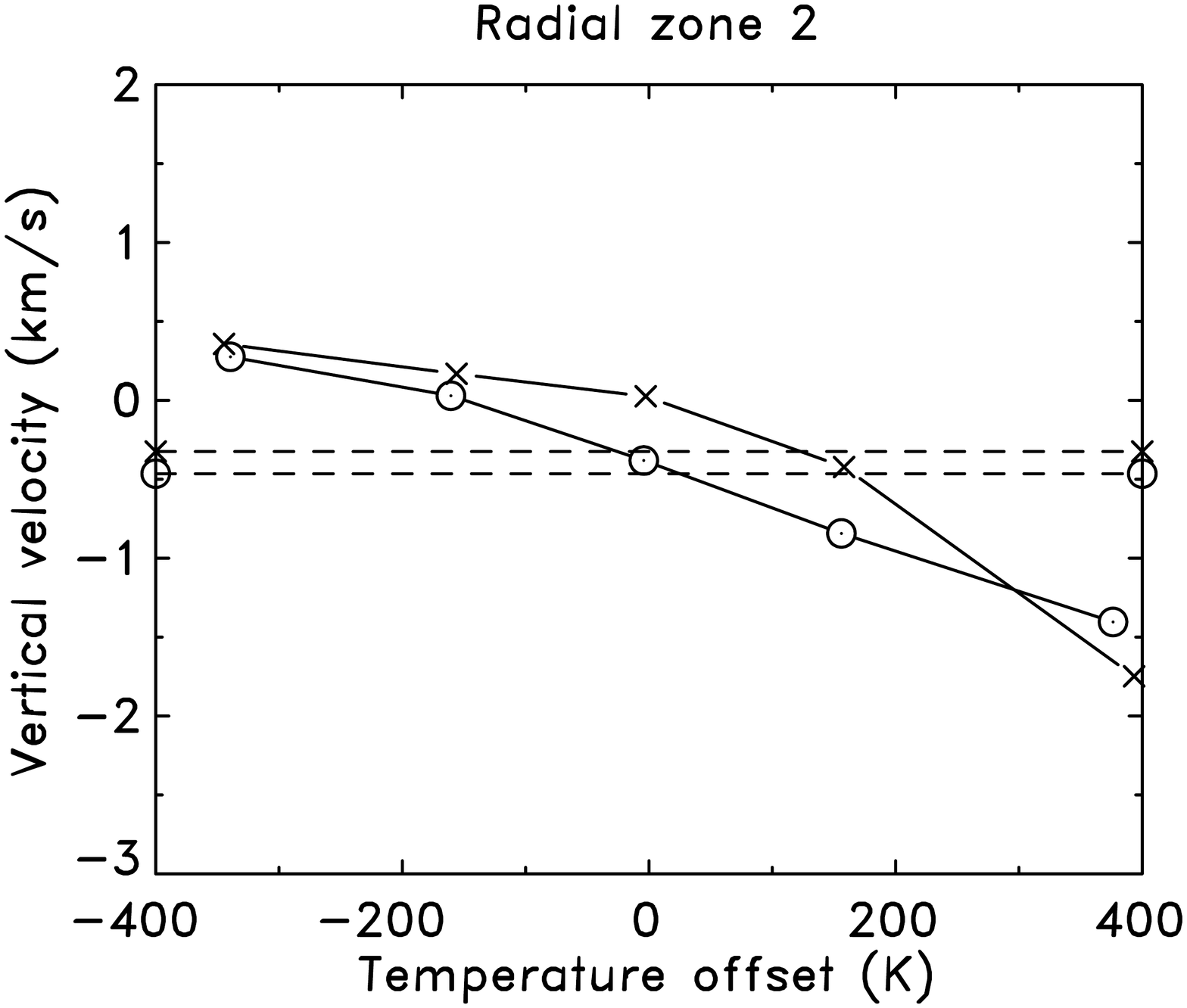}
\includegraphics[bb=24 95 509 709, angle=-90,width=0.24\textwidth,clip]{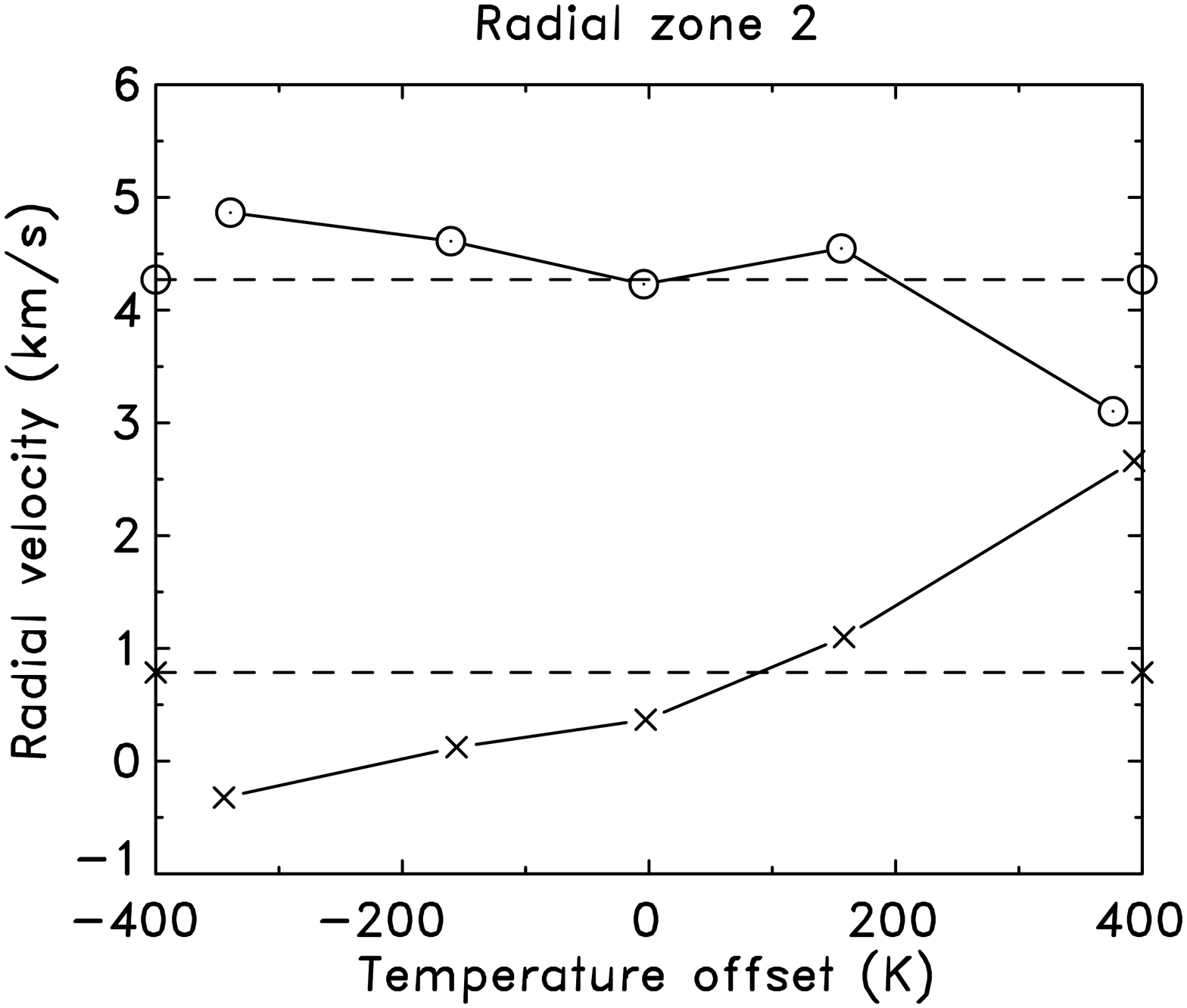}
 \includegraphics[bb=24 95 509 709, angle=-90,width=0.24\textwidth,clip]{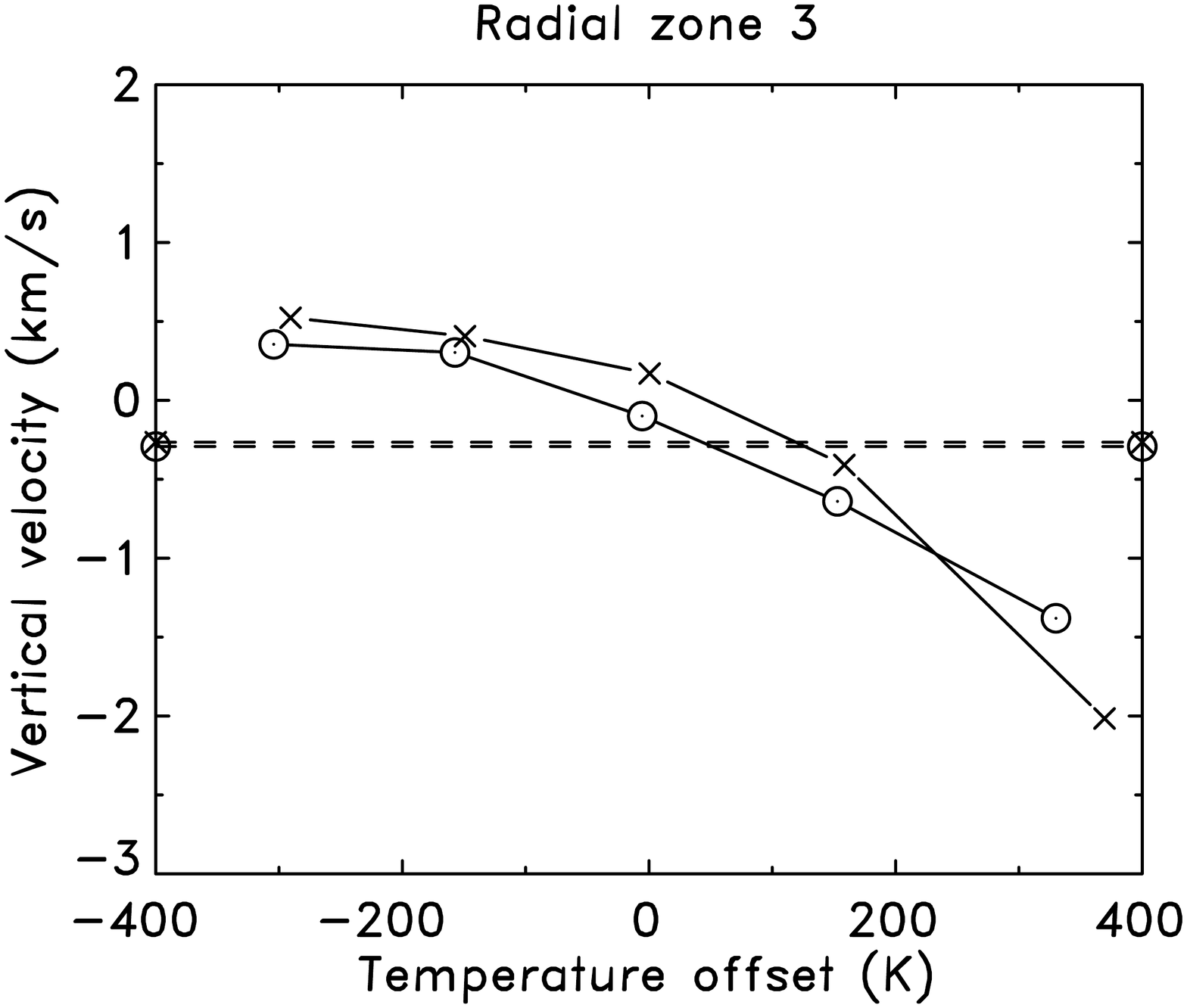}
\includegraphics[bb=24 95 509 709, angle=-90,width=0.24\textwidth,clip]{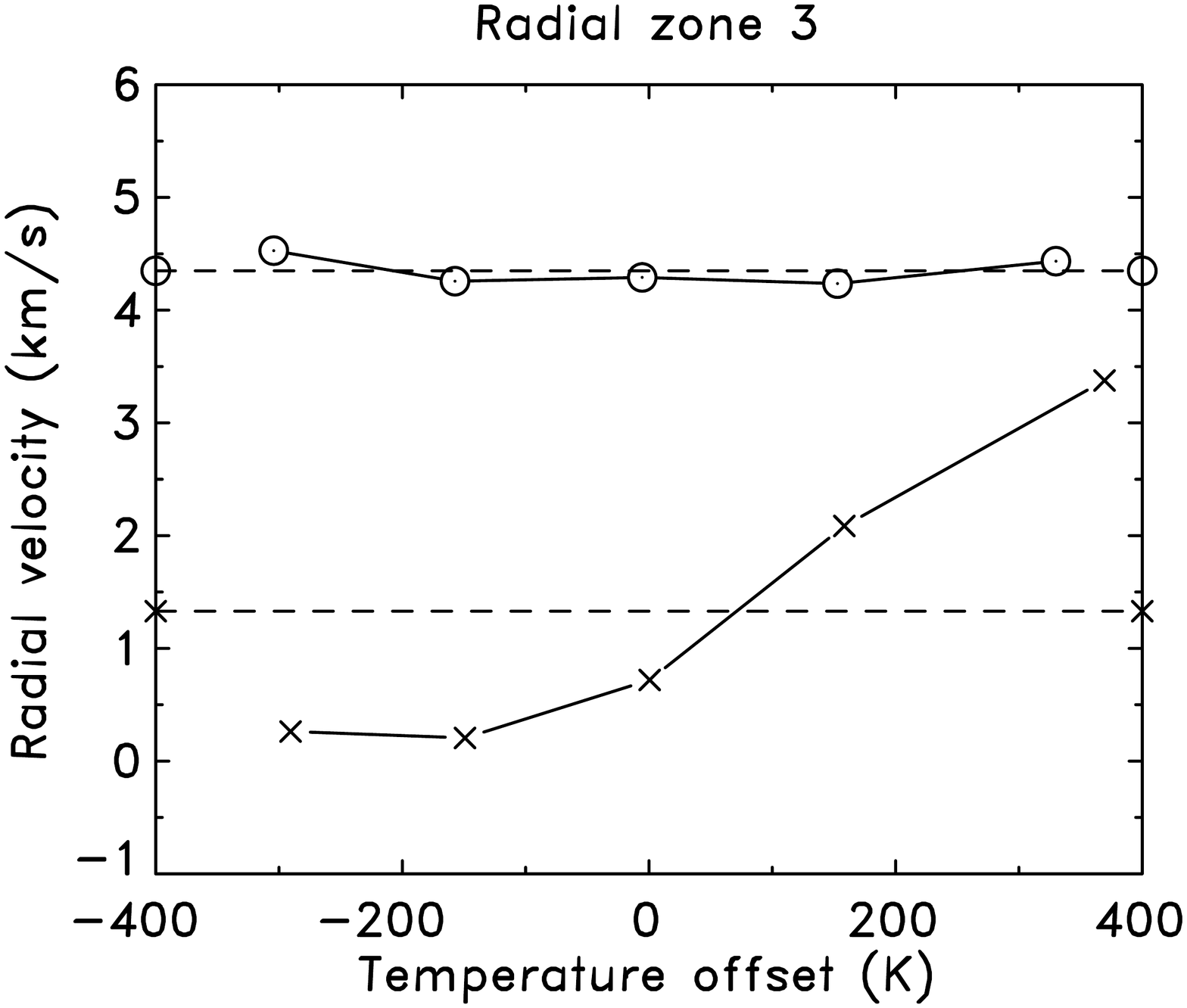}
 \includegraphics[bb=24 95 509 709, angle=-90,width=0.24\textwidth,clip]{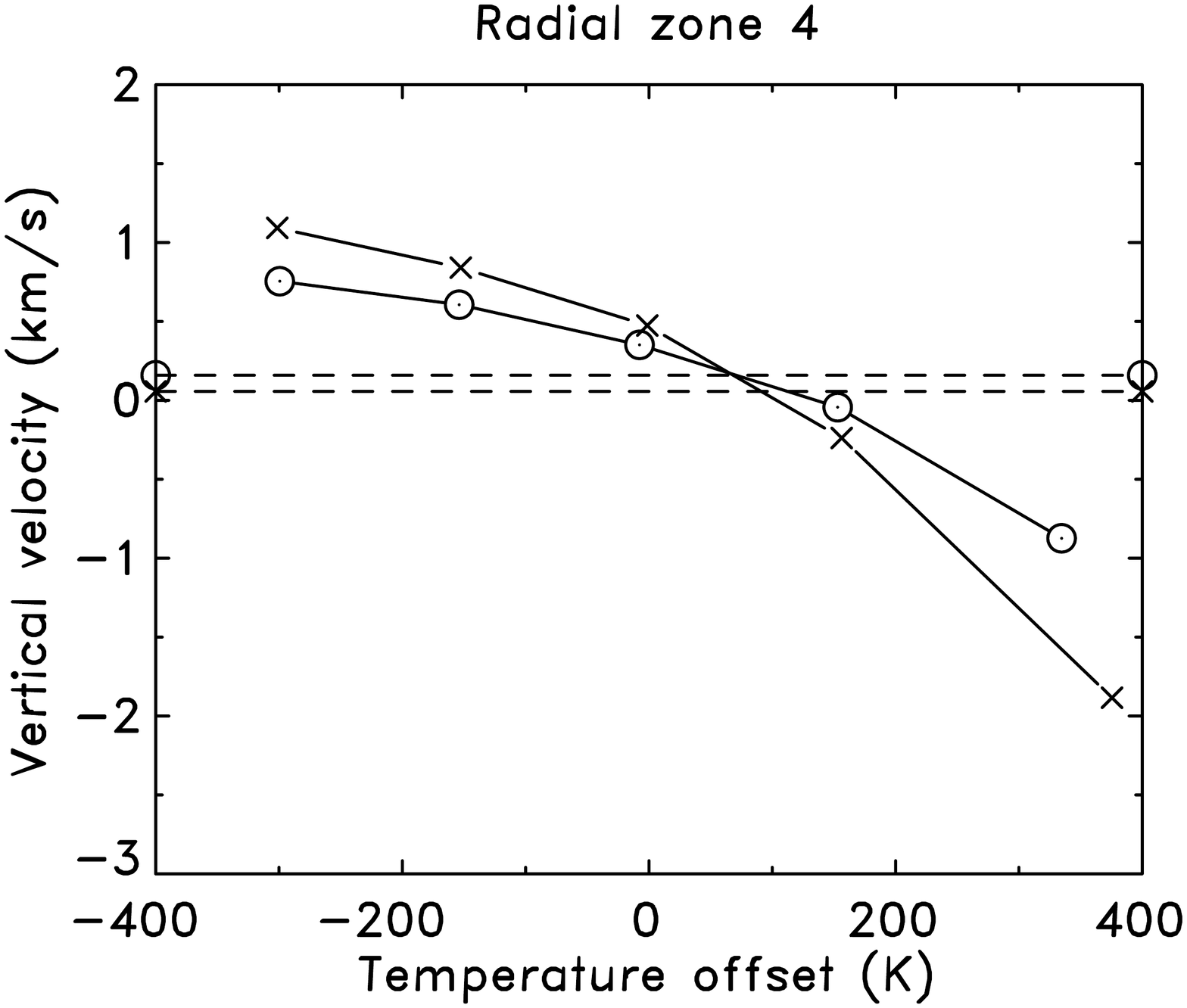}
\includegraphics[bb=24 95 509 709, angle=-90,width=0.24\textwidth,clip]{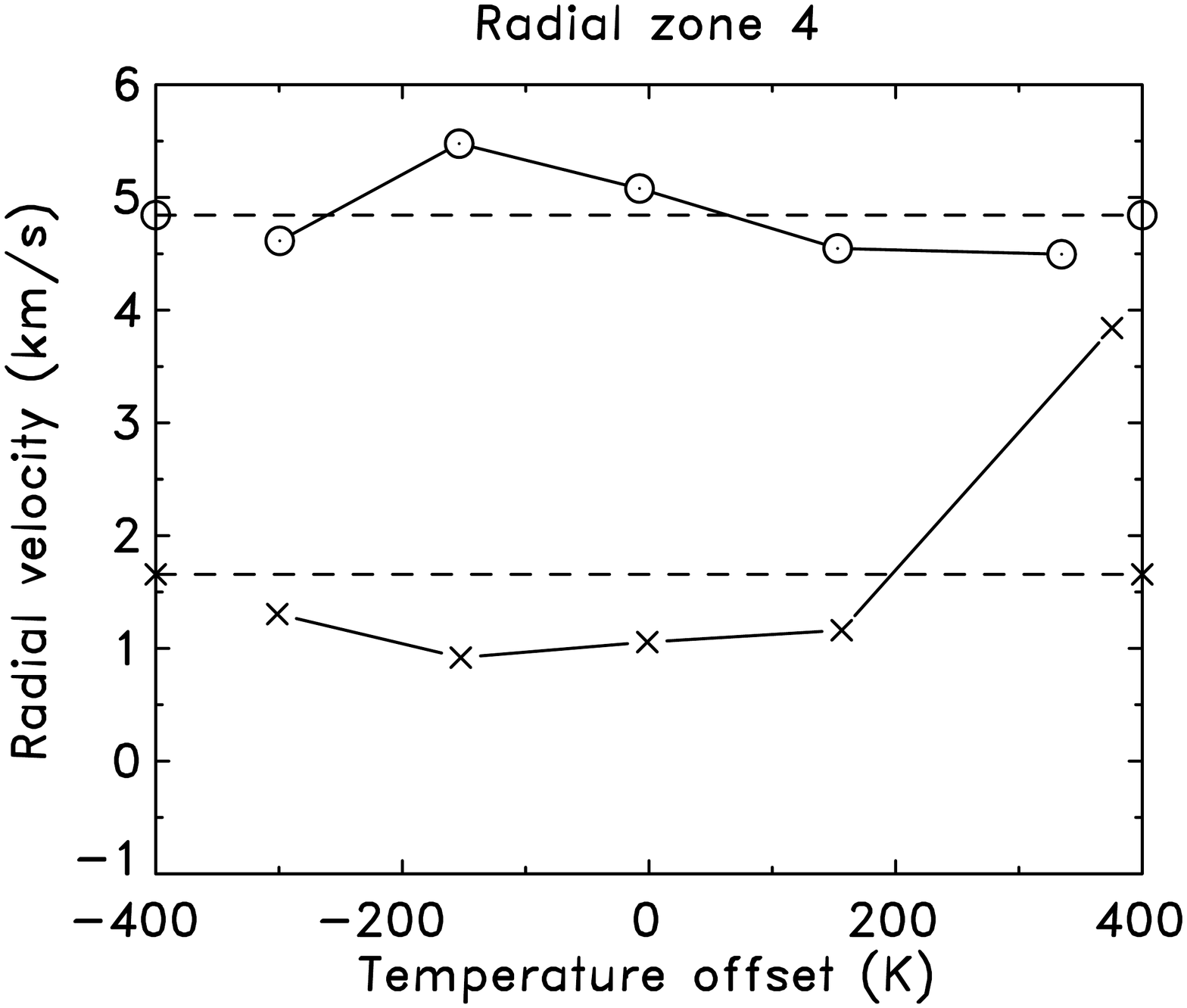}
 \includegraphics[bb=24 95 579 709, angle=-90,width=0.24\textwidth,clip]{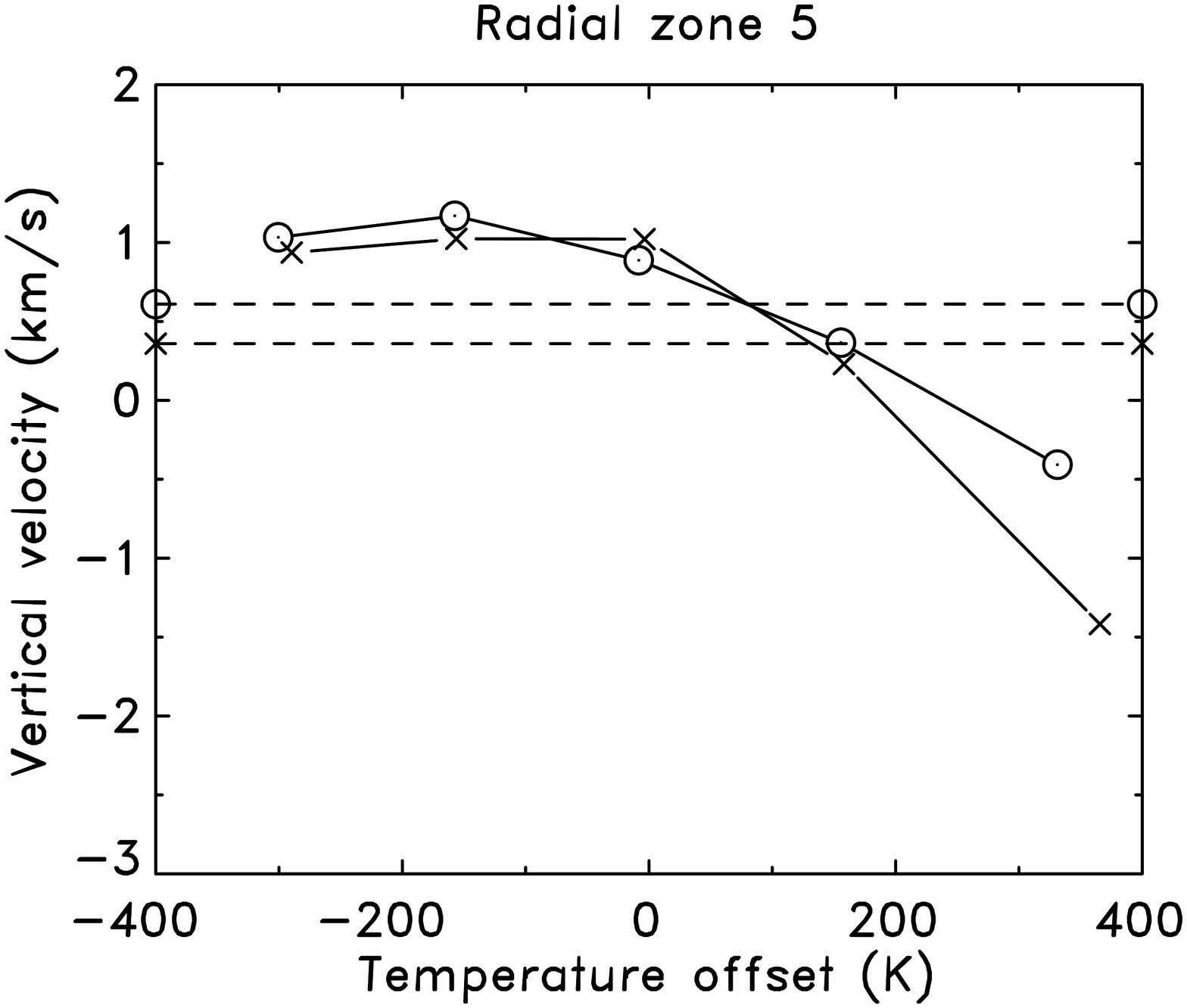}
\includegraphics[bb=24 95 579 709, angle=-90,width=0.24\textwidth,clip]{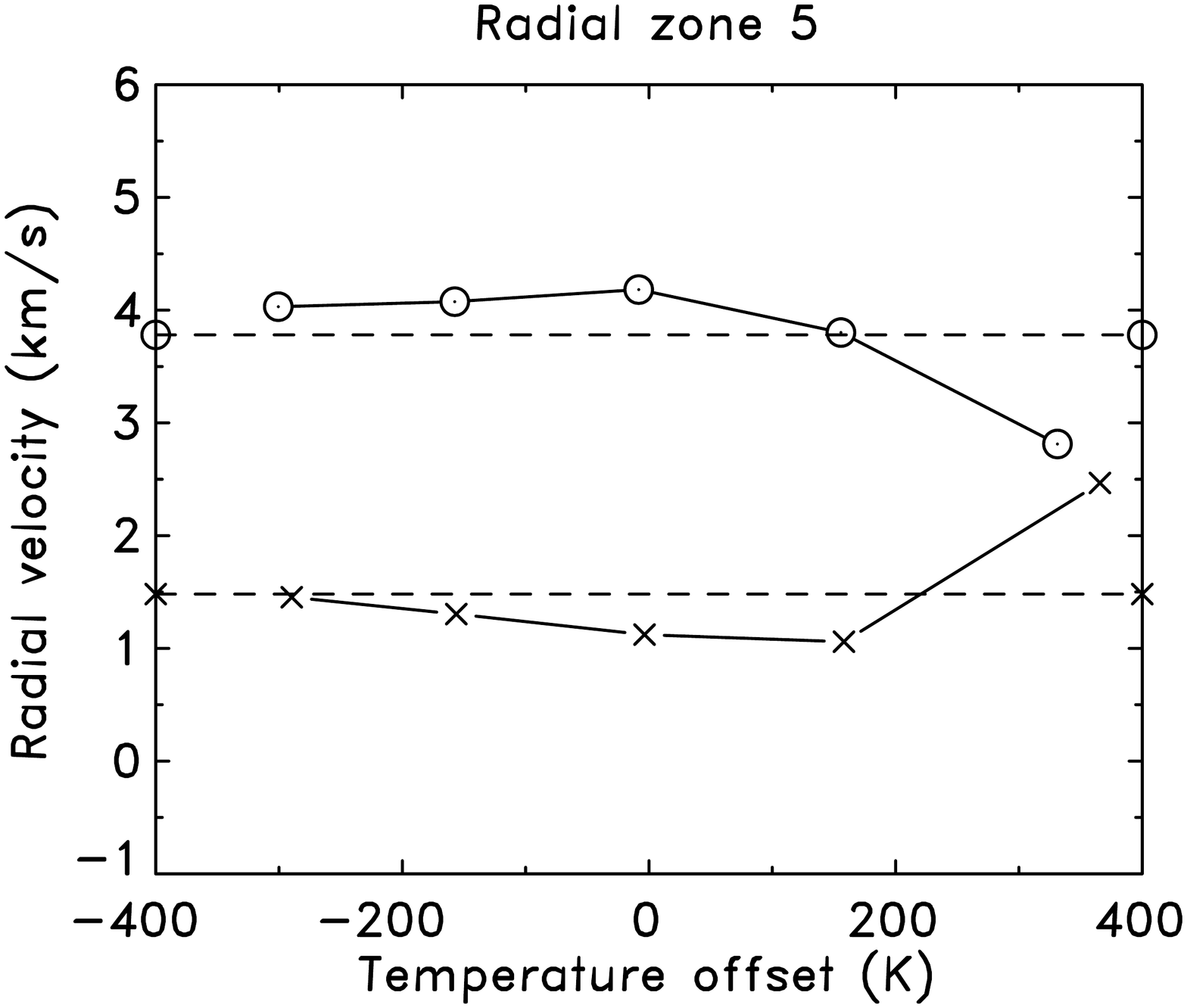}
\caption{The variation of vertical (left column) and radial outflow (right column) velocities with local (high-pass spatially filtered) temperature fluctuations, obtained by fitting the azimuthal variations of the inferred LOS velocities at $\tau_c=$~0.95, and shown for radial zones 2--5. The vertical velocities show the expected convective signature: the relatively cool component has downflows of up to 1.2~km~s$^{-1}$, the relatively hot gas upflows of up to -1.3~km~s$^{-1}$. Radial velocities are strong in the intra-spines (circles) and weak in the spines (X-symbols), but the hottest spine structures show radial outflows in excess of 3~km~s$^{-1}$. }
\label{fig:velocities2}
\end{figure}

\begin{figure}[htbp]
\centering
\includegraphics[width=0.49\textwidth,angle=0, clip]{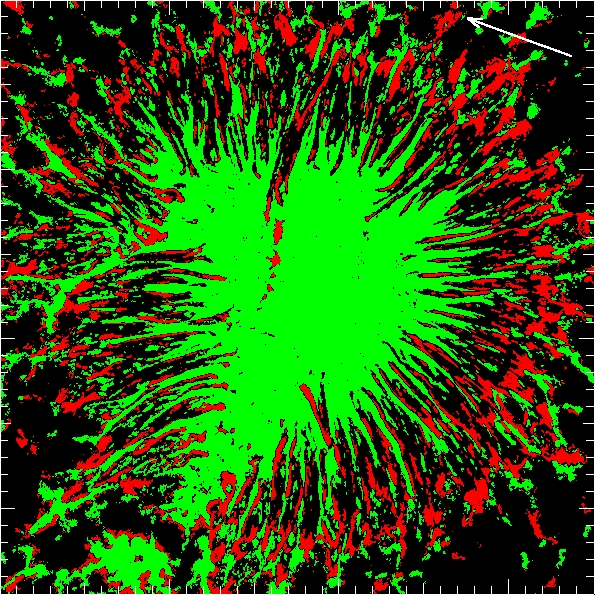}
  \caption{Composite masks showing the magnetic field with inclination less than 60\degr at $\tau_c=$~0.95 (in the solar frame) in green and vertical field in the solar frame of less than -100~G (opposite polarity field) in red. In the penumbra, the green color corresponds to the magnetic \emph{spines}, and the black and red colors correspond to the \emph{intra-spines}. \emph{The opposite polarity field in the penumbra is mostly found adjacent to the spines.} Opposite polarity field is found also near some strong flux concentrations (plage and pores), and in a few cases near network field. Areas with field strength smaller than 350~G are shown black. The FOV outlined with the white box and tick marks is $\sim 35\times 19$ \arcsec{} and corresponds to that shown in Figs.~\ref{fig:bz} and \ref{fig:bt}. Tick marks are at 1\arcsec{} intervals.}
  \label{fig:opp_pol}
\end{figure}

\subsubsection{Pore}
The lower-left part of the FOV in Fig.~\ref{fig:bt} shows a pore-like structure. Along its \emph{magnetic} boundary, which is well outside the visible boundary in the temperature map at $\tau_c=$~0.95, Fig.~\ref{fig:bz} shows several patches of opposite polarity field, one of which is outlined with an arrow. These opposite polarity patches are absent at $\tau_c=$~0.09 and 0.01. Figure \ref{fig:bt} (panel d) shows strong downflows at the same locations, in agreement with earlier observations of downflows adjacent to pores \citep{1999ApJ...510..422K, 2003A&A...405..331H} as well as with simulations \citep{2007A&A...474..261C}, and leading to the interpretation that some of the field is dragged down by the downflows. A high-quality movie of the same spot shows vigorous convection in the region between the pore and the sunspot, and also intensity patterns suggesting \emph{inflows} toward the pore \citep[as reported earlier by e.g.][]{1999ApJ...511..436S, 2002A&A...395..249R}, consistent with the (strong) downflows seen at its perimeter. Panel c in Fig.~\ref{fig:bt} shows a ring of transverse field surrounding the pore, indicating a strongly inclined field (a canopy) at the locations of these downflows. This transverse field is stronger on the limb side than on the disk center side of the pore, as is generally also the case for the sunspot penumbra, which is in agreement with expectations.

In the interior of the pore is a bright granular-like structure. The LOS magnetic field is strongly reduced near its center at $\tau_c=$~0.95, but at $\tau_c=$~0.01 the LOS magnetic field is almost the same as in the surroundings. This is consistent with a magnetic field folding over this ``granule'', having (strongly) reduced field strength in the deeper layers, as is the case for the umbral dots in the sunspot umbra and the light bridge. 

\subsection{Orientation of the horizontal magnetic field}

Figure~\ref{fig:bt}c shows the magnitude of the transverse field (in the observers frame) and orientation of the horizontal field (in the local frame) for every 20th pixel. As is evident, the horizontal field is approximately parallel to the penumbral filaments and there are no indications of major systematic errors from the calibration of, and compensation for, the telescope polarization. 

\begin{figure}
\centering
\includegraphics[angle=0,width=0.485\textwidth,clip]{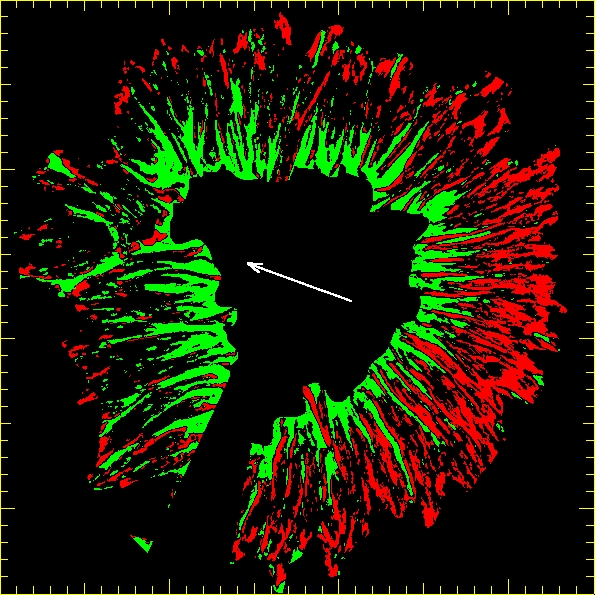}
  \caption{Composite masks showing the vertical magnetic field with $B_z >$~1~kG at $\tau_c=$~0.95 in the \emph{observers} frame shown in green, and the field with $B_z <$~-150~G (opposite polarity field) in red. Note that opposite polarity field is seen adjacent to the spines also in many locations at the \emph{disk center side}. Tick marks are at 1\arcsec{} intervals.}
  \label{fig:opp_pol2}
\end{figure}

\begin{figure}
\center
 \includegraphics[bb=55 45 510 700,angle=-90,width=0.24\textwidth,clip]{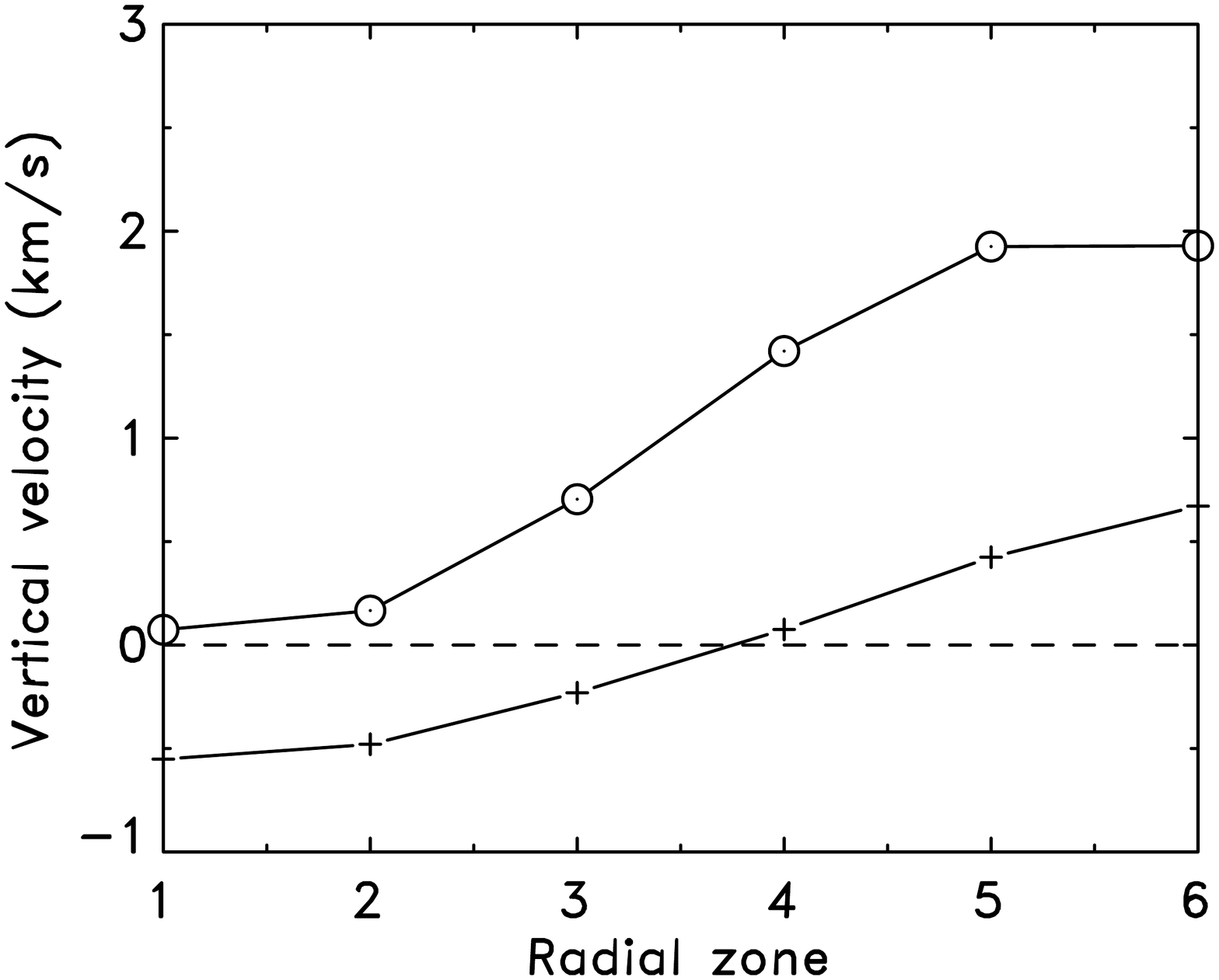}
 \includegraphics[bb=55 45 510 700,angle=-90,width=0.24\textwidth,clip]{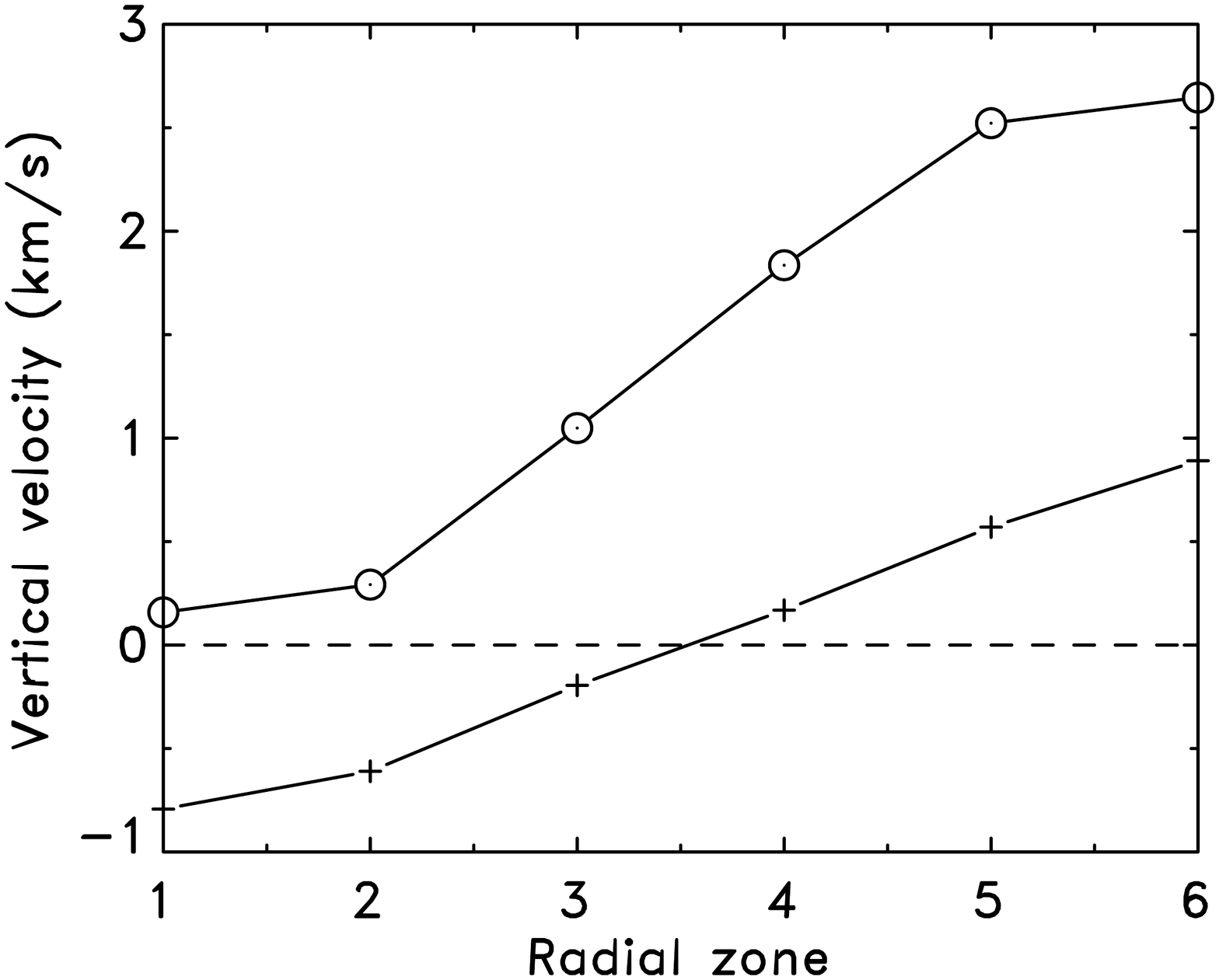}
 \includegraphics[bb=55 45 510 700,angle=-90,width=0.24\textwidth,clip]{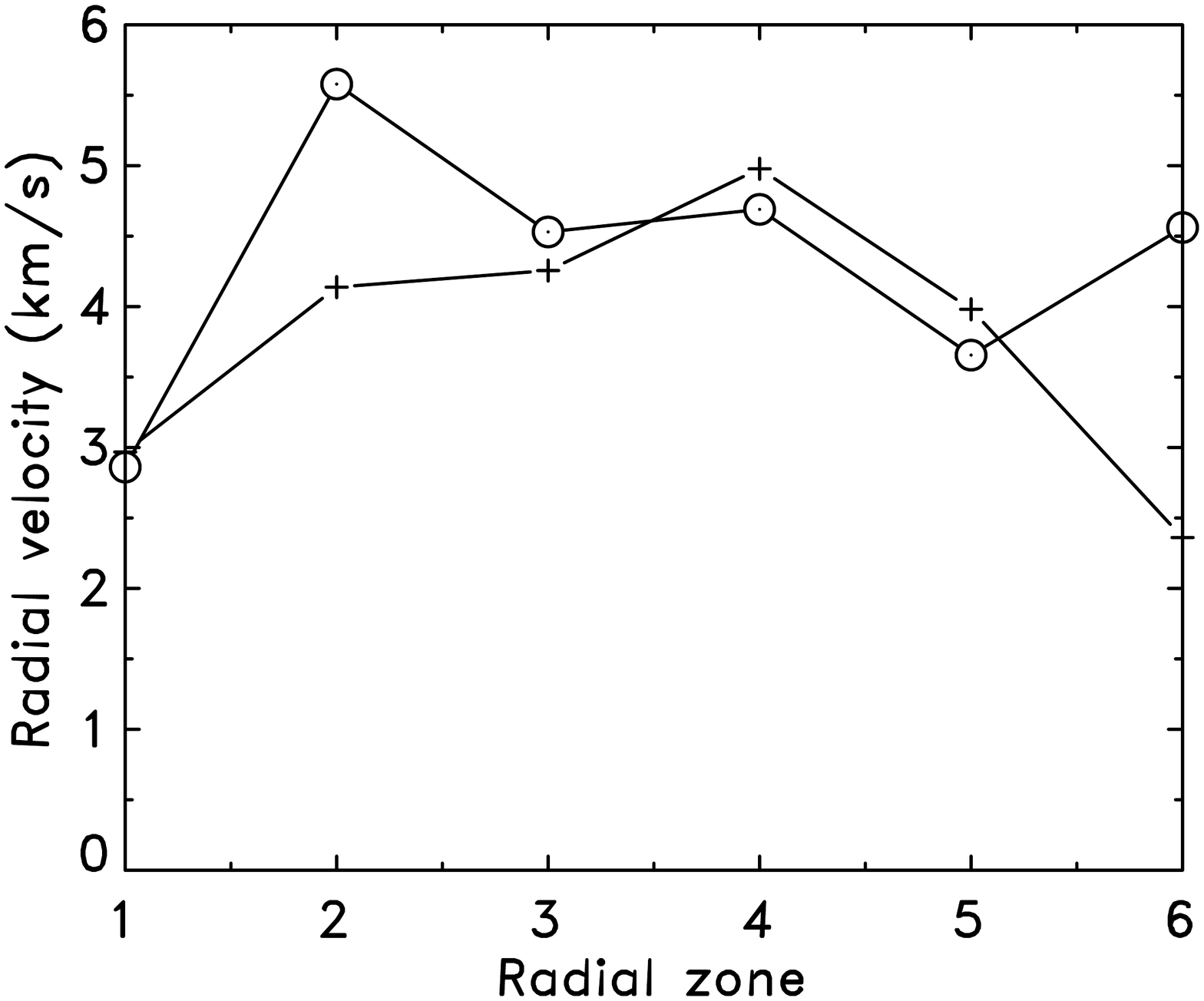}
\includegraphics[bb=55 45 510 700,angle=-90,width=0.24\textwidth,clip]{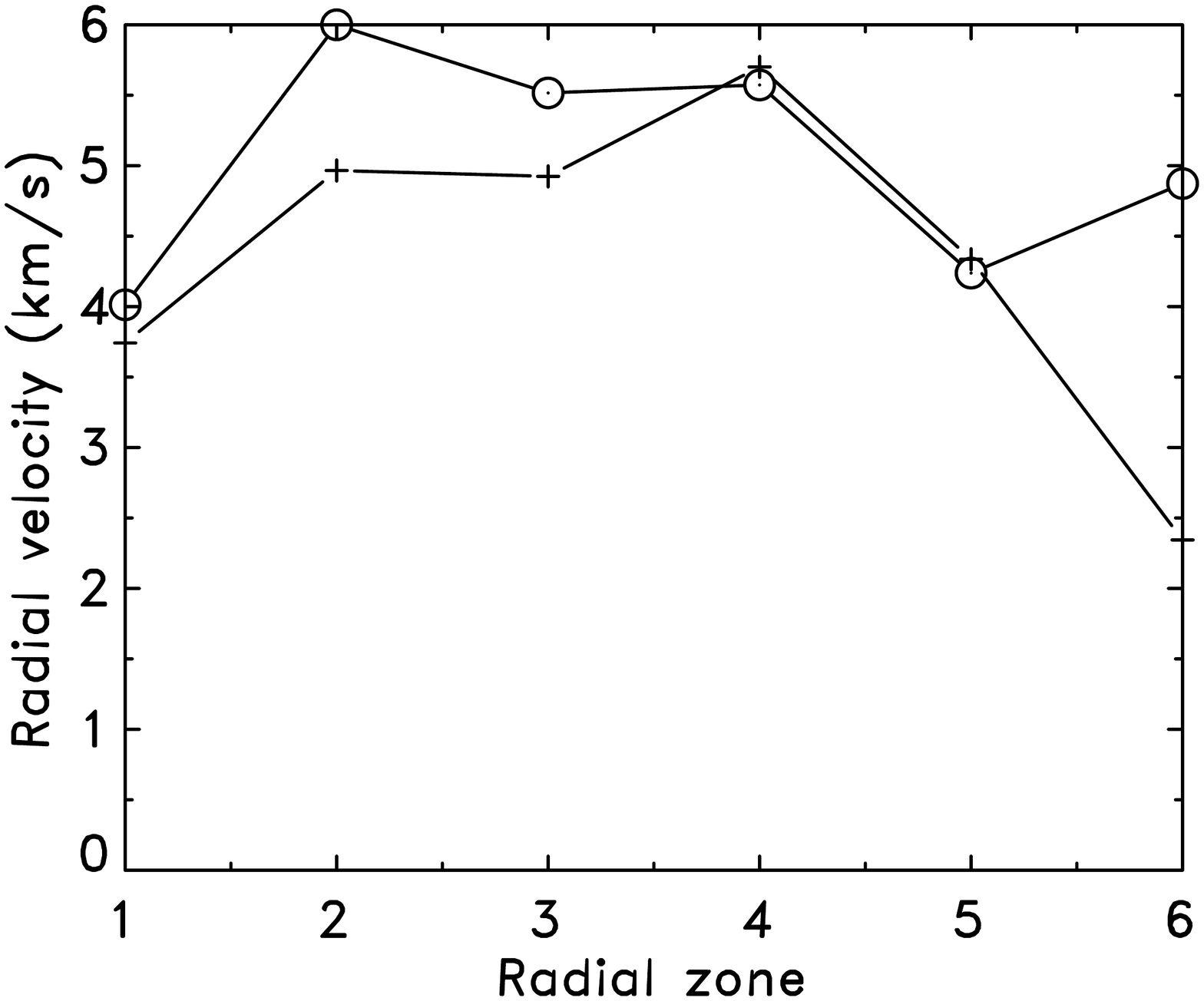}
\includegraphics[bb=55 45 574 700,angle=-90,width=0.24\textwidth,clip]{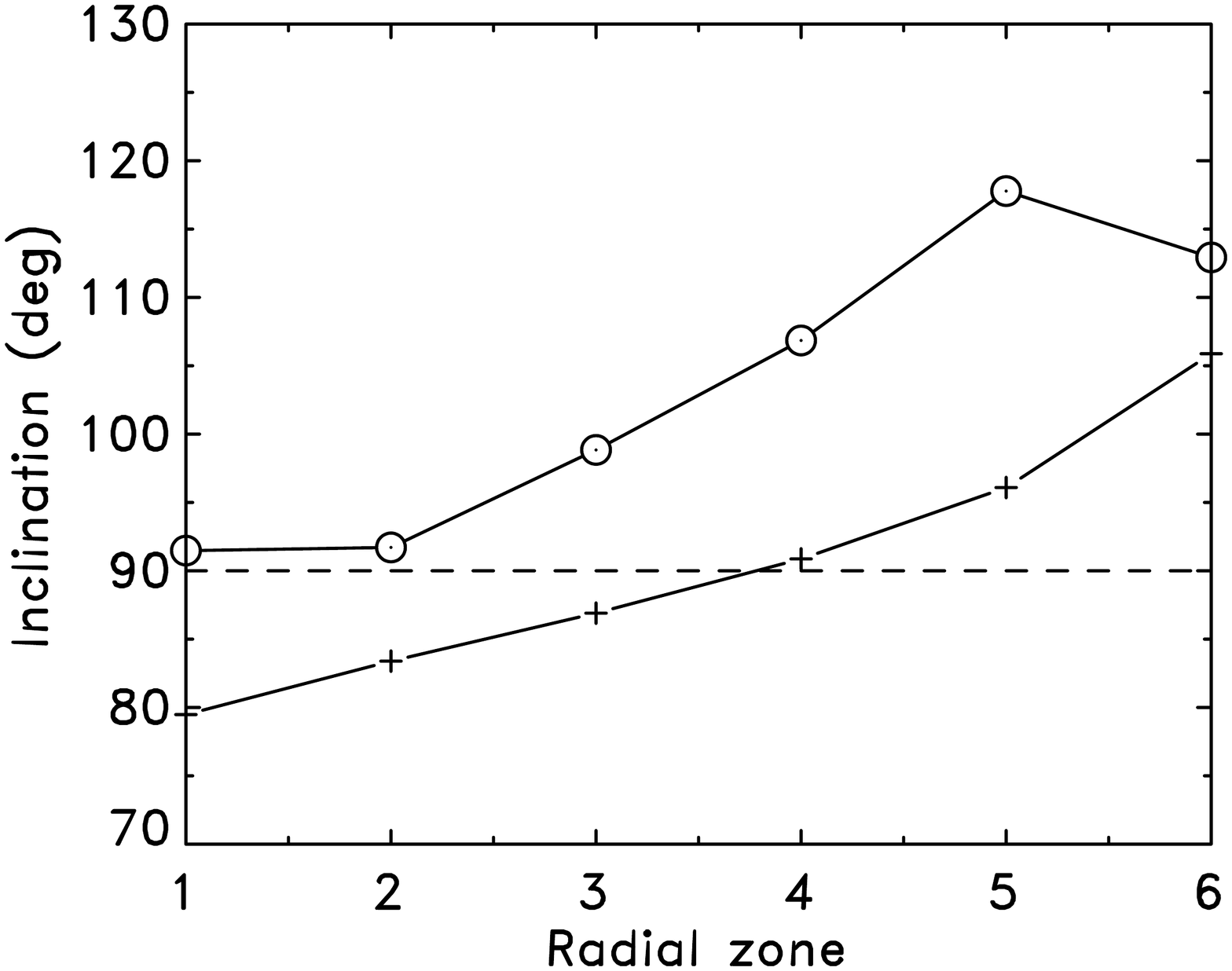}
\includegraphics[bb=55 45 574 700,angle=-90,width=0.24\textwidth,clip]{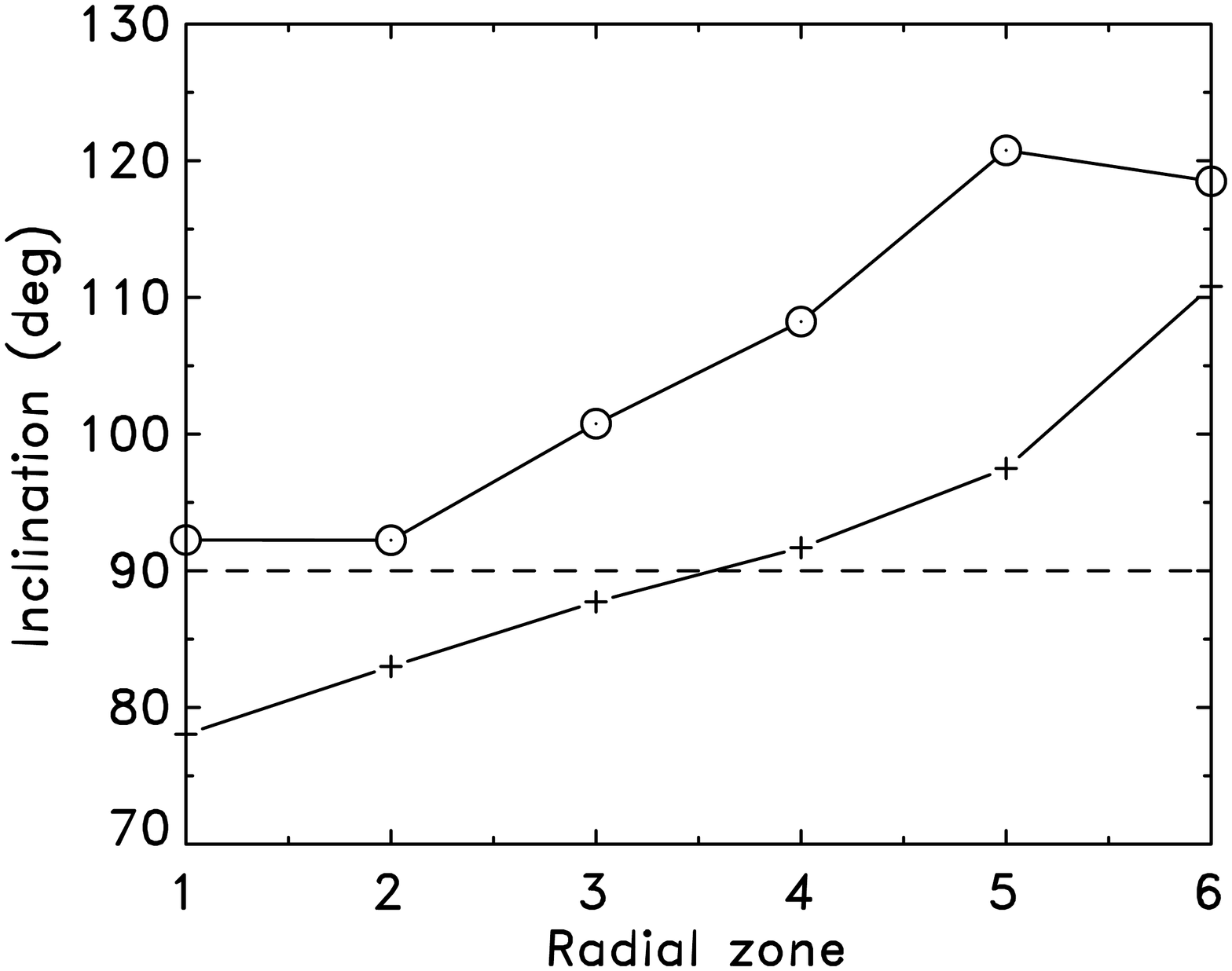}

\caption{Average flow properties of opposite polarity patches in the penumbra, obtained by making azimuthal fits of the inferred LOS velocities at $\tau_c=$~0.95 within the red patches (circles), and for all other (i.e., excluding the opposite polarity field) intra-spine pixels (plus symbols) shown in Fig.~\ref{fig:opp_pol}. The left column shows the results obtained with inversions using 2 nodes for the LOS velocity, the right column with 3 nodes. The penumbra is divided into 6 radial zones, where zone 1 corresponds to the innermost penumbra. On the average, the \emph{opposite polarity patches show genuine downflows} throughout the penumbra, corresponding to a downflow of 0.6--1.3~km~s$^{-1}$ relative to the average of all intra-spine pixels, and an increased inclination by 10--20\degr, compared to the intra-spines. The radial outflow velocity of the opposite polarity field is similar to that of other intra-spine pixels.} 
\label{fig:downflows}
\end{figure}

\begin{figure}
\centering
\begin{overpic}[width=0.495\textwidth,angle=0]{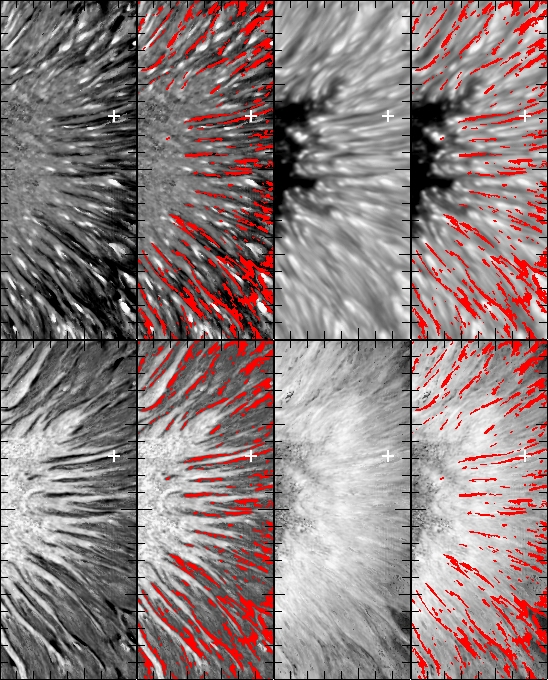} 
\put(1,52){\textcolor{white}{\textbf{\large a)}}}
\put(42,52){\textcolor{white}{\textbf{\large b)}}}
\put(1,2){\textcolor{white}{\textbf{\large c)}}}\index{\footnote{}}
\put(42,2){\textcolor{white}{\textbf{\large d)}}}
\end{overpic}

  \caption{
  Panels a-d show properties of the limb-side penumbra (excluding the outermost penumbra) at $\tau_c=$~0.95. Pixels with opposite polarity field in the \emph{solar} frame are indicated with red color. Panel a shows the LOS velocity (clipped at (-2.5, 2.5)~km~s$^{-1}$), panel b the temperature (clipped at (4500, 6500)~K), panel c the vertical magnetic field (in the solar frame; clipped at (-1000, 2200)~G), and panel d the horizontal magnetic field strength (in the solar frame; clipped at 1500~G. The FOV shown is 8$\times$20\arcsec. Tick marks are at 1\arcsec{} intervals.} 
  \label{fig:limb_side}
\end{figure}

\begin{figure}
 \centering
\includegraphics[bb=201 35 508 705,angle=-90,width=0.235\textwidth,clip]{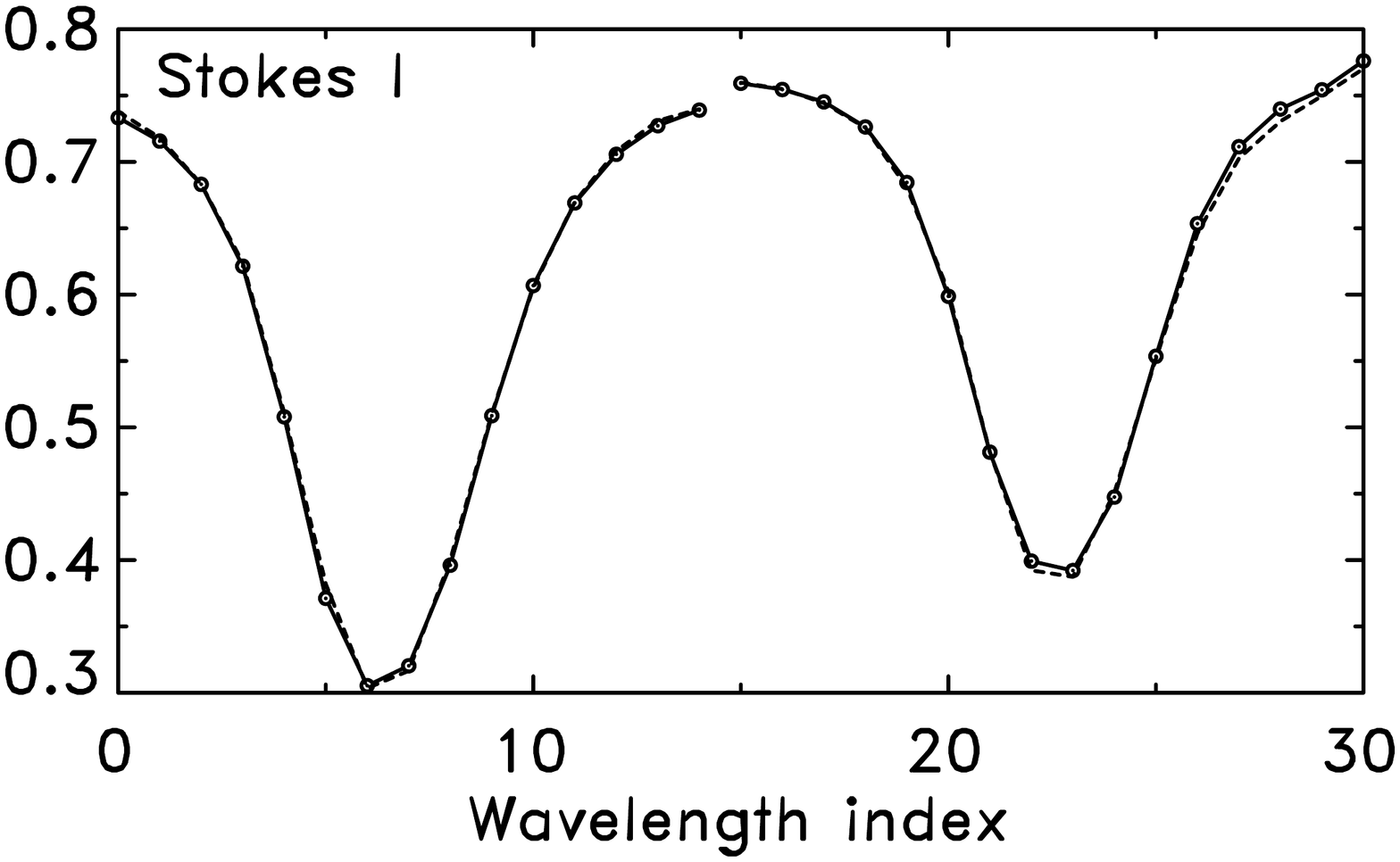}
\includegraphics[bb=201 35 508 705,angle=-90,width=0.235\textwidth,clip]{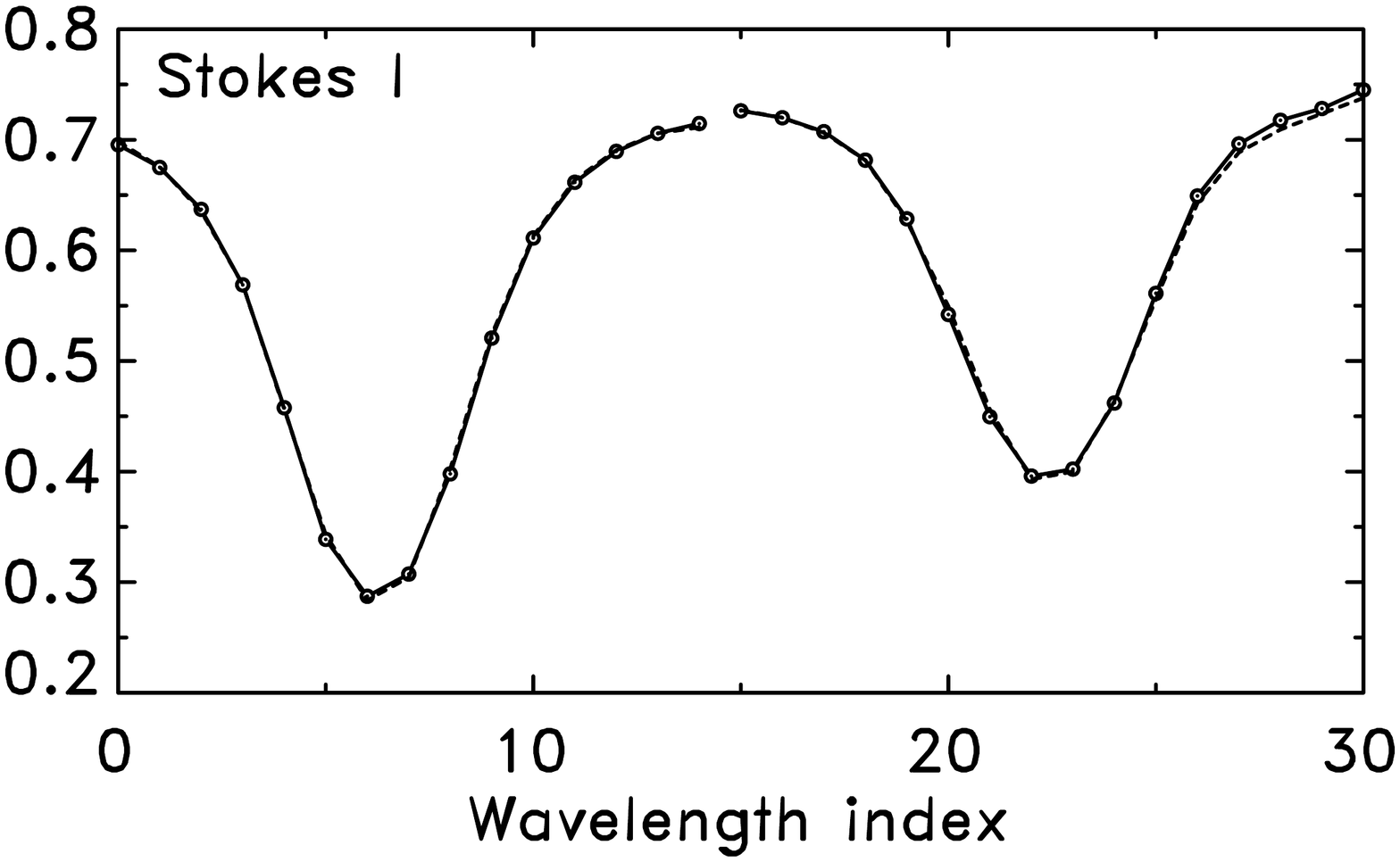}
\includegraphics[bb=201 35 508 705,angle=-90,width=0.235\textwidth,clip]{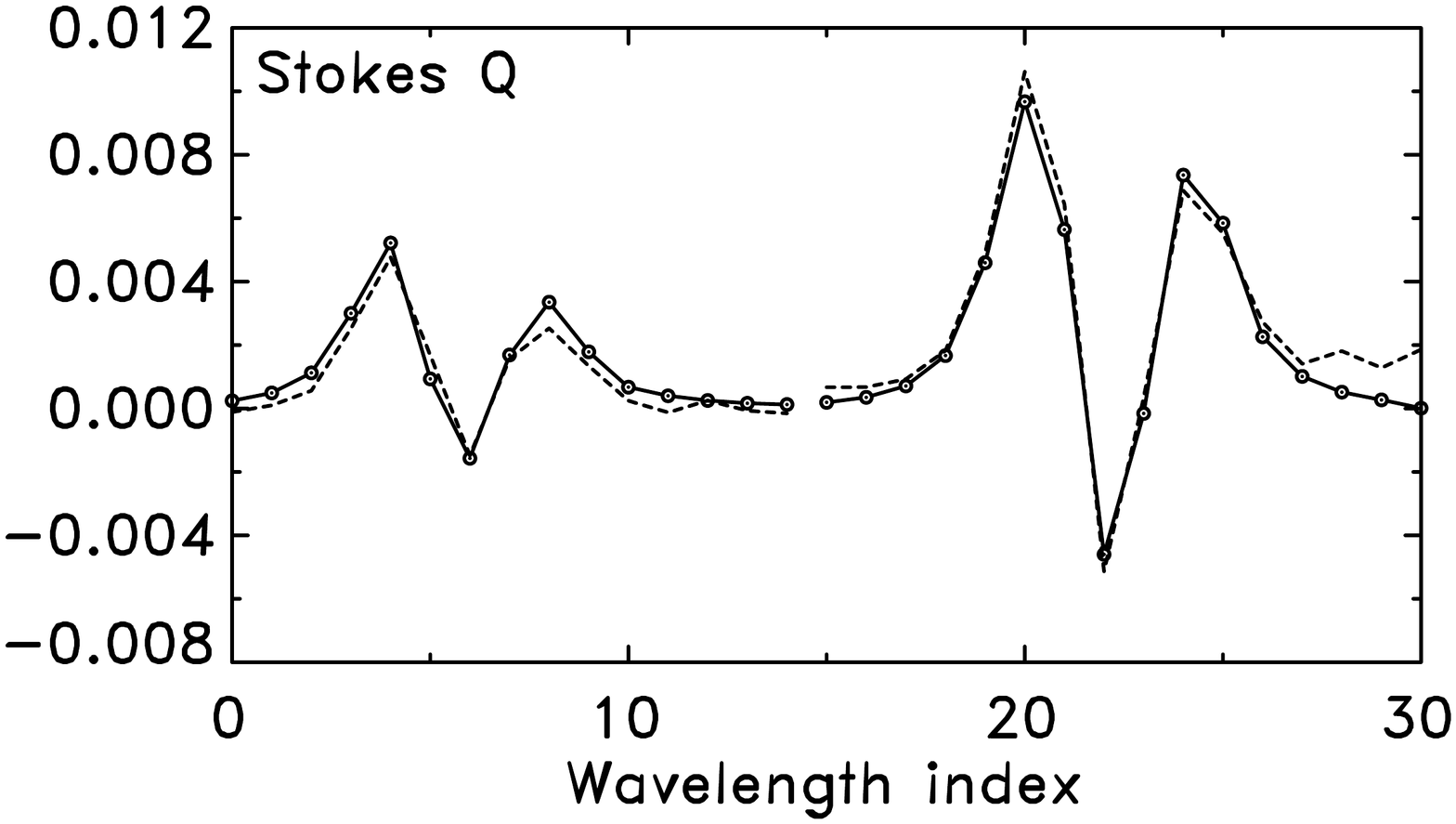}
\includegraphics[bb=201 35 508 705,angle=-90,width=0.235\textwidth,clip]{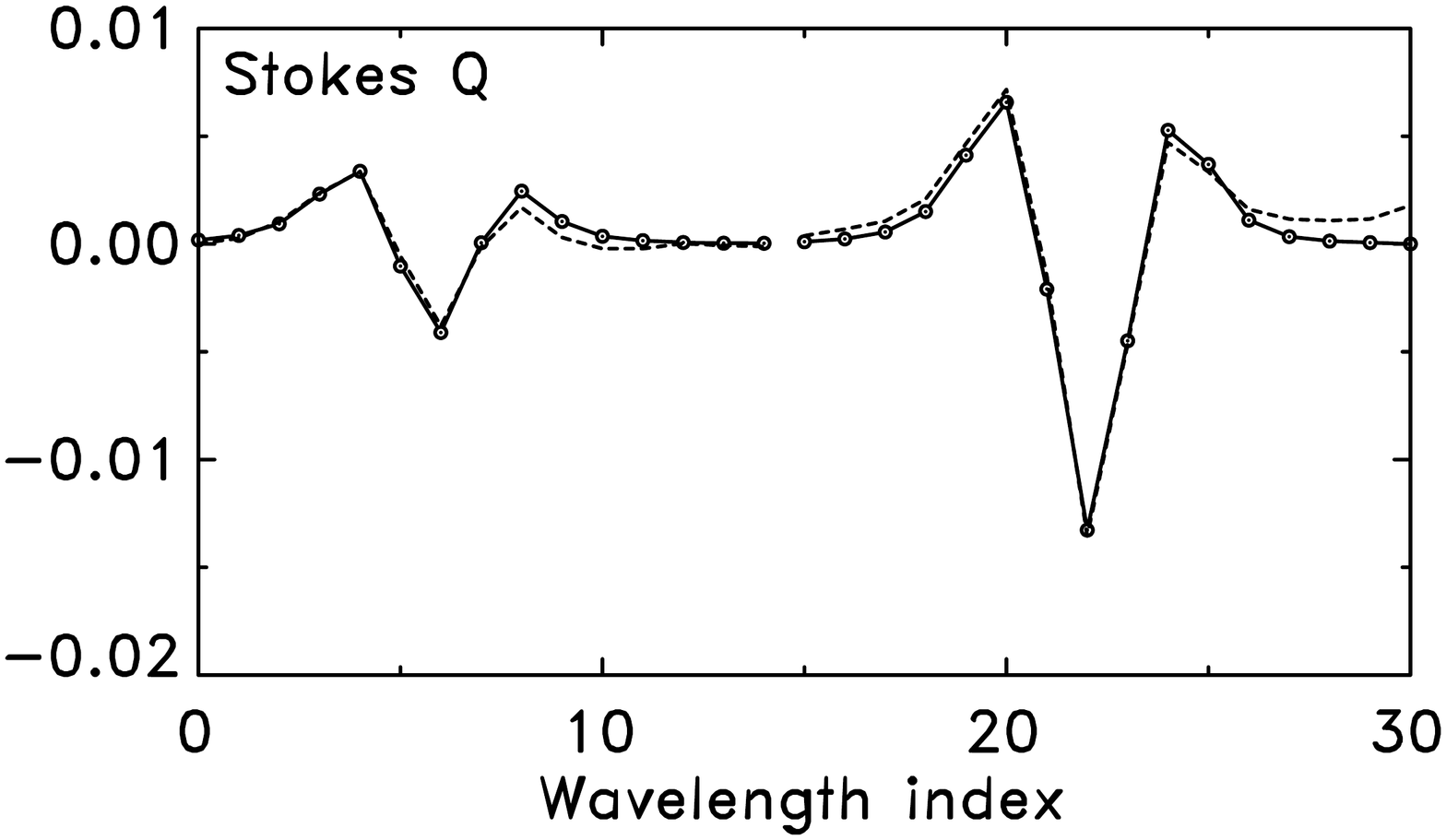}
\includegraphics[bb=201 35 508 705,angle=-90,width=0.235\textwidth,clip]{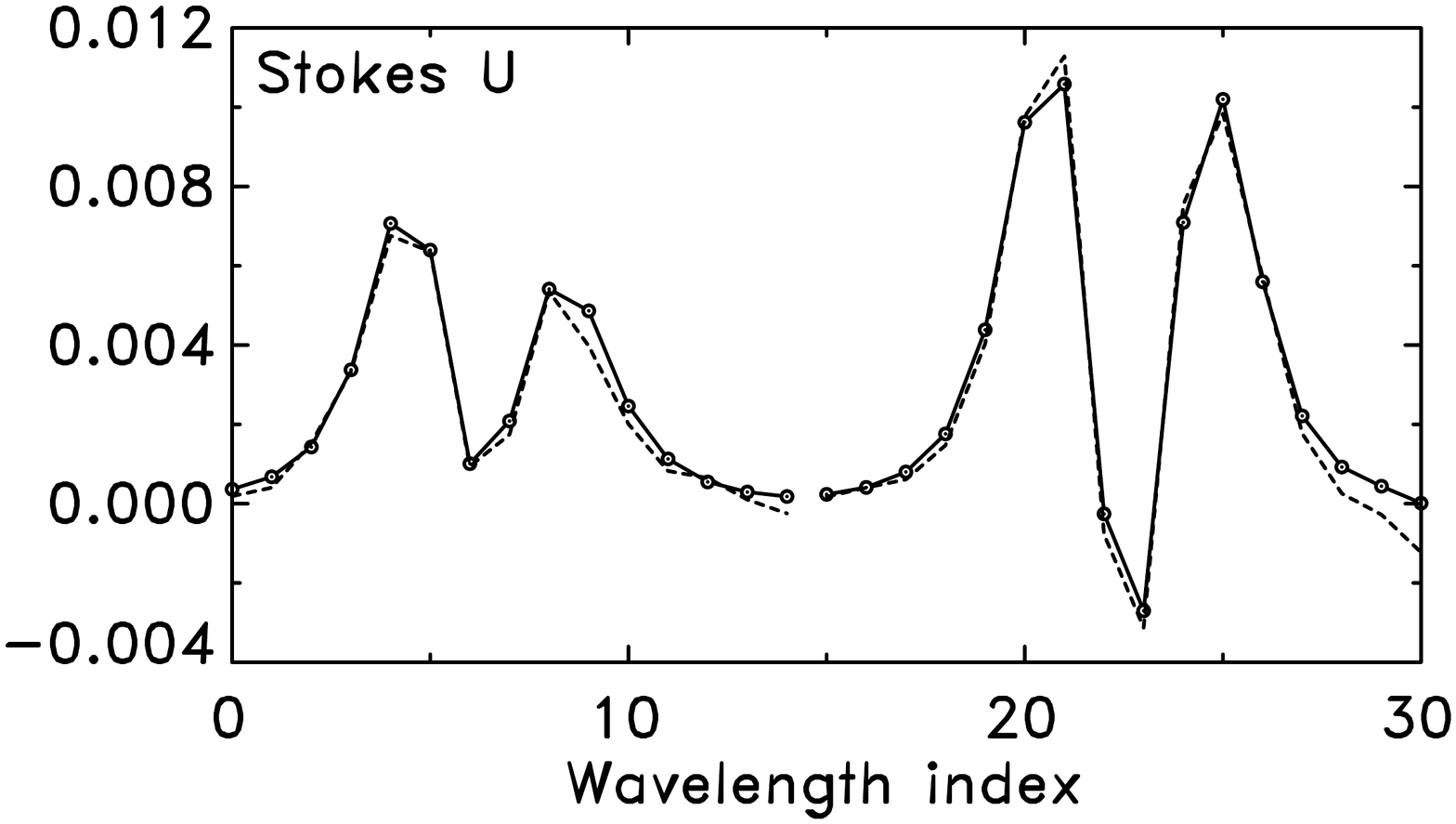}
\includegraphics[bb=201 35 508 705,angle=-90,width=0.235\textwidth,clip]{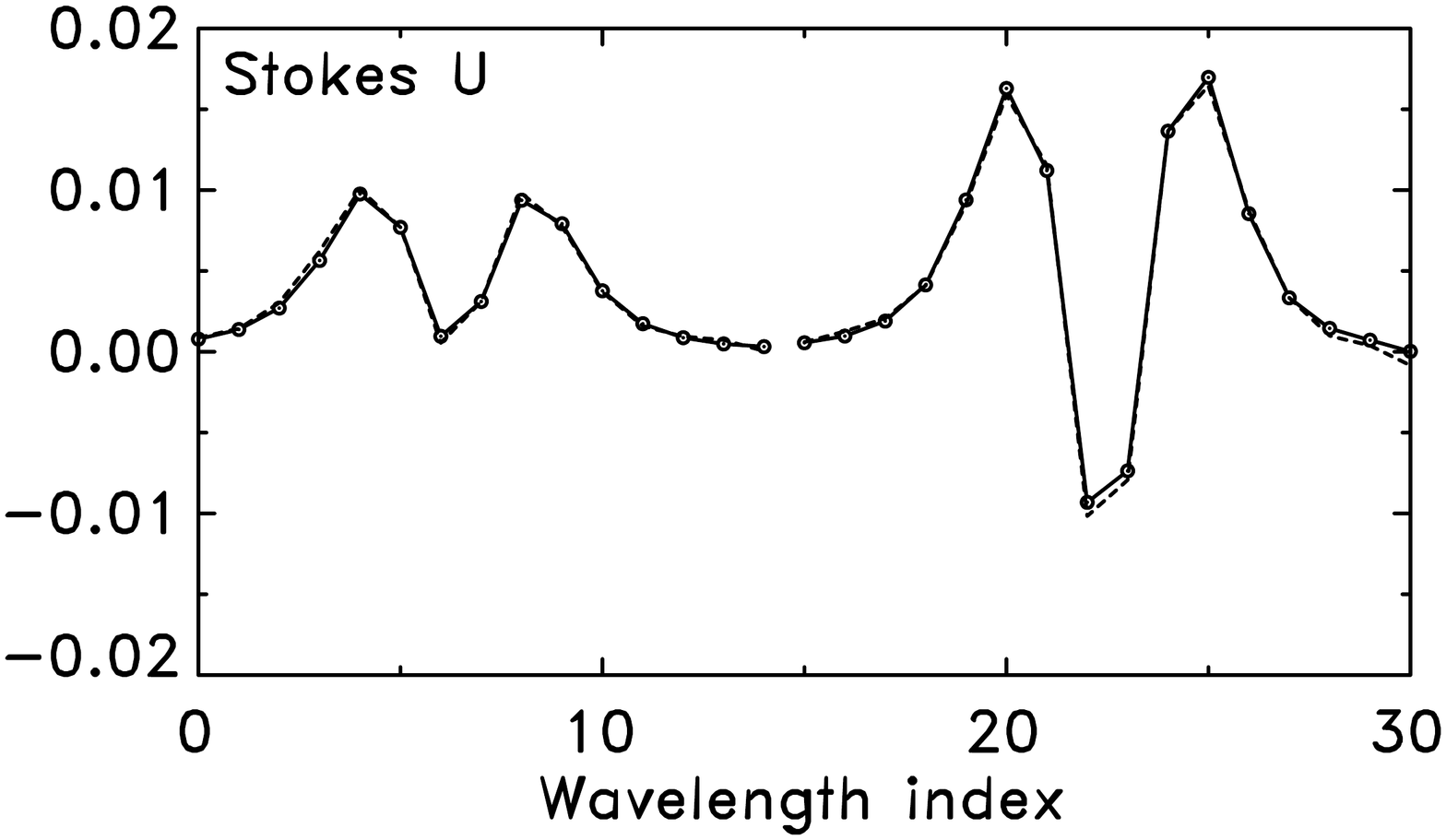}
\includegraphics[bb=201 35 575 705,angle=-90,width=0.235\textwidth,clip]{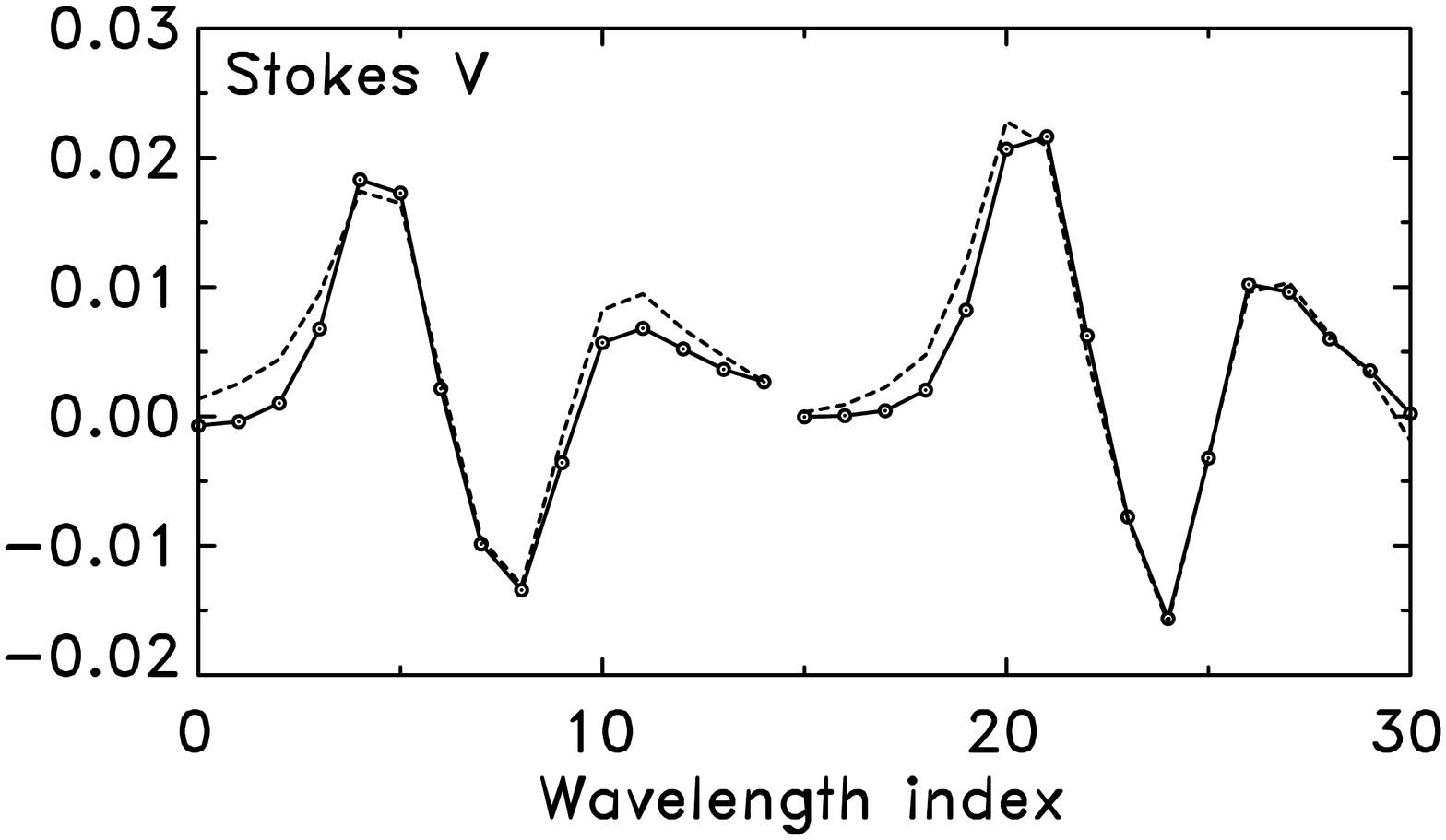}
\includegraphics[bb=201 35 575 705,angle=-90,width=0.235\textwidth,clip]{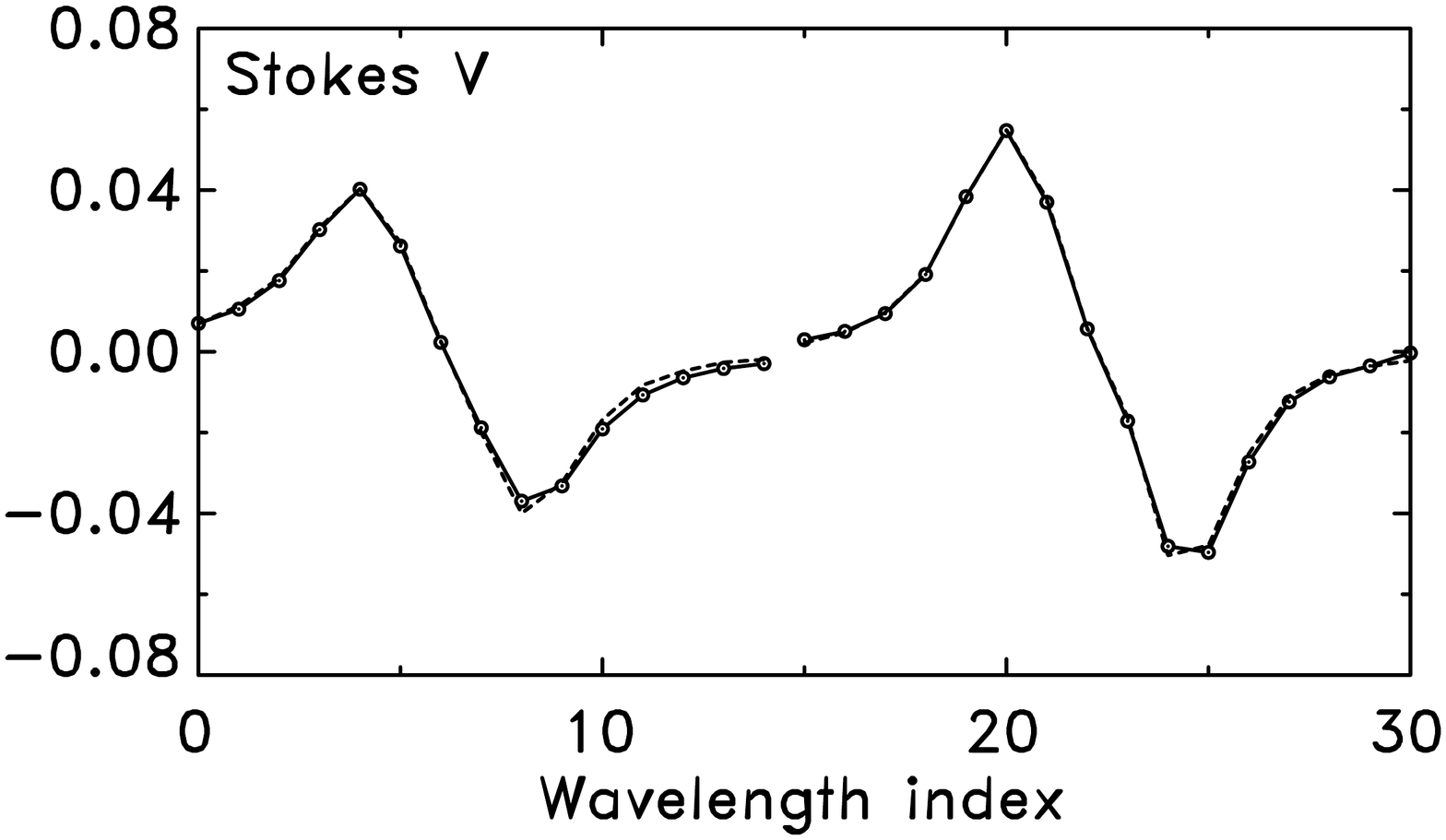}

\caption{Observed (solid) and fitted synthetic (dashed) 630.15/630.25~nm Stokes profiles from within the limb side penumbra shown in Fig.~\ref{fig:limb_side}. The left column shows the \emph{averaged profiles from all opposite polarity pixels} (marked with red color in Fig.~\ref{fig:limb_side}). Note the abnormal Stokes $V$ profile with 3 lobes, where the third ``extra'' lobe appears in the \emph{red wings} of the lines, and that the abnormal Stokes $V$ profile is fitted reasonably well by the inversions. The right column shows the averaged profiles from all other pixels within the same subfield, for which the average Stokes $V$ profile is normal. The line profiles (plotted side by side) are normalized to the average continuum intensity outside the spot.
}
\label{fig:plots}
\end{figure}

\begin{figure}
 \centering
\includegraphics[bb=201 54 510 705,angle=-90,width=0.235\textwidth,clip]{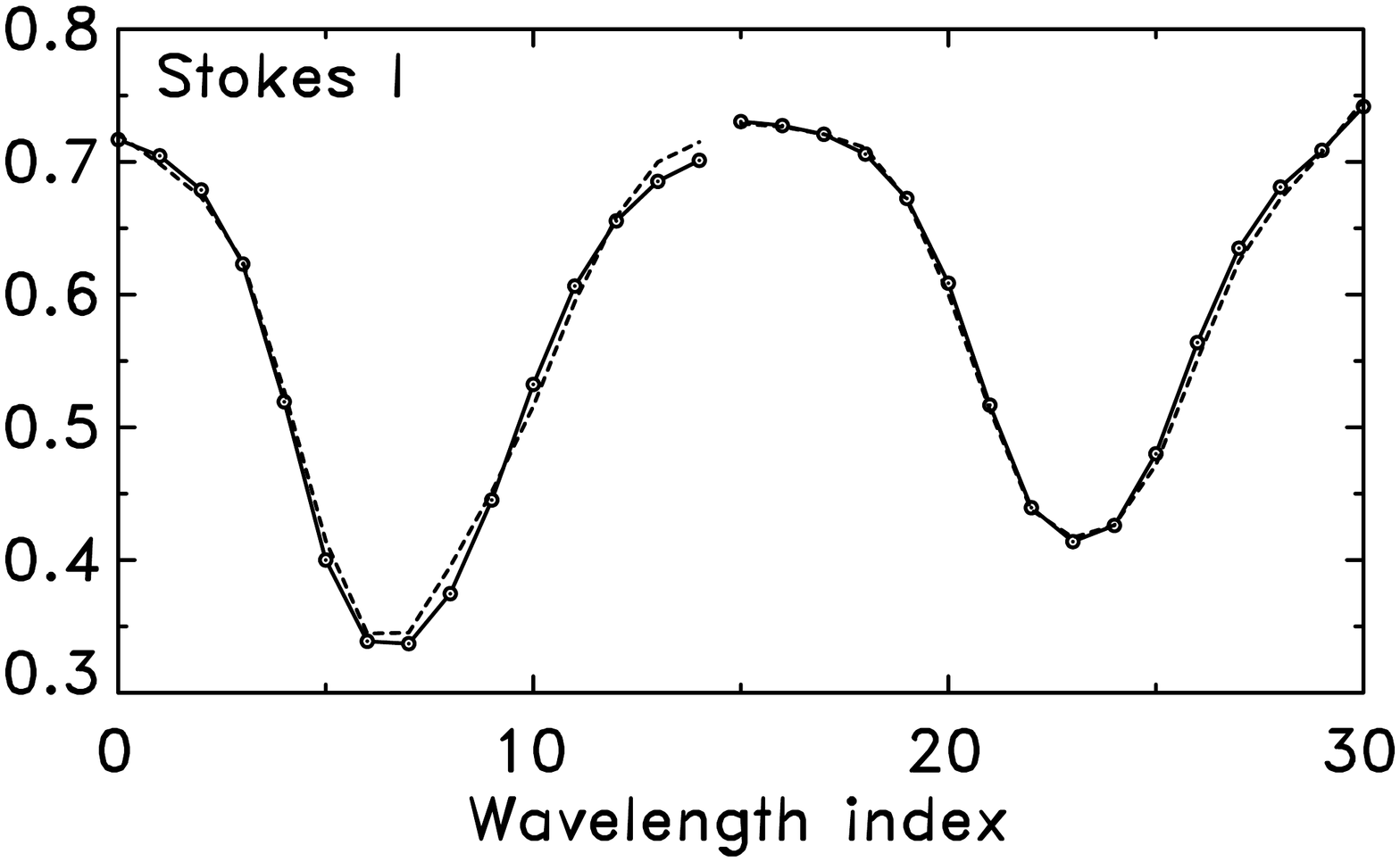}
\includegraphics[bb=201 54 510 705,angle=-90,width=0.235\textwidth,clip]{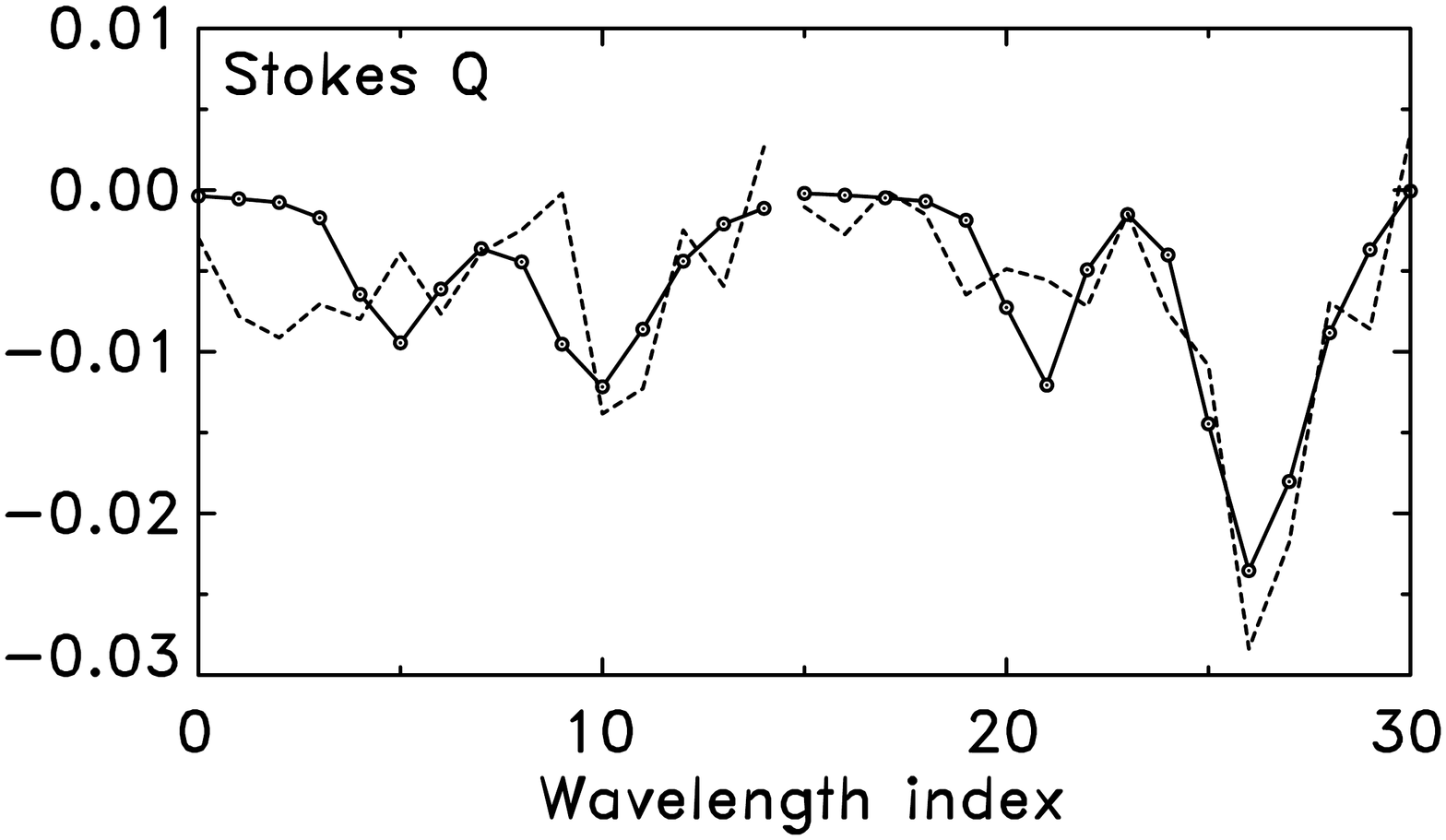}
\includegraphics[bb=201 54 580 705,angle=-90,width=0.235\textwidth,clip]{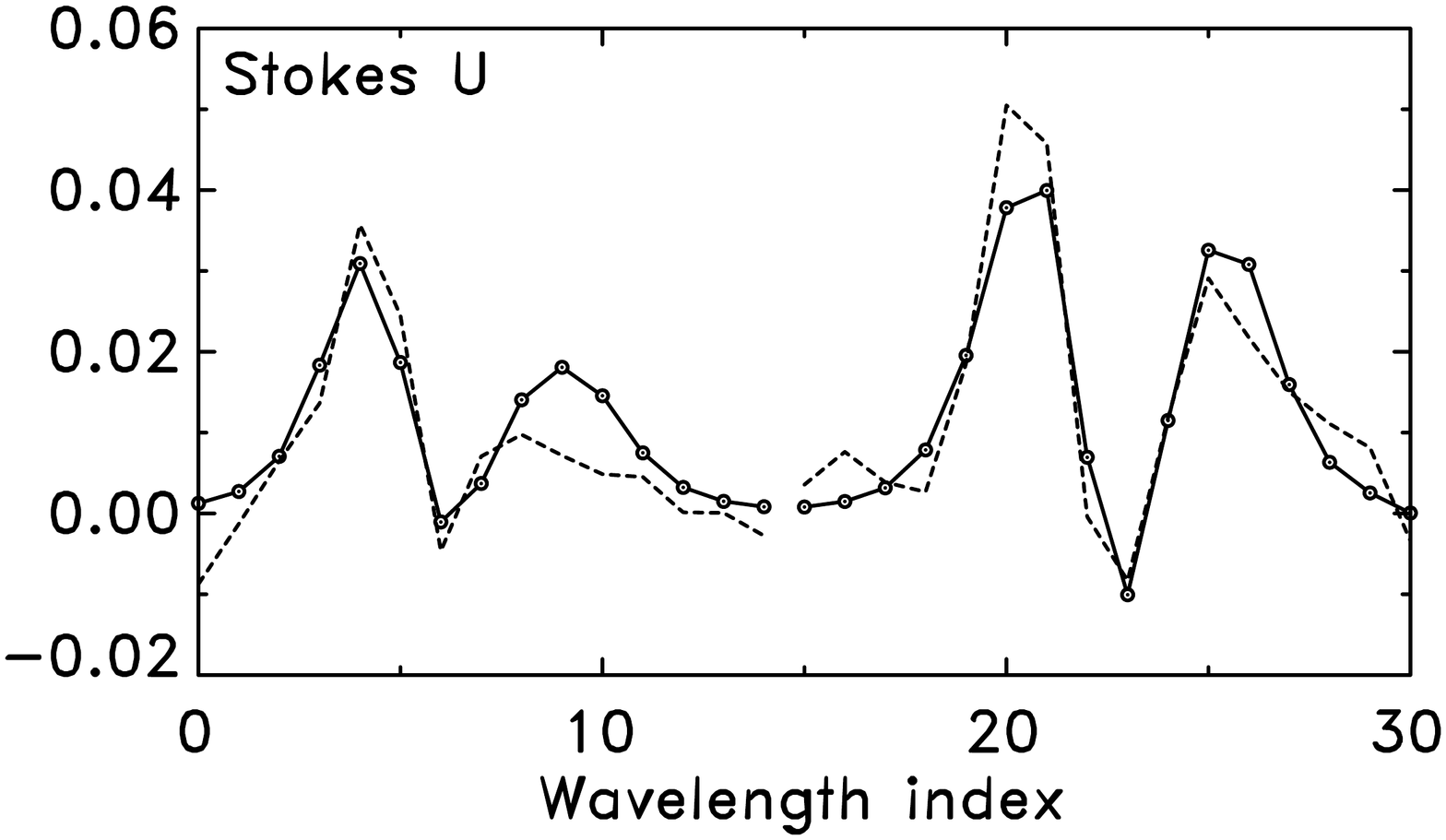}
\includegraphics[bb=201 54 580 705,angle=-90,width=0.235\textwidth,clip]{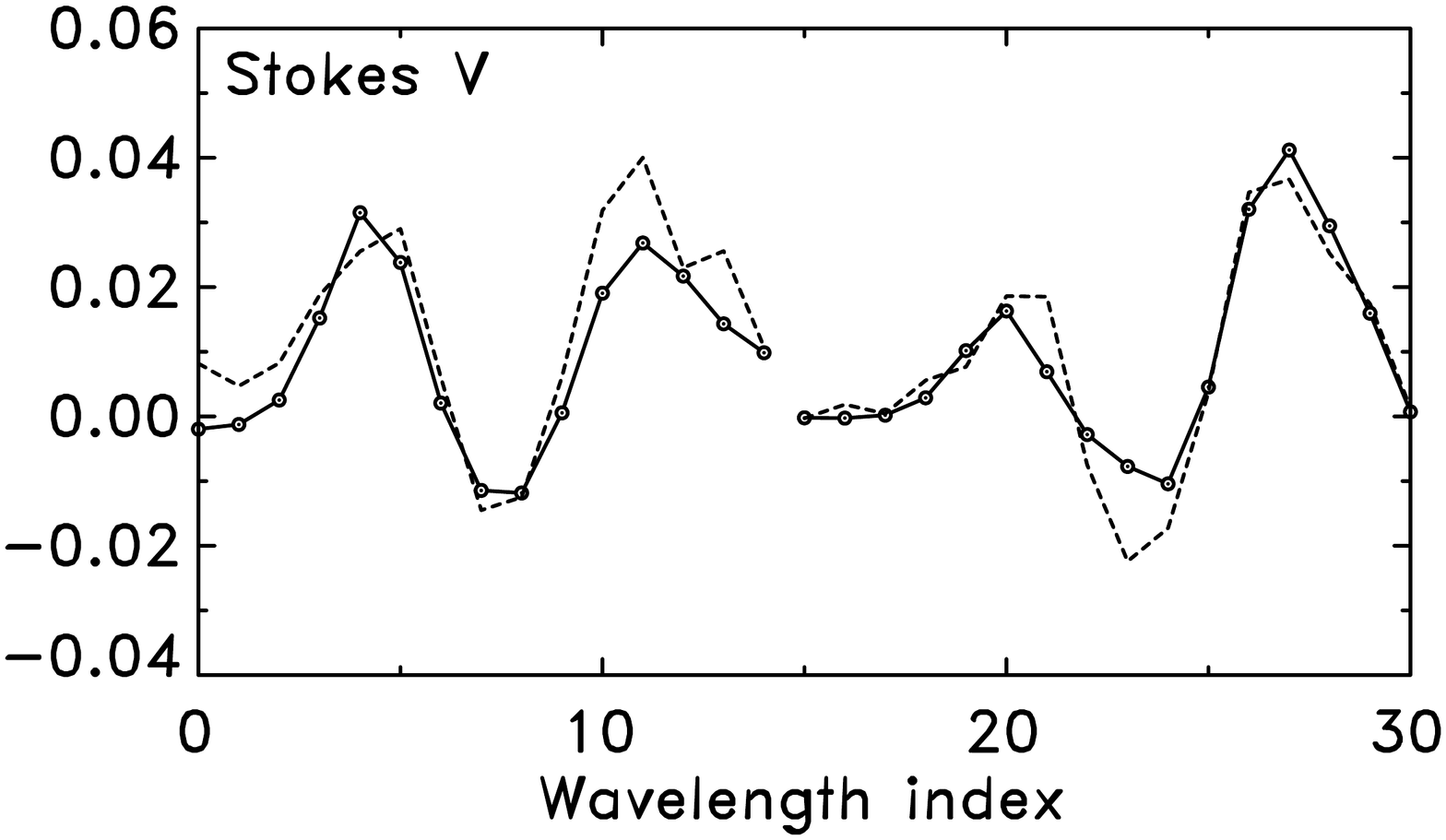}
\includegraphics[bb=201 54 580 705,angle=-90,width=0.235\textwidth,clip]{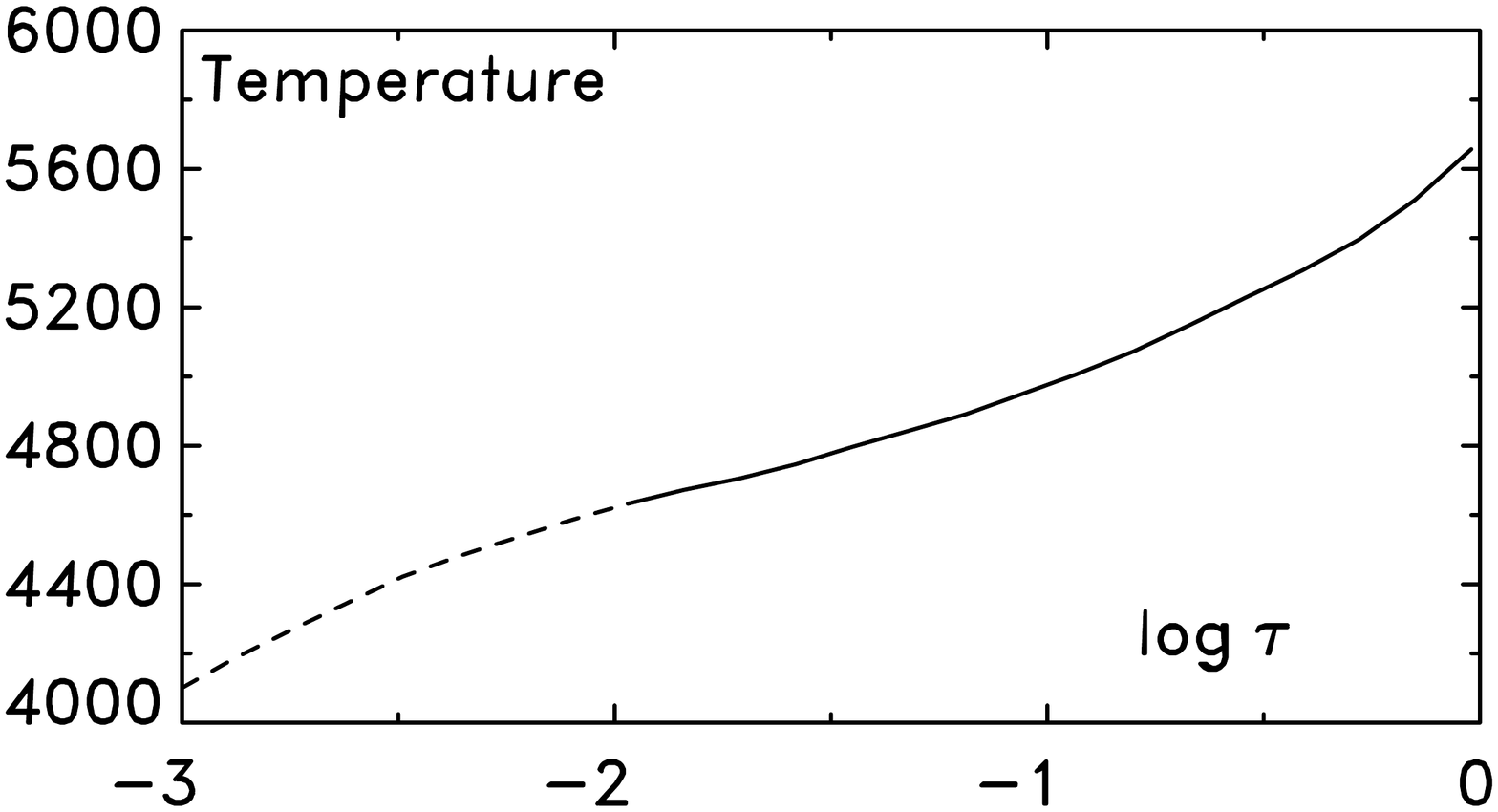}
\includegraphics[bb=201 54 580 705,angle=-90,width=0.235\textwidth,clip]{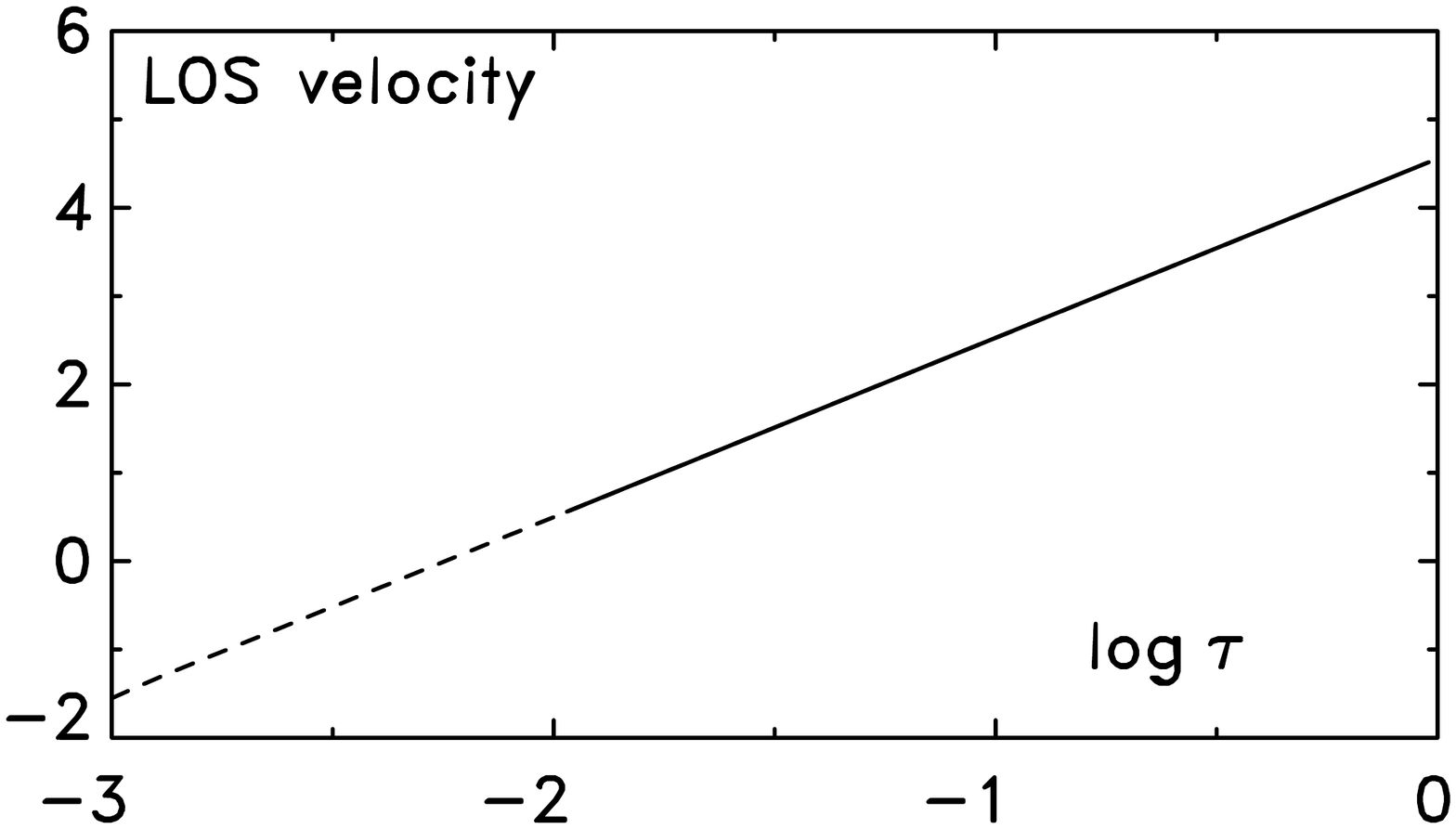}
\includegraphics[bb=201 54 580 705,angle=-90,width=0.235\textwidth,clip]{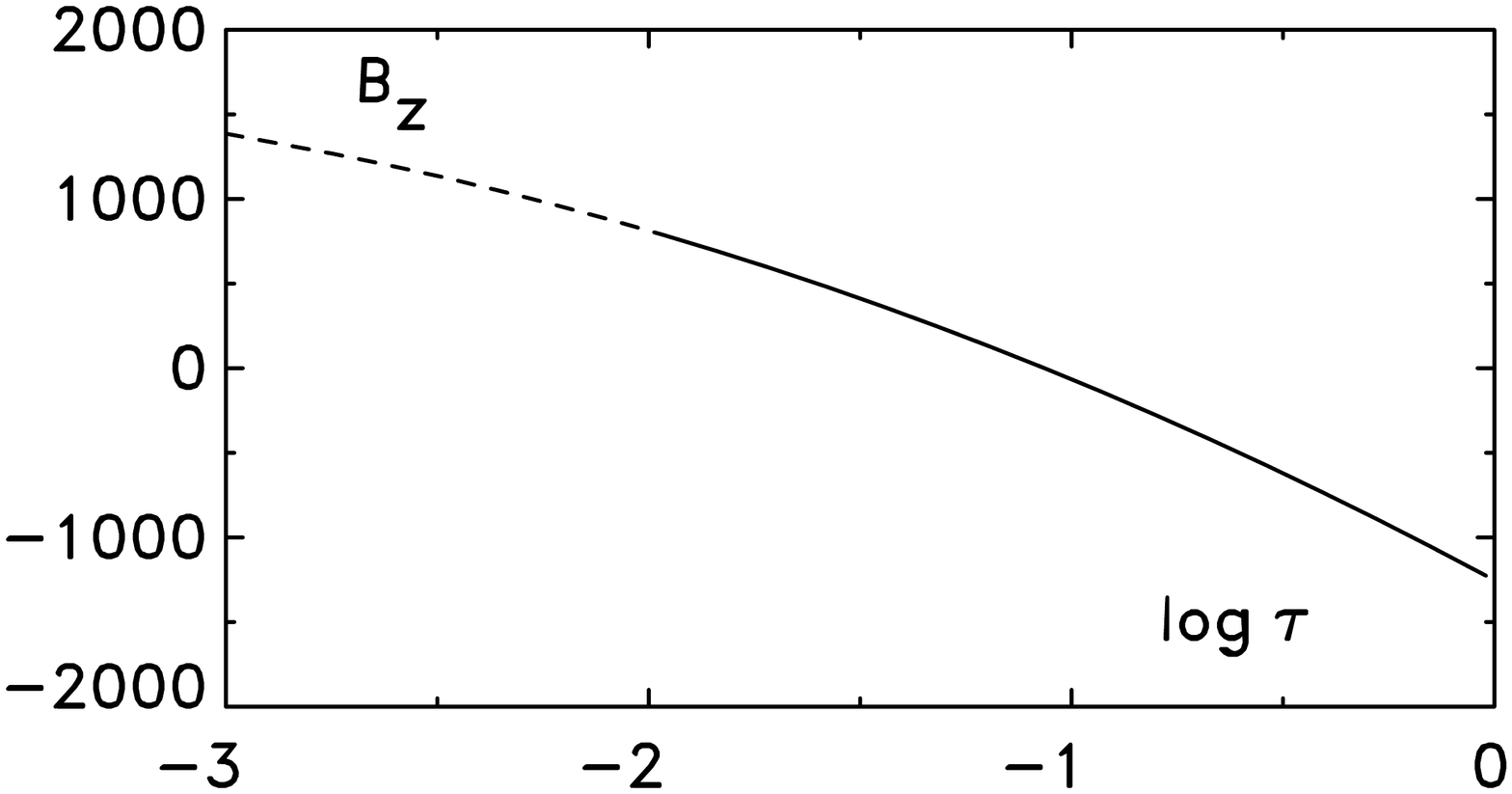}
\includegraphics[bb=201 54 580 705,angle=-90,width=0.235\textwidth,clip]{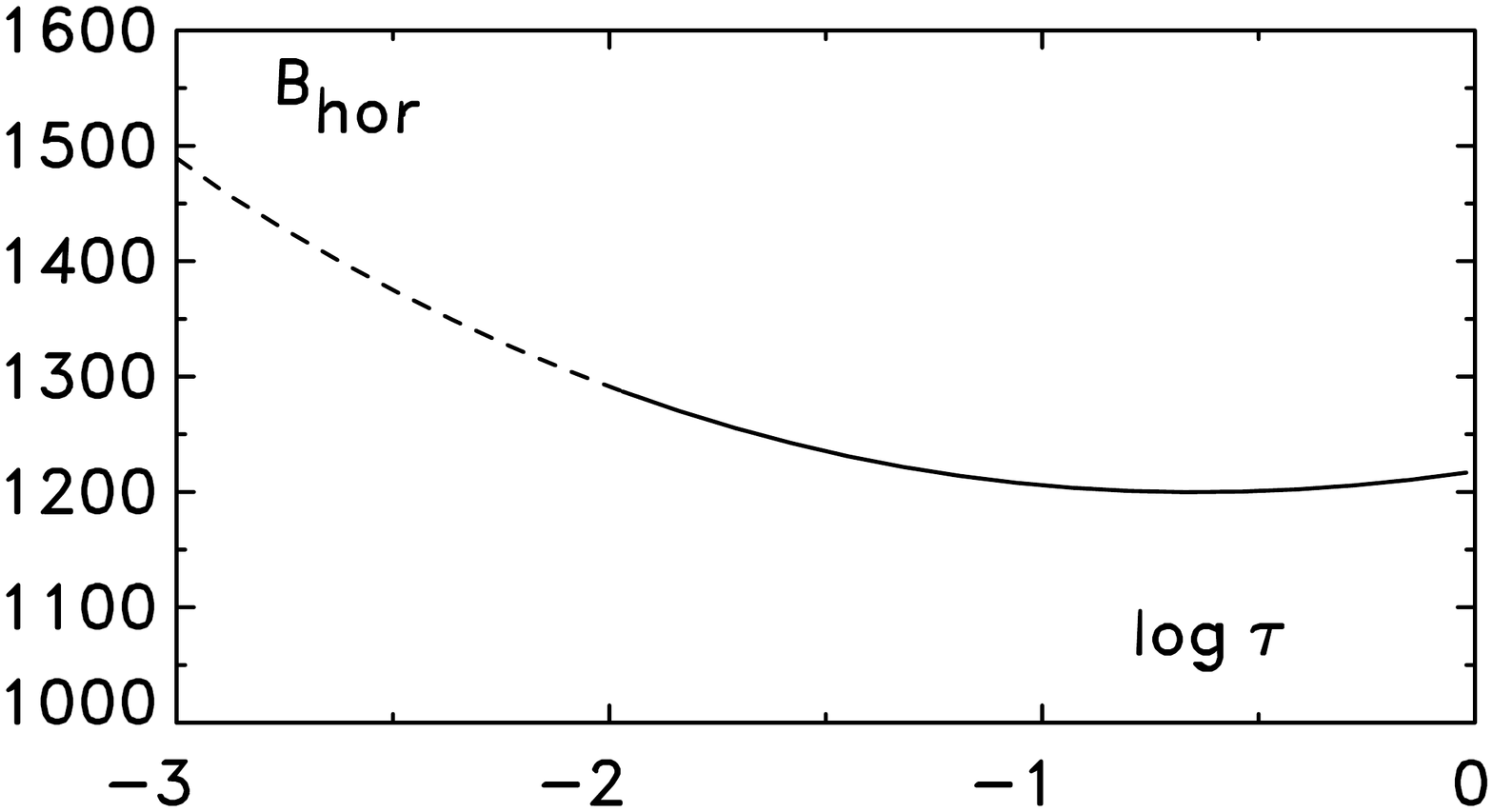}

\caption{The upper 4 panels show the observed (dashed) and fitted synthetic (full) Stokes profiles from the pixel marked with ``+'' in Fig.~\ref{fig:limb_side}. This pixel is adjacent to a spine and associated with a strong redshift (4 km~s$^{-1}$ at $\tau_c=$~0.95). Note the abnormal Stokes $V$ profile with 3 lobes. The synthetic fitted profile shows the correct shape, but does not reproduce all details of the observed profile. The 630.15/630.25~nm line profiles (plotted side by side) are normalized to the average continuum intensity outside the spot. The lower 4 panels show the variations with continuum optical depth of the temperature, LOS velocity, and vertical and horizontal magnetic field (in the solar frame) returned by NICOLE. 
}
\label{fig:plots2}
\end{figure}

\begin{figure}
\centering
\begin{overpic}[width=0.495\textwidth,angle=0]{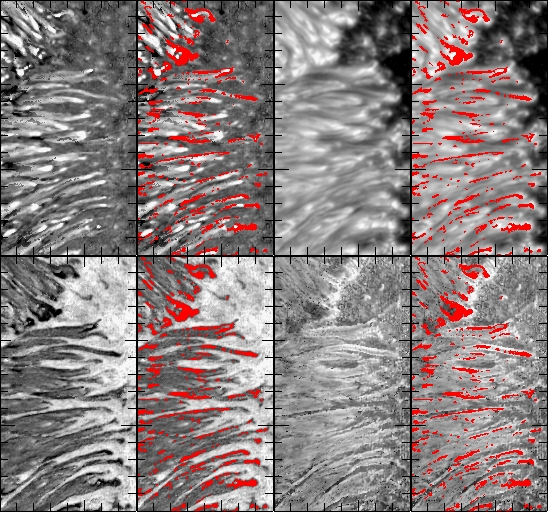} 
\put(1,50){\textcolor{white}{\textbf{\large a)}}}
\put(52,50){\textcolor{white}{\textbf{\large b)}}}
\put(1,2.5){\textcolor{white}{\textbf{\large c)}}}\index{\footnote{}}
\put(52,2.5){\textcolor{white}{\textbf{\large d)}}}
\end{overpic}

  \caption{
  Panels a-d show properties of the disk center side penumbra (excluding the outermost penumbra) at $\tau_c=$~0.95. Pixels with opposite polarity field in the \emph{solar} frame are indicated with red color. Panel a shows the LOS velocity (clipped at (-4.0, 2.5)~km~s$^{-1}$), panel b the temperature (clipped at (4500, 6500)~K), panel c the vertical magnetic field (in the solar frame; clipped at (-1000, 2200)~G), and panel d the horizontal magnetic field strength (in the solar frame; clipped at 1500~G). The FOV shown is 8$\times$15\arcsec. Tick marks are at 1\arcsec{} intervals.} 
  \label{fig:disk_center}
\end{figure}

\section{Properties of penumbral convection}
\subsection{Convective flows in spines and intra-spines}

Figure~\ref{fig:masks} shows the temperature at $\tau_c=$~0.95 overlaid by a mask identifying the penumbral part of the sunspot, and subdividing this into 6 radial zones, counting from the innermost zone and outwards. Figure~\ref{fig:incl_hist} shows a histogram of the magnetic field inclinations in the radial zones 1--5, excluding the end points of filaments protruding into the surrounding quiet Sun. The histogram is clearly bimodal, with peaks at approximately 45\degr{} and 80\degr{}, suggesting the existence of two components with fundamentally different properties. This is in agreement with simulations, showing a similar bimodal distribution of field inclinations with peaks around 45\degr{} and 85\degr{} (Rempel, private communication). In the following, we identify the magnetic spines and intra-spines as having inclinations less than and larger than 60\degr, respectively.

Figure~\ref{fig:velocities} shows the LOS velocities obtained with NICOLE at $\tau_c=$~0.95 in the intra-spines (top) and spines (bottom). Within the intra-spines, the LOS velocity shows a clear signature of horizontal (Evershed) flows: blue-shifts in the disk center direction and red-shifts in the limb direction. This signature appears entirely absent in the intra-spines. Thus it is clear that the flow properties of spines and intra-spines must be fundamentally different for this sunspot, as found already by \citet{2011Sci...333..316S} and \citet{2012A&A...540A..19S}, using a spatially high-pass filtered LOS magnetic field map to identify spines and intra-spines. That the Evershed flow resides in the nearly horizontal component of the penumbral magnetic field is consistent with earlier direct measurements \citep[e.g.][]{1993ApJ...403..780T} and inferences from 2-component inversions of spectropolarimetric data obtained at much lower spatial resolution than here \citep[e.g.][]{1993A&A...275..283S, 2000A&A...361..734M, 2002A&A...381..668S, 2003A&A...403L..47B, 2004A&A...427..319B, 2007ApJ...666L.133B, 2007ApJ...671L..85T}. Here, we distinguish between spines and intra-spines using measurements of the field inclination from spectropolarimetric data recorded at extremely high spatial resolution, by relying on 1-component inversions with NICOLE. 

To quantify properties of the vertical and radial flows in the spines and intra-spines, we follow the approach of \citet{2011Sci...333..316S} and \citet{2012A&A...540A..19S} and fit the variations of the LOS velocity $v_{\rm LOS}$ with the azimuth angle $\phi$ (set to zero in the disk center direction) to 
\begin{equation}
 v_{\rm LOS} = -v_r \cos \phi~\sin \theta + v_z \cos \theta,
\end{equation}
\citep{1952MNRAS.112..414P}, where $v_r$ and $v_z$ are the radial and vertical velocities and $\theta$ the heliocentric distance (for this sunspot, 15\degr). The sign conventions are that outward radial flows, downward flows, and LOS flows away from the observer are counted positive. Applying a high-pass spatial filter to the temperature map at $\tau_c=$~0.95, we remove any large-scale radial and azimuthal variations in temperature and obtain a map showing only small-scale ``local'' temperature fluctuations, $\delta T$. We divide these data into 5 temperature bins and make the azimuthal fits separately for each temperature bin and radial zone. The corresponding fitted data show large scatter \citep[see][their Fig. 7, for plots showing similar data]{2012A&A...540A..19S}, but the fitted parameters in Eq. (2) are well determined, as demonstrated in particular by the smoothness of their radial variations. The results are summarized in Fig.~\ref{fig:velocities2}, showing the variation of the vertical (left column) and radial (right column) average velocities with $\delta T$ for radial zones 2--5. The corresponding average velocity obtained when including all pixels in each radial zone is shown with dashed lines.

The results agree qualitatively with those found earlier \citep{2011Sci...333..316S,2012A&A...540A..19S}. \emph{The vertical flows show a clear convective signature everywhere in the penumbra}, cool structures on the average show downflows and hot structures upflows. These signatures are similar for spines and intra-spines, except that the hottest structures show stronger upflows in spines than in intra-spines. The radial flows are strong in intra-spines and very weak in spines, as expected from Fig.~\ref{fig:velocities}, except that the hottest structures in the spines show radial velocities that are nearly as strong as in the intra-spines. However, the overall variations in vertical velocity with temperature are smaller than found earlier \citep{2011Sci...333..316S,2012A&A...540A..19S}. Redoing the NICOLE inversions with 3 nodes for velocity instead of 2 nodes shows stronger downflows and upflows, clearly suggesting that using only 2 nodes for the velocity leads to underestimates of the LOS velocities in the deep layers of the penumbra. However, using 3 nodes also leads to strongly enhanced noise in the obtained LOS velocities, such that we restrict most of the following analysis to inversions made with 2 nodes for the LOS velocity.

The azimuthal fits also allow an estimate of the RMS vertical convective velocities in the penumbra. Removing the contributions from the fitted radial velocities in Eq.~(2), gives $v_z^{\rm rms}=$~(1.1, 1.1, 1.1, 1.2, 1.4, 1.4)~km~s$^{-1}$ in the intra-spines, and (0.7, 1.1, 1.4, 1.5, 1.6, 1.6)~km~s$^{-1}$ in the spines, for radial zones 1--6, respectively. These RMS convective velocities are similar to our previous estimate of 1.2~km~s$^{-1}$ and certainly sufficiently large to heat the penumbra \citep{2011Sci...333..316S}. The larger RMS velocities found in the spines come from the contributions of the very strong LOS velocities in the hottest structures.

\subsection{Spatial locations of penumbral opposite polarity field}
The histogram for $\tau_c=$~0.95 in Fig.~\ref{fig:incl_hist} shows an extended tail for inclinations larger than 90\degr, corresponding to a significant fraction of opposite polarity field in the penumbra. We find that (28, 20, 19, 26, 38, 47) percent of the pixels in the intra-spines have an inclination larger than 95\degr, and (21, 14, 13, 19, 31, 40) percent larger than 100\degr in radial zones 1--6 resp. Note that these percentages exclude pixels corresponding to the spines.

Figure~\ref{fig:opp_pol} shows in red color locations where $B_z <$~-100~G at $\tau_c=$~0.95, i.e. corresponds to opposite polarity field in the solar frame. In green color, we show also the magnetic field for which the inclination is less than 60\degr{}, corresponding to our definition of the magnetic spines. In the quiet Sun, the noise in the Stokes $Q$ and $U$ data results in a mean strength of the horizontal field of nearly 200~G.

Figure~\ref{fig:opp_pol} shows many examples of opposite polarity field far inside the outer boundary of the penumbra. Away from the outermost parts of the penumbra, it is evident that \emph{most of these opposite polarity patches are located close to and are aligned with the boundaries of the penumbral spines}. 

Figure~\ref{fig:opp_pol2} shows the opposite polarity field in the \emph{observers frame}, i.e., before transforming the field to the solar frame. Green pixels show locations where $B_{\rm{LOS}}$ at $\tau_c=$~0.95 is larger than 1~kG, red pixels where $B_{\rm{LOS}}$ is less than -150~G, i.e., has opposite polarity. Also on the \emph{disk center side penumbra}, many spines are flanked by (small) patches of opposite polarity field, and we must conclude that some of this opposite polarity field dives down at a rather steep angle with respect to the horizontal plane.

\subsection{Vertical and radial flows in opposite polarity patches}
Figure~\ref{fig:downflows} summarizes the results of azimuthal fits made for opposite polarity patches having a vertical field (in the solar frame) of less than -100~G (corresponding to opposite polarity field), shown in red color in Fig.~\ref{fig:opp_pol}. The results obtained with inversions using 2 nodes for the LOS velocity are shown in the left column, and with 3 nodes in the right column. We show the average vertical and radial velocities and the average flow field inclination for these patches (circles) and for all remaining intra-spine pixels in each radial zone (plus symbols). Relative to the average vertical velocity for each radial zone, \emph{the opposite polarity patches show downflows of about 0.6--1.3~km~s$^{-1}$ in all radial zones}. We note that our wavelength calibration (Sect.~2) gives average velocities for the umbra that are only 60~m~s$^{-1}$ at $\tau_c=$~0.95, such that this is a robust result. The bottom panels in Fig.~\ref{fig:downflows} shows that the opposite polarity downflows are directed more downward by about 10--20\degr{} relative to their surroundings, except in the outermost radial zone. The outward radial velocities of the opposite polarity patches are very similar to those of the surroundings.

We note that the inclination of the opposite polarity downflows is around 100--120\degr{} in the mid and outer penumbra, such that they are well aligned with our LOS angle of 105\degr{} on the limb-side penumbra (at this heliocentric angle of 15\degr{}). In addition, upflows show a strongly reduced LOS velocity at the limb side. This makes the limb side ideal for seeing opposite polarity patches harboring downflows. Figure~\ref{fig:limb_side} shows in panel a the LOS velocity at $\tau_c=$~0.95, without (left) and with (right) the red mask showing the locations of the opposite polarity field. It is obvious that \emph{the opposite polarity patches are located in narrow radially extended downflow lanes}, as confirmed by the azimuthal variations of their LOS velocities. The vertical magnetic field map (panel c) shows the strong tendency of these opposite polarity patches (and thus also the downflow lanes) to be adjacent to the spines. Panel b illustrates that opposite polarity patches cannot easily be associated with cold or hot structures at $\tau_c=$~0.95, rather they seem to be located at flanks of the filamentary structures. Finally, panel c shows the absence of spine structure in the horizontal magnetic field map. 

In Fig.~\ref{fig:plots} we show the \emph{average} Stokes profiles for all opposite polarity patches (left column) and all remaining pixels (right column), within the FOV in Fig.~\ref{fig:limb_side}. The Stokes $V$ profile is weak and abnormal in the opposite polarity patches, and have a third ``extra'' lobe in the red wings of both lines. Figure~\ref{fig:plots} also shows (as dashed lines) the correspondingly averaged fitted synthetic profiles returned by NICOLE\footnote{Note: we first fit the profiles with NICOLE, and then average the fitted profiles.}. Quite evidently, \emph{the fitted profiles reproduce even the observed abnormal $V$ profiles rather well.} 

Figure~\ref{fig:plots2} shows the observed and fitted synthetic Stokes profiles for a downflow lane pixel marked with ``+'' in Figs.~\ref{fig:limb_side}. The Stokes V profile is abnormal with 3 lobes instead of 2, but is nevertheless fitted by the variations of the LOS velocity and \emph{vertical} magnetic field (in the \emph{local} frame) with $\tau_c$, shown in the lower panels. The polarity of the vertical magnetic field reverses below $\tau_c=0.1$, and the magnitude of the horizontal field increases with height.

Abnormal and 3-lobed penumbral Stokes $V$ profiles have been reported recently \citep[cf.][]{2011arXiv1107.2586F, 2013A&A...550A..97F}, and were found to be associated with downflows and opposite polarity field in the deep layers, but most such profiles were found in the \emph{outermost} parts of the penumbra and none in the inner penumbra. \emph{Here, we find clear evidence of opposite polarity field even in the innermost penumbra.} 

Figure~\ref{fig:disk_center} shows the same layout as in Fig.~\ref{fig:limb_side}, but for the disk center side of the spot. This confirms the strong tendency for the opposite polarity patches to be located adjacent to the boundaries of the spines. The strongest Doppler signatures at the disk center sides are from the bright upflows, and panel a shows several examples where the opposite polarity field is located at the sides (flanks) of the bright upflows, which is according to expectations.

\section{Conclusions}
By calibrating the SST/CRISP transmission profile, and feeding this information pixel by pixel to the inversion code NICOLE, we have been successful in simultaneously estimating LOS velocities, the magnetic field vector and their gradients in a sunspot penumbra from combined 630.15~nm and 630.25~nm Stokes spectra, at a spatial resolution close to 0\farcs15. The inversions return good fits to the observed Stokes profiles, even when the $V$ profiles are abnormal with 3 lobes.

Using these data, we validate the inversions through the returned variations of temperature, LOS velocity and magnetic field with optical depth for different magnetic and non-magnetic structures within the FOV. We find that that the magnetic field in the penumbral spines expands over the intra-spines with height, in agreement with predictions of the convective gap model \citep{2006A&A...447..343S,2006A&A...460..605S} and numerical simulations \citep[][his Fig. 22]{2012ApJ...750...62R}. Moreover, our data show the ubiquitous presence of narrow and \emph{radially extended} opposite polarity patches in the entire penumbra. 
We find that these patches are located in the intra-spines (where the magnetic field is nearly horizontal), and predominantly close to the boundaries to the magnetic spines, where the magnetic field has a strong vertical component. \citet{2013A&A...550A..97F} recently reported on penumbral opposite polarity field that appears ``patchy'' rather than radially elongated, and considered this to support an interpretation of the associated flows in terms of arching magnetic flux tubes. Our finding that the opposite polarity patches are radially extended and preferentially located close to the magnetic spines rules out such an explanation.

Analyzing the azimuthal variations of the LOS velocity for the opposite polarity patches, we find that they harbor downflows of typically 1~km~s$^{-1}$ near $\tau_c=$~0.95, consistent with convective downflows dragging down some field-lines. The locations of the opposite polarity downflows agree perfectly with predictions of the convective gap model \citep{2006A&A...460..605S} and recent simulations \citep[][his Fig. 22]{2012ApJ...750...62R}.

The \emph{horizontal} field inferred from our data is found to be nearly as strong in the intra-spines as in the spines. This is in agreement with the simulations of \citet{2012ApJ...750...62R}, showing a layer of strong horizontal magnetic field near $\tau_c=$~1, and weak-field gaps located below the visible surface. Neither these nor other observations allow conclusions about the sub-photospheric layers, but the distinctly different morphologies of the penumbral vertical and horizontal magnetic field at the limb side, shown in Figs.~\ref{fig:bz} and ~\ref{fig:bt}, are remarkably similar to those of the simulations \citep[][his Figs. 6b and 6c]{2012ApJ...750...62R}.

Our results also confirm earlier measurements in the deeply formed \ion{C}{I} line \citep{2011Sci...333..316S} and the \ion{Fe}{I} 630.15~nm line \citep{2012A&A...540A..19S}. Throughout the entire penumbra, we detect a vertical velocity field that shows the expected correlation with temperature: hot gas moves up and cool gas down. Removing the contribution from the radial flows to the measured LOS velocities, we estimate an RMS velocity of 1.2~km~s$^{-1}$ at $\tau_c=$~0.95 for this vertical component of the penumbral convection. These vertical velocities are sufficiently large to explain the penumbral radiative heat loss \citep{2011Sci...333..316S}.

\begin{acknowledgements}
We thank an anonymous referee for valuable comments and suggestions.
 The Swedish 1-m Solar Telescope is operated on the island of La Palma by the Institute for Solar Physics of the Royal Swedish Academy of Sciences in the Spanish Observatorio del Roque de los Muchachos of the Instituto de Astrof\'isica de Canarias.
\end{acknowledgements}


\end{document}